\documentclass[preprint,number,sort&compress]{elsarticle}
\usepackage{color}
\usepackage{amsmath}
\usepackage{graphicx}

\journal{Computer Physics Communications}

\makeatletter

\newcommand{\noun}[1]{\textsc{#1}}

\makeatother

\begin{document}

\begin{frontmatter}
\title{LAMMPS Framework for Dynamic Bonding and an Application Modeling
DNA}

\author[zqex]{Carsten Svaneborg\corref{cor}}
\cortext[cor]{Corresponding author}

\ead{science@zqex.dk}

\address[zqex]{Center for Fundamental Living Technology, Department of Physics
and Chemistry, University of Southern Denmark, Campusvej 55, DK-5320
Odense, Denmark}
\begin{abstract}
We have extended the Large-scale Atomic/Molecular Massively Parallel
Simulator (LAMMPS) to support directional bonds and dynamic bonding.
The framework supports stochastic formation of new bonds, breakage
of existing bonds, and conversion between bond types. Bond formation
can be controlled to limit the maximal functionality of a bead with
respect to various bond types. Concomitant with the bond dynamics,
angular and dihedral interactions are dynamically introduced between
newly connected triplets and quartets of beads, where the interaction
type is determined from the local pattern of bead and bond types.
When breaking bonds, all angular and dihedral interactions involving
broken bonds are removed. The framework allows chemical reactions
to be modeled, and use it to simulate a simplistic, coarse-grained
DNA model. The resulting DNA dynamics illustrate the power of the
present framework.
\end{abstract}
\end{frontmatter}

\section{Introduction}

When performing molecular dynamics simulations, we distinguish between
bonded and non-bonded interactions.\cite{leach2001molecular,frenkel2002understanding}
Effectively, this means that the interactions have been coarse-grained
on the energy scale of the simulation. Certain degrees of freedom
are frozen, and we describe them as being permanent bonded. Other
degrees of freedom remain dynamic, and we describe them with relatively
weak non-bonded interactions. However, this situation is less clear
when simulating systems undergoing chemical reactions where bonds
are created or broken. Another example is DNA molecules where hybridization
bonds are broken at high temperatures and reformed when cooling the
system. For such systems, it can be computationally more efficient
to model these degrees of freedom as being dynamically bonded.

The problem of bond dynamics is closely related to the question of
how to represent chemical reactions in a molecular dynamics simulation.
Reactive force fields such as ReaxFF and empirical valence bond (EVB)
can be used to model chemical reactions.\cite{van2001reaxff} Bond
order potentials are interesting since they allow three body interactions
in the neighborhood of a bond to modify the strength of the bond.\cite{pettifor1989new}
When coarse-graining systems capable of chemical reactions, it is
important to note that the reaction radius and probability also has
to be appropriately coarse-grained.\cite{RosaEveraers} When the
bonds become dynamic, this also induces a dynamic for the angular
and dihedral interactions. When breaking a bond, all angular and dihedral
interactions involving that bond become invalid, and should be removed.
Similarly, when creating a bond, we have to identify which angular
and dihedral interactions to create in the bond neighborhood. This
ensures that after melting and renaturing of a system, it is again
governed by the same set of interactions and return to the same equilibrium
structure.

DNA molecules are comprised of the four bases adenine (A), cytosine
(C), guanine (G), and thymine (T). The bases are attached to a 2-deoxyribose
sugar ring. For naturally occurring DNA, sugar rings are linked to
each other through phosphodiester bonds, that connect the 3' to 5'
carbons in consecutive sugar rings. This builds a molecular directionality
into the back bone of a DNA strand, which will have a 3' and a 5'
end. The strand is also characterized by a specific sequence of bases.
Together the phosphate backbone, the sugar and the base is denoted
a nucleotide, which is the repeat unit of a single DNA strand. A-T
and C-G are Watson-Crick pairs and can form hydrogen bonds with each
other. The energetically favorable stacking interactions allow two
complementary single strands to form 3'-5'/ 5'-3' anti-parallel aligned
double strands. Double stranded DNA can be melted and renatured by
repeated cycling the temperature around the melting point or by varying
solvent conditions.

DNA is a very complex molecule and numerous models exists to describe
behavior from atomistic properties to mesoscopic mechanical properties.
The molecular structural details of short DNA oligomers can be studied
with atomistic molecular dynamics simulations such as Amber\cite{case2010amber,cheatham2000molecular}
and Charmm\cite{CHARMM2009,mackerell2000development}. However, when
we want to understand the large scale properties of DNA molecules
or materials in which DNA molecules are a component, coarse-grained
DNA models are essential. Coarse-graining is the statistical mechanical
process by which uninteresting microscopic details are systematically
removed, leaving a coarse-grained, effective model that is described
by an effective free energy functional.\cite{moller2002coarse,nielsen2004coarse,langowski2006polymer,de2011polymer}
A major advantage of coarse-grain models are that we can use them
to simulate the interesting large scale dynamics of a system directly
without wasting time on uninteresting details. This allows larger
systems to be studied for longer times which paves the way for studying
e.g. the properties materials rather than single molecules.

A number of coarse-grain DNA molecular dynamics models exists. In
the {}``three site per nucleotide'' model of de Pablo and co-workers,
a single nucleotide is represented by a phosphate backbone site, a
sugar group site, and a base site, respectively\cite{Knotts3sitemodel2007,Sambriski2009,sambriski2009sequence,FlorescuJoyeux2011}.
The model uses an implicit representation of counter ions at the level
of Debye-Hückel theory, but has recently been generalized to explicit
counter ions.\cite{FreemanHinckleyPablo_CoarseGrainDNA3site} A version
of this model has also been generalized to include non-Watson-Crick
base pairing such as Hoogsteen pairing.\cite{LinakTourdotDorfman2011}
There is also a number of {}``two site per nucleotide'' models where
one site represents the back bone and the sugar ring. The other site
represents the base.\cite{drukker2000model,KenwardDorfman2009,ouldridge2010dna,LinakDorfman2010,OulridgeLouisDoye2011}
One challenge to {}``one site per nucleotide'' models are to represent
the DNA double helix. Savelyev and Papoian\cite{SavelyevPapoian2011,Savelyev2012}
does this by special {}``fan'' shaped pair-interactions between
a bead and a large number of beads on the opposite strand. This model
does not allow for DNA melting. Trovato and Tozzini\cite{TrovatoTozzini2008}
produce a helical structure using angular and dihedral interactions
along the double strand. In the case where the large scale DNA mechanical
properties are of interest, it can be advantageous to coarse-grain
a whole base-pair to a single rigid ellipsoidal or plate-shaped object
and regard DNA as a latter-like chain of such objects.\cite{MergelEveraers_PhysRevE.68.021911,BeckerEveraers_PhysRevE.76.021923}
Here the coarse-graining has eliminated the melting and renaturation
dynamics all together. Other types of coarse-grain DNA models are
applied to study behavior of DNA functionalized nano-particles. The
DNA molecules can e.g. be modeled as rigid rods with a single sticky
site on one end and tethered to the surface of the nano-structure
by the other end\cite{LeunissenFrenkel2011}, as semi-flexible polymers
with attractive sites on the monomers\cite{hsu2010theoretical},
or the whole DNA molecule can be modeled as a single sticky site that
can be hybridized to complementary free sticky sites.\cite{Martines-VeracoecheaPRL2011}
Here the coarse-graining has completely eliminated the chemical structure,
while the melting, renaturing, and sequence specificity has been retained
in the dynamics.

The two most prevalent statistical mechanical models of RNA and DNA
melting are the Poland-Scheraga\cite{poland1966phase,jost2009unified}
(PS) and the Dauxois-Peyrard-Bishop\cite{peyrard2008modelling} (DPB)
models. The Poland-Scheraga model describes DNA as a 1D lattice model
where a base-pair can either be hybridized or open. The free energy
expression for the PS model contains empirical stacking free energies
each stack of hybridized base-pairs as well as contributions from
the strand configuration entropy due to internal bubbles, frayed ends
and empirical initiation terms. The DPB model also describes DNA as
a 1D lattice model, but where each base-pair is characterized by a
continuous base-pair distance. Contrary to the PS model, the DPB model
has a Hamiltonian where the base-base potential is described by an
anharmonic potential representing hydrogen bonding, and deviations
between nearest neighbor base-pair extensions are penalized by a harmonic
term. A generalization of the PS model exists, where the strand conformations
are represented explicitly as lattice polymers. This provides a conceptual
simplification since the conformational entropy of bubbles and frayed
ends emerges naturally from the polymer model. This real space lattice
PS model has been studied using exact enumeration techniques\cite{EveraersKumarSimm2007},
a version of the model has also been applied to study RNA folding
using Monte Carlo simulations.\cite{JostEveraers2010} 

The dynamic bonding framework allows us to study classes of DNA models
where hybridization bonds, angular bonds, and dihedral bonds are created
and broken dynamically. These dynamic bonding DNA models are intermediates
between the real space lattice PS models, the coarse-grained molecular
dynamics models, and the sticky DNA models described above. In the
PS model, base pairs can either be hybridized or open and are characterized
by a corresponding free energy. In a dynamic bonding model, base pairs
will be either hybridized or open and a free energy will also characterize
this transition. In the coarse-grained molecular dynamics models and
the DPB model, base pairs are represented by a continuous non-bonded
pair-potential. In the dynamic bonding DNA models, base pairs are
characterized by a continuous bond potential. The dynamic bond DNA
models can also be regarded as being off-lattice generalizations of
the real space lattice PS model, where a single strand is described
as a semi-flexible bead-spring polymer
where complementary monomers will form hybridization bonds when they
are close. The dynamic bonded DNA models are {}``one site per nucleotide''
models, but we can also lump sequence of nucleotides into a single
coarse-grained bead. In this case, we can as a first approximation
assume that only beads representing complementary sequences can hybridize,
and that the breaking of a hybridization bond corresponds to the creation
of a DNA bubble. This would be a {}``many nucleotides per site''
dynamic bonding DNA model more akin to the sticky site DNA models
used to study DNA functionalized nano-particles.

The dynamic bonded DNA models ensure anti-parallel strand alignment
in the double strand state, through the interplay between the dihedral
interactions and the directional bonds. Such degrees of freedom are
absent from both the PS and DPB 1D lattice models. The coarse-grained
models use angular and dihedral interactions to ensure a structure
resembling the real chemical structure of DNA molecules. In dynamic
bonded DNA models, the angular and dihedral interactions are dynamically
introduced when hybridization bonds are formed to promote a zipper-like
closing dynamic. Similarly angular and dihedral interactions are dynamically
removed as hybridization bonds are broken to promote zipper-like opening
dynamic. Hence in dynamic bond DNA model, we utilize the interplay
between dynamic bonded, angular, and dihedral interactions to model
cooperative effects in the DNA bubble and zippering dynamics, rather
than to model chemical structure. 

The simplicity and success of the PS model in predicting sequence
specific DNA melting temperatures suggests that the essential physics
of DNA hybridization, melting and renaturing can, in fact, be accurately
captured in a model without chemical details, and where the key property
is the dynamics of hybridization. This is our motivation for developing
the dynamic bonding framework. We will use it to develop and apply
models to study the properties of hybrid materials containing both
DNA molecules and soft-condensed matter.

We have implemented directional bonds and dynamic bonding in the Large-scale
Atomic/Molecular Massively Parallel Simulator\cite{Lammps}
(LAMMPS). LAMMPS is a versatile, parallel, highly optimized, open
source code for performing Molecular Dynamics (MD) and Dissipative
Particle Dynamics (DPD) simulations of coarse-grained models. Due
to the modular design, LAMMPS is easy to extend with new interactions
and functionality. The dynamic bonding implementation is also modular
and easy to extend with new functionality. Our extension is by no
means limited to modeling DNA, but could equally well be used for
simulations of chemical reactions such as living polymerization, cross-linking
of stiff polymers, coarse-grained dynamics of worm-like micelles and
active driven materials. A snapshot of the LAMMPS code with the directional
bonds and dynamic bonding implementation can be obtained from the
CPC Program Library. Included with the code is also the documentation
necessary for porting the directional and dynamic bonding framework
to future LAMMPS versions. 

Sect. \ref{sec:Implementation1} is a summary of the implementation
of directional bonds and the dynamic bonding framework. We present
a simplified DNA model based on the dynamic bonding framework in sect.
\ref{sec:DNA-model}, which is provides the examples of DNA dynamics
shown in sect. \ref{sec:Illustrations}. We conclude with our conclusions
in sect. \ref{sec:Conclusions}, and present the details of the directional
bonds and dynamic bonding implementation in an appendix.

\section{Implementation\label{sec:Implementation1}}

Double stranded DNA only exists in a state where the two strands are
aligned anti-parallel 3'-5'/5'-3'. In order distinguish between parallel
and anti-parallel strand alignment, we regard the 3'-5' back bone
structure as a property of the back bone bonds, which become directional.
This is necessary since the chemical structure of the nucleotides
have been coarse-grained to a single structureless site. The directional
bonds will also play a crucial role when introducing angular and dihedral
interactions in a double stranded DNA molecule, since this affects
the stability, zippering dynamics, and mechanical properties.

To implement directional bonds in LAMMPS, we make use of the fact
that Newtons 3rd law is optional when calculating bond forces. When
Newtons 3rd law is enabled, each bond force is only calculated once,
but subsequently has to be communicated to the bond partner. When
it is disabled, LAMMPS calculates the bond force twice, once for each
of the two bond partners. In this case, each of the two bond partners
store information about the bond type and the identity of the other
bond partner. We can denote this situation by $A\substack{t\\
\rightarrow
}
B$ and $A\substack{t\\
\leftarrow
}
B$, which shows that the $A$ bead stores $t$ as the type of the bond
to $B$, and the $B$ bead stores $t$ as the type of the bond to
$A$. With a few modifications, LAMMPS will load and store different
bond types in the two bond partners. Hence, we can have $A\substack{t\\
\rightarrow
}
B$ and $A\substack{s\\
\leftarrow
}
B$, where the bond type $s$ from $B$ to $A$ and the bond type $t$
from $A$ to $B$ differs. When the two bond types refer to the same
bond potential, Newtons 3rd law still applies, and the dynamics is
unaffected. However, we can interpret the pattern of bond types as
the directionality of the strand. Note that if we instead use different
bond potentials in the two directions or only a {}``half'' bond,
the result would be a net force along the bond, which can be used
to model driven active matter. We shall not pursue this situation
further in the present paper.

\begin{figure}
Define a dynamic bond fix\noun{:}\\
\noun{fix} fixid beadgroup \noun{bonddynamics} \emph{everystep} {[}\noun{paircheck}13{]}
{[}\noun{paircheck}14{]} <list of rules>\\
\\
Each dynamic bonding rule is one of:\\
\noun{createbond} \emph{bondtype} \emph{beadtype1} \emph{beadtype2}
maxdistance probability\\
\noun{createdirbond} \emph{bondtype1} \emph{bondtype2} \emph{beadtype1}
\emph{beadtype2} maxdistance probability\\
\noun{breakbond} bondtype mindistance probability\\
\noun{convertbond} bondtype1 bondtype 2 probability\\
\noun{killbond} \emph{bondtype} mindistance\\
\noun{createangle} \emph{angletype} \emph{beadtype1} \emph{beadtype2}
\emph{beadtype3} \emph{bondtype1} \emph{bondtype2}\\
\noun{createdihedral} \emph{dihedraltype} \emph{beadtype1} \emph{beadtype2}
\emph{beadtype3} \emph{beadtype4} \emph{bondtype1} \emph{bondtype2}
\emph{bondtype3}\\
\noun{maxbond} \emph{bondtype} \emph{beadtype} \emph{maxnumber}

\caption{\label{fig:Syntax}LAMMPS syntax for the dynamic bonding fix, and
the types of rules currently implemented.}
\end{figure}

The dynamic bonding framework allows a number of rules to be specified,
that completely define the bond dynamics. These rules are applied
to a specified group of reactive beads with a specified frequency.
The application of the rules is conditional on the types of beads,
types bonds, distance between beads and length of bonds involved.
In particular, we have implemented rules for stochastic creation of
symmetric and directional bonds within a certain reaction distance,
stochastic removal of symmetric bonds larger than a breaking distance,
removal of all symmetric bonds exceeding a certain length, and stochastic
conversion of a symmetric bond from one type to another. Furthermore,
all bond creation rules ensure that a bead can never have more than
a specified number of bonds of a given type. The implementation is
structured such that it is easy to implement new types of rules.

Besides the bond dynamics, the consistency of the angular and dihedral
interactions should be ensured at all times. After bonds have been
broken, all invalid potential angular and dihedral interactions involving
broken bonds should also be removed. After bonds have been formed,
all triplets or quartets of beads that could be connected by at least
one new bond are checked to see if they require the creation of an
angular or dihedral interaction. We discard cyclic triplets and quartets
where the same bead appears more than once.

An angular creation rule specifies which angular interaction can be
introduced between a triplet of connected beads $A$, $B$, and $C$.
Since the triplets are not ordered, the rule should match either $ABC$
or $CBA$. To test if the $ABC$ bead order matches, we first compare
the types of the $ABC$ beads with the bead types the rule specifies.
We then compare the two bond types $t$ and $s$ with the bond types
the rule specifies, where the bond types are defined directionally
as $A\substack{t\\
\leftarrow
}
B$ and $B\substack{s\\
\rightarrow
}
C$. If the $ABC$ bead order did not match, it is repeated $CBA$ bead
order, where the bonds types are defined directionally as $C\substack{t\\
\leftarrow
}
B$ and $B\substack{s\\
\rightarrow
}
A$. If a rule match, then the specified angular interaction is introduced
between the three beads. A creation rule for a dihedral interaction
specifies four bead types and three bond types. Again we test both
$ABCD$ and $DCBA$ ordered bead quartets. First the bead types of
the quartet are compared to the bead types specified by the rule,
subsequently the bond types are compared, the bond types are defined
directionally as $A\substack{r\\
\leftarrow
}
B$, $B\substack{s\\
\leftarrow
}
C$, $B\substack{s'\\
\rightarrow
}
C$, and $C\substack{t\\
\rightarrow
}
D$. The bond types match if $r$, $s$ or $s'$, and $t$ matches the
three bond types specified by the rule. If the $ABCD$ bead order
did not match, it is repeated $DCBA$ order. If a dihedral rule match
a quartet of beads, the specified dihedral interaction is introduced
between the four beads. These rules allows us to selectively and dynamically
introduce angular and dihedral interactions taking both bead types
and directional bond types into account. Note that the same directionality
applies to matching the bead type and bond type patterns. 

To have an efficient parallel implementation, we implement the bond
creation and breaking by an pair matching algorithm inspired from
the bond/break and bond/create fixes already implemented in LAMMPS.
In the dynamic bonding fix, preferred bond creation/breakage partners
are identified in each simulation domain. This information is communicated
between and aggregated across neighbor simulation domains. Afterwards,
the bonds selected for breakage are removed. The local neighborhood
of all reactive beads are checked for angular and dihedral interactions,
that should be removed because they cross broken bonds. Then bonds
are created between partners selected for bonding. Again, we check
the local neighborhood of all reactive beads to introduce angular
and dihedral interactions. After this final step, we broadcast bond
statistics to all simulation domains. Note that due to the pair matching
algorithm, each bead can maximally have one bond created and broken
at each call to the dynamic bonding fix. All rules are applied to
a bead pair (in the specified order) when identifying if they are
eligible for matching. If multiple rules apply to the same bead pair,
the last matching rule will be always be chosen. Hence, if this last
rule has a very low reaction probability, it will completely shadow
more probable rules specified earlier. These shadowing issues does
not apply to the DNA model below, and will not play a role at low
concentrations of reacting beads. The details of the implementation
and shadowing issues are discussed in Appendix \ref{sec:Appendix-A}.

The LAMMPS syntax of the dynamic bond fix is shown in fig. \ref{fig:Syntax}.
The first line defines the name of the particular instance of the
fix, the group of reactive beads (\emph{beadgroup}), and how often
the bond dynamics fix is applied (\emph{everystep}). By default creation
rules only apply to potential bonding bead pairs, that are further
than 4 bonds apart or not bonded. The optional\emph{ Paircheck13}
and \emph{Paircheck14} switches includes 1-3 and 1-4 chemically distant
beads in the search of potential bonding partners. The line is followed
by a number of dynamic bonding rules. \emph{Createbond} rules specify
pairs of bead types, that can be bonded, if they are within a certain
maximum reaction distance from each other. If a bead has more than
one potential bond partners, the closest partner is chosen, and a
bond with the specified type is then created with the given probability.
\emph{Createdirbond} rules does the same as \emph{createbond}, but
creates a directional bond with the two specified bond types between
the two bead types. \emph{Breakbond} rules identifies bonded bead
pairs with bonds longer than the specified minimum distance and breaks
the bond with the specified probability. If a bead has more than one
potential bond break partner, then the most distant partner is chosen.
Since only a single bond can be removed per bead per call to the dynamic
bonding fix, a breakbond rule with unit probability does not ensure
that all bonds longer than the minimum distance are broken. Hence,
we have also implemented \emph{killbond} rules. These rules operate
directly on the bond structures, and are not limited by the pair matching
algorithm. \emph{Convertbond} rules stochastically convert symmetric
bonds of one type into another type. This is implemented as nominating
the bond pair for removal of the old bond, followed by creation of
the new bond. The dynamic bonding framework ensures that angular and
dihedral interactions across the bond are also converted accordingly.
\emph{Createangle} and \emph{Createdihedral} rules defines which angular
and dihedral interaction types should be created between triplets
and quartets of beads with the specified types of bead, and types
of bonds between the beads as discussed above. \emph{Createangle}
and \emph{createdihedral} rules do not specify a probability, since
they are created as required by the local neighborhood around new
bonds. Note that angular and dihedral interactions are only introduced
as a consequence bond creation events, they are not introduced between
already bonded beads even though the bead types and bond types match
the rule. When checking potential beads for bond creation, all \emph{Maxbond}
rules are checked to discard beads that already have the maximal number
of the specified bond types.

\section{DNA model\label{sec:DNA-model}}

We have chosen the present DNA model has been chosen because it produces
a simple ladder like equilibrium structure, which allows us to illustrate
the power of the dynamic bonding framework, and to visualize all the
interactions that are dynamically introduced and removed. Real DNA
molecules performs a whole twist every 10.45 base pairs, and to model
twist we need a somewhat more complex force field, but exactly the
same dynamic bonding rules. Because we are interested in studying
DNA programmed self-assembly, we choose to use Dissipative Particle
Dynamics (DPD)\cite{hoogerbrugge1992simulating,espanol1995statistical}.
DPD is given by a force field comprising a conservative soft pair-force
$F^{C}$, a dissipative friction force $F^{D}$, and a stochastic
driving force $F^{R}$ given by

\[
{\bf F}_{ij}=\left(F^{C}+F^{R}+F^{D}\right)\frac{{\bf r}_{ij}}{r}\quad\mbox{for}\quad r=|{\bf r}_{ij}|<r_{c}
\]
where the forces contributions are given by

\[
F^{C}=aw(r)\quad\quad F^{D}=-\frac{\gamma w^{2}(r)}{r}\left({\bf r}_{ij}\cdot{\bf v}_{ij}\right)\quad\quad F^{R}=\frac{\sigma w(r)\xi}{\sqrt{\Delta t}}.
\]

Here ${\bf r}_{ij}={\bf r}_{i}-{\bf r}_{j}$ and ${\bf v}_{ij}={\bf v}_{i}-{\bf v}_{j}$
denotes the separation and relative velocity between two interacting
beads $i$ and $j$, respectively. $\xi$ denotes a Gaussian random
number with zero mean and unit variance, and the thermostat coupling
strength is $\sigma=\sqrt{2k_{B}T\gamma}$. The weighting function
is $w(r)=1-\frac{r}{r_{c}}$. We integrate the DPD dynamics with a
Velocity Verlet algorithm with a time step $\Delta t=0.01\tau$. The
unit of energy is $\epsilon=k_{B}T$, where we chose to set Boltzmann's
constant to unity, such that temperature is measured in energy units.
We use $T=1\epsilon$ in all of the simulations except the DNA bubble
simulation where $T=5\epsilon$. The unit of length $\sigma$ is defined
by the pair force cut-off $r_{c}=1\sigma$. The mass is $m=1$ for
all beads, this allows us to define the unit of time as $\tau=\sigma\sqrt{m/\epsilon}$.
The DPD pair-force parameter is $a=25\epsilon\sigma^{-1}$ between
all species of beads. The viscosity is $\eta=100\epsilon\tau\sigma^{-2}$.
Non-bonded pair interactions are switched off between beads in molecules
that are less than 3 bonds apart. The DNA molecule is simulated in
an explicit solvent at a density $\rho=3\sigma^{-3}$.

We represent a nucleotide by a single DPD bead, and let the four ATCG
nucleotides correspond to bead types 1-4, respectively. They are colored
red, green, blue, and magenta, respectively, in figures below. Red
and green beads (A-T) are complementary as are blue and magenta (C-G)
beads. A single strand of DNA is represented as a string of beads
joined by permanent directional back bone bonds. The two 3' to 5'
and 5' to 3' backbone bond potentials (bond type 2 and 3, respectively,
colored green and blue in the bond visualizations) are given by the
same potential

\[
U_{backbone}(r)=\frac{U_{min}}{(r_{l}-r_{0})^{2}}\left((r-r_{l})^{2}-(r_{0}-r_{l})^{-2}\right),
\]
with $U_{min}=10.0\epsilon$, $r_{l}=0.3\sigma$, and $r_{0}=0.6\sigma$.
The hybridization bond potential (bond type 1, colored red in the
bond visualizations) is given by

\[
U_{hyb}(r)=\begin{cases}
\frac{U_{min}}{(r_{h}-r_{0})^{2}}\left((r-r_{0})^{2}-(r_{h}-r_{0})^{-2}\right) & \quad\mbox{for}\quad r<r_{c}\\
0 & \quad\mbox{for}\quad r\geq r_{c}
\end{cases},
\]
with $U_{min}=1.0\epsilon$, $r_{h}=0.6\sigma$, and $r_{c}=1.0\sigma$. 

Besides the DNA interactions, we need to define the bonding dynamics
of the DNA beads. The corresponding dynamic bonding fix command is
shown in fig \ref{fig:Fix-DNA-dynamics}. Hybridization bonds are
created with probability one when two complementary beads are within
a distance of $r_{h}$. Bead type $2$ and $3$ are able to form a
5' 3' backbone bond when they are within a distance of $r_{l}=0.3\sigma$
from each other. The probability of creation of a back bone bond is
$0.1$. This is a simplification for the oligomer-template simulation
below. Only hybridization bonds can be broken, and they are removed
if they are longer than $r_{c}=1\sigma$. To control hybridization,
we only allow all bead types ($*$) to have maximally one hybridization
bond (type 1), one 3' end (type 2) and one 5' end (type 2) of a back
bone bond. In the model all nucleotides has the same interactions,
hence use $*$ for all the bead types rule specifications.

\begin{figure}
1: fix dnadyn dna bonddynamics 1 paircheck14

2: \noun{createbond} \textcolor{red}{1} \textbf{\textcolor{red}{1}}
\textbf{\textcolor{green}{2}} 0.6 1.0

3: \noun{createbond} \textcolor{red}{1} \textbf{\textcolor{blue}{3}}
\textbf{\textcolor{magenta}{4}} 0.6 1.0

4: \noun{createdirbond }\textcolor{green}{2} \textcolor{blue}{3} \textbf{\textcolor{green}{2}}
\textbf{\textcolor{blue}{3}} 0.3 0.1

5: \noun{killbond} \textcolor{red}{1} 1.0

6: \noun{maxbond} 1 {*} \textcolor{red}{1}

7: \noun{maxbond} 1 {*} \textcolor{green}{2}

8: \noun{maxbond} 1 {*} \textcolor{blue}{3}

9: \noun{createangle} \textit{\textcolor{red}{1}} {*} {*} {*} \textcolor{green}{2}
\textcolor{blue}{3}

10: \noun{createangle} \textit{\textcolor{green}{2}} {*} {*} {*} \textcolor{red}{1}
\textcolor{green}{2},\textcolor{blue}{3}

11: \noun{createdihedral} \textit{\textcolor{red}{1}} {*} {*} {*}
{*} \textcolor{red}{1} \textcolor{green}{2},\textcolor{blue}{3} \textcolor{red}{1}

12: \noun{createdihedral} \textit{\textcolor{green}{2}} {*} {*} {*}
{*} \textcolor{green}{2} \textcolor{red}{1} \textcolor{blue}{3}

13: \noun{createdihedral} \textit{\textcolor{blue}{3}} {*} {*} {*}
{*} \textcolor{green}{2} \textcolor{red}{1} \textcolor{green}{2}

14: \noun{createdihedral} \textit{\textcolor{blue}{3}} {*} {*} {*}
{*} \textcolor{blue}{3} \textcolor{red}{1} \textcolor{blue}{3} 

\caption{\label{fig:Fix-DNA-dynamics}LAMMPS dynamic bonding fix for producing
the DNA dynamics shown in figs. \ref{fig:example-hybridization}-\ref{fig:example-bubble}.
Bond types are shown with plain digits (hybridization: red 1, back
bone 3' bonds: green 2, and back bone 5' bonds: blue 3). Bead types
are shown with bold digits represent nucleotides (A:red 1, T:green
2, C:blue 3, G: magenta 4). Angular and dihedral bond types are shown
italic digits corresponding to the interaction type numbers. The bead
and interaction type colors correspond to those used in the visualizations.
{*} is the wild card and is used to match any bead or bond type.}
\end{figure}

The model has two angular interactions, which are described by the
potential $U(\theta)=K(\theta-\theta_{0})^{2}$, where $K$ defines
the angular spring constant and $\theta_{0}$ the equilibrium angle.
The first angle interaction (type 1) promotes a straight angle between
back bone bonds. This interaction is shown as red angles in the angle
visualizations, and it has parameters $K=20\epsilon$ and $\theta_{0}=180\text{º}$.
Type 1 angles are dynamically introduced for bonding patterns $A\substack{3'\\
\leftarrow
}
B$, $B\substack{5'\\
\rightarrow
}
C$ and $A\substack{5'\\
\leftarrow
}
B$, $B\substack{3'\\
\rightarrow
}
C$ (i.e. for model bonds types $2$ $3$, since $CBA$ order matches
$3$ $2$). The second angle interaction (type 2) promotes a right
angle between back bone and hybridization bonds. This interaction
is shown as green angles in the angle visualizations, and it has $K=1\epsilon$
and $\theta_{0}=90\text{º}$. Type 2 angles are dynamically introduced
for bonding patterns $A\substack{H\\
\leftarrow
}
B$, $B\substack{3'/5'\\
\rightarrow
}
C$ and $A\substack{3'/5'\\
\leftarrow
}
B$, $B\substack{H\\
\rightarrow
}
C$ (i.e. model bond types $1$ and $2,3$, since $CBA$ order matches
the reverse pattern).

The DNA model has three dihedral interactions, which are described
by the potential $U(\phi)=K(1+d\cos(\phi))$. We use dihedral spring
constant $K=1.0\epsilon$, and $d=+1$ ($-1$) for promoting trans
(cis) conformations. The first dihedral interaction (type 1, shown
red in dihedral visualizations) promotes a cis conformation when a
back bone bond connects two hybridized nucleotide pairs. This corresponds
to the bonding patterns $A\substack{H\\
\leftarrow
}
B$, $B\substack{3'\\
\leftarrow
}
C$, $B\substack{5'\\
\rightarrow
}
C$, $C\substack{H\\
\rightarrow
}
D$ and $A\substack{H\\
\leftarrow
}
B$, $B\substack{5'\\
\leftarrow
}
C$, $B\substack{3'\\
\rightarrow
}
C$, $C\substack{H\\
\rightarrow
}
D$, where $H$ denotes a hybridization bond (i.e. model bond numbers
$1$ $2,3$ $1$). The second dihedral interaction (type 2, shown
green in the dihedral visualizations) promotes a cis conformation
of the two beads that are connected by back bone bonds to a hybridized
bead pair and is located on the same side of the bead pair. The bonding
pattern is $A\substack{3'\\
\leftarrow
}
B$, $B\substack{H\\
\leftarrow
}
C$, $B\substack{H\\
\rightarrow
}
C$, $C\substack{5'\\
\rightarrow
}
D$ (i.e. model bond numbers $2$ $1$ $3$). The third interaction (type
3, shown blue in the dihedral visualizations) promotes a trans conformation
of the two bead that are connected by back bone bonds to a hybridized
bead pair but are localized on opposite sides of the bead pair. The
bonding patterns are $A\substack{3'\\
\leftarrow
}
B$, $B\substack{H\\
\leftarrow
}
C$, $B\substack{H\\
\rightarrow
}
C$, $C\substack{3'\\
\rightarrow
}
D$ and $A\substack{5'\\
\leftarrow
}
B$, $B\substack{H\\
\leftarrow
}
C$, $B\substack{H\\
\rightarrow
}
C$, $C\substack{5'\\
\rightarrow
}
D$ (i.e. model bond numbers $2$ $1$ $2$ and $3$ $1$ $3$). Note
that without the directional bond, we would be unable to distinguish
between these two last types of dihedrals. The examples belows are
included as test cases with the dynamic bonding code submitted to
the CPC Program Library, and require less than a CPU hour of
computational effort.

\section{Example DNA dynamic\label{sec:Illustrations}}

\begin{figure}
\includegraphics[bb=0bp 0bp 976bp 622bp,width=0.33\columnwidth]{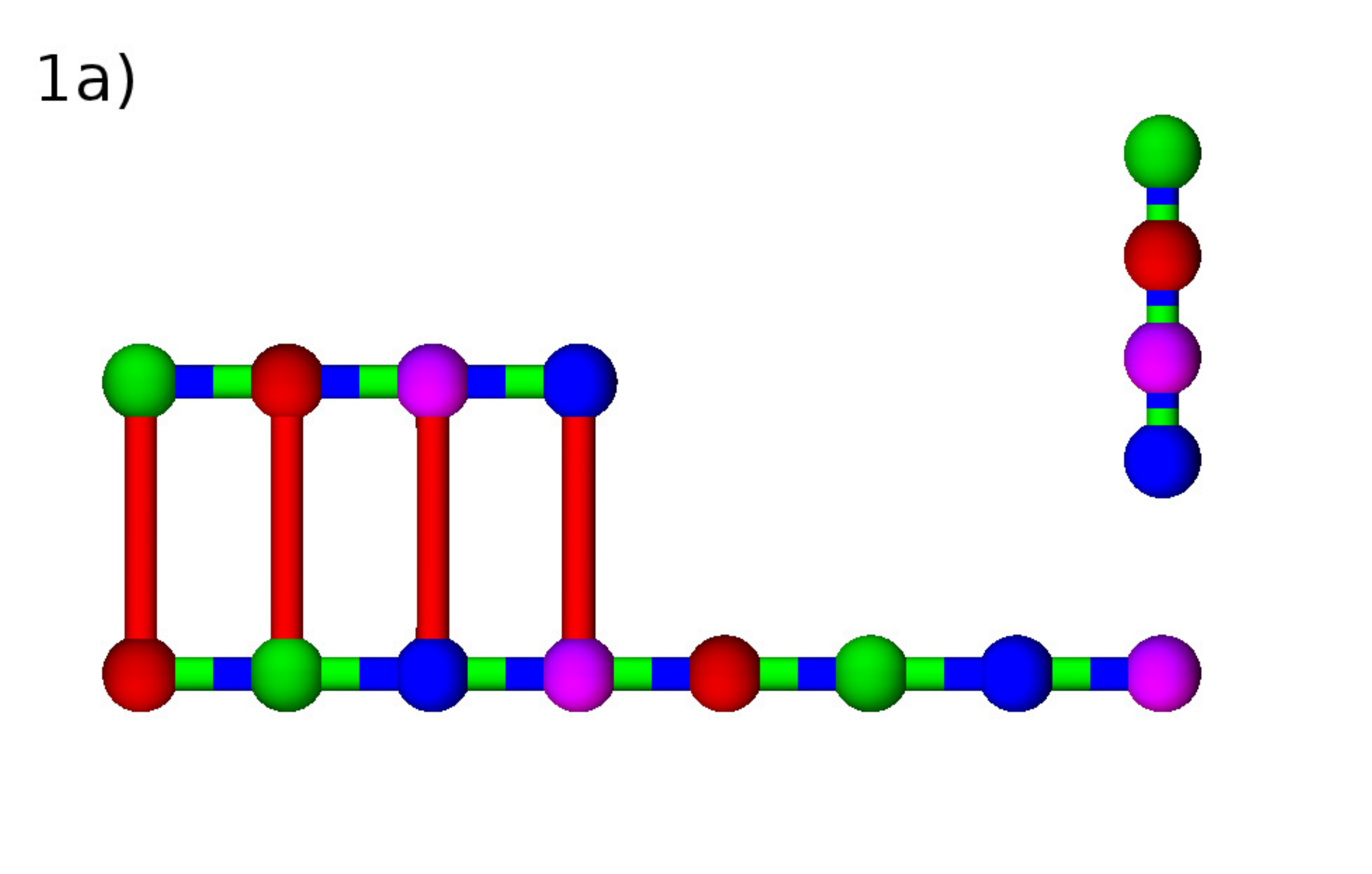}\includegraphics[bb=0bp 0bp 976bp 622bp,width=0.33\columnwidth]{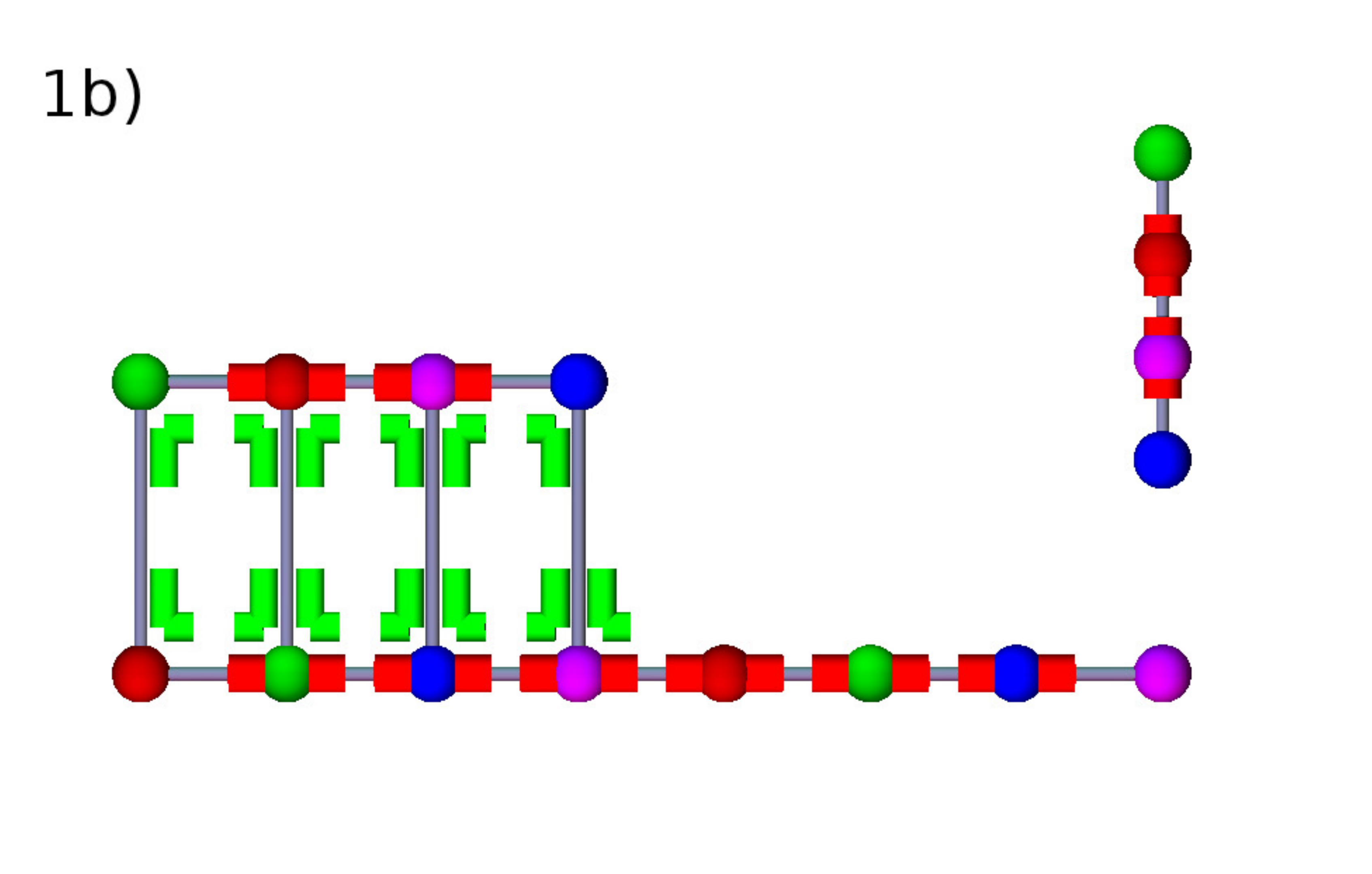}\includegraphics[bb=0bp 0bp 976bp 622bp,width=0.33\columnwidth]{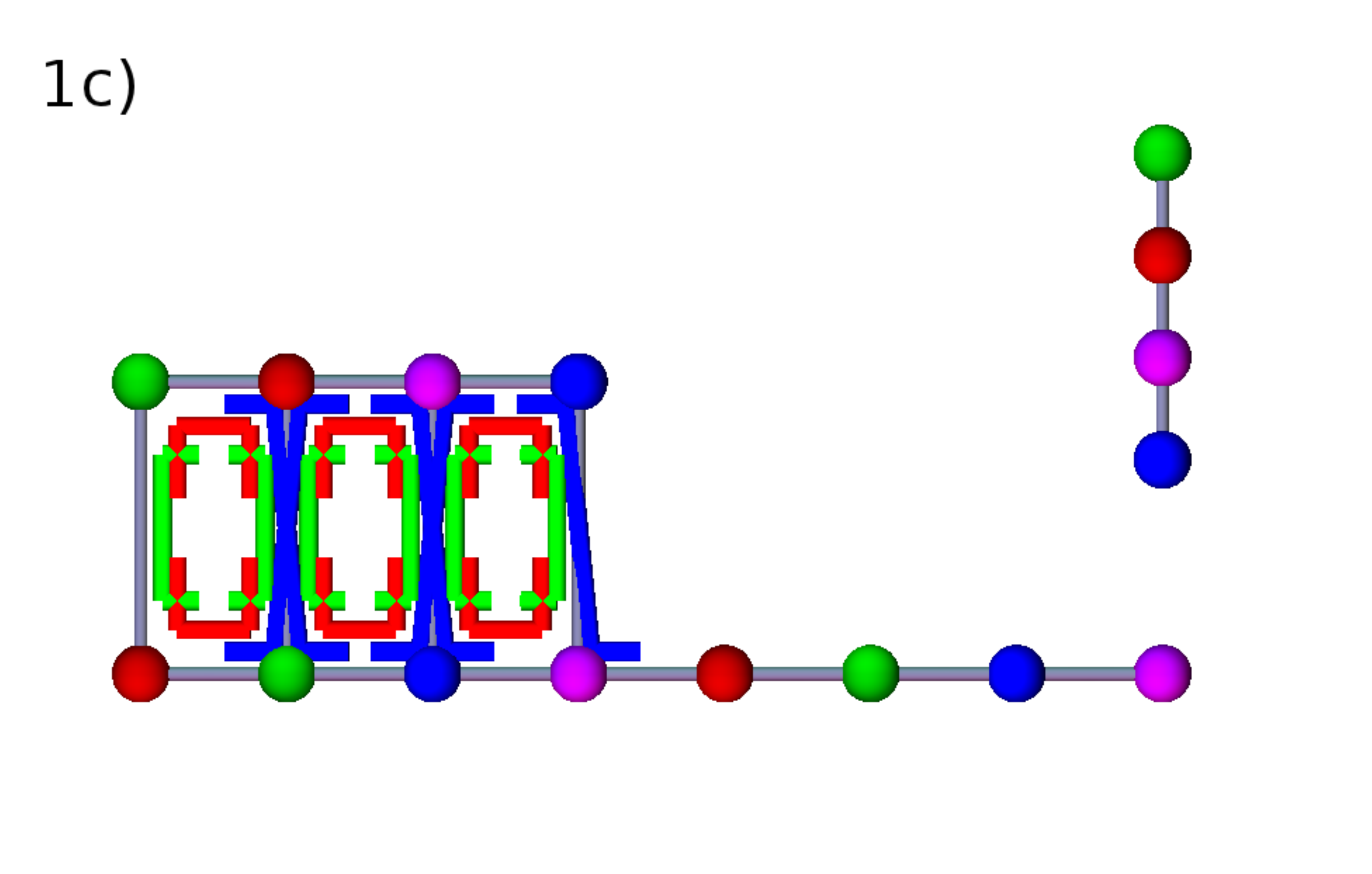}

\includegraphics[bb=0bp 0bp 976bp 622bp,width=0.33\columnwidth]{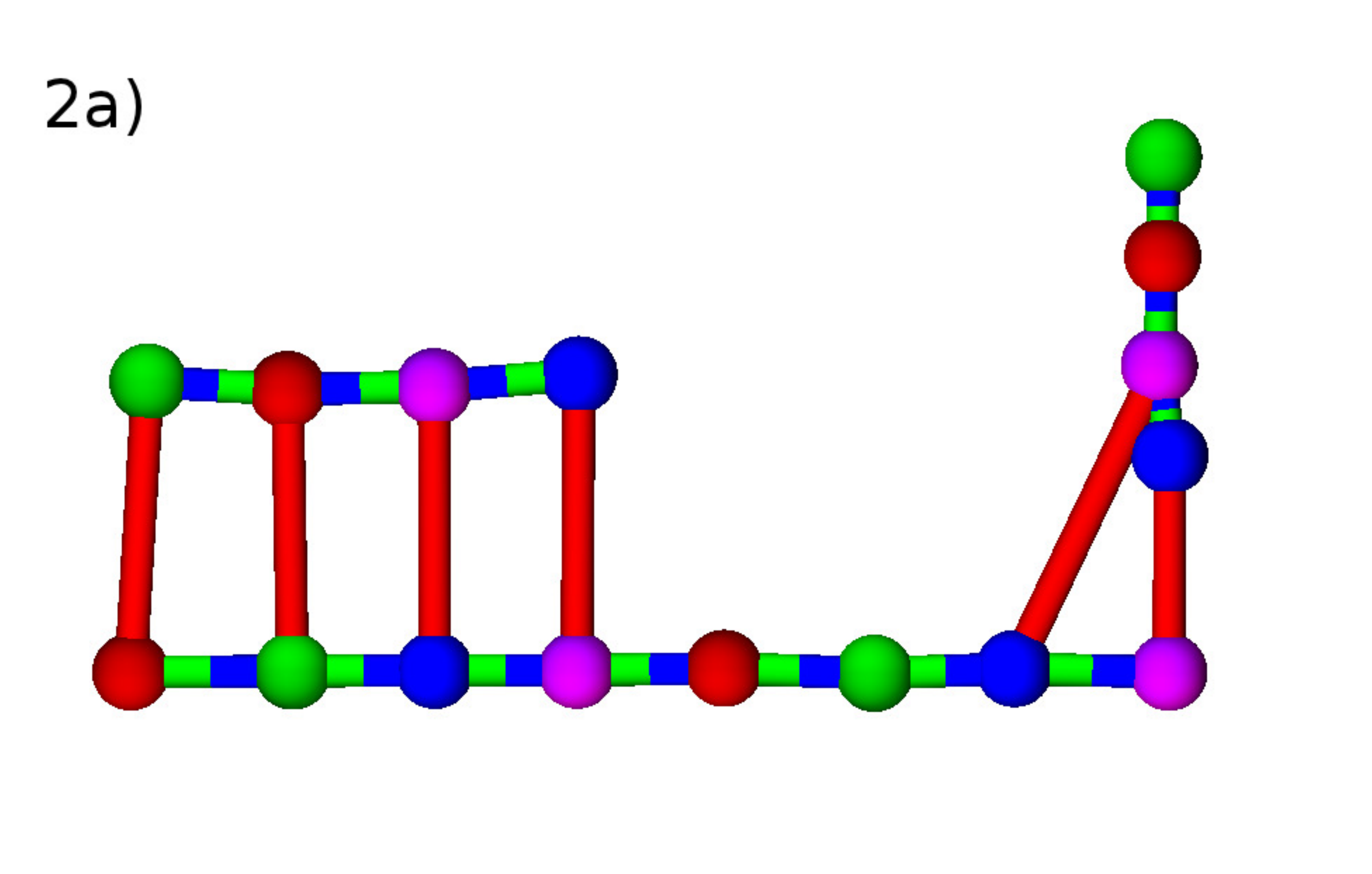}\includegraphics[bb=0bp 0bp 976bp 622bp,width=0.33\columnwidth]{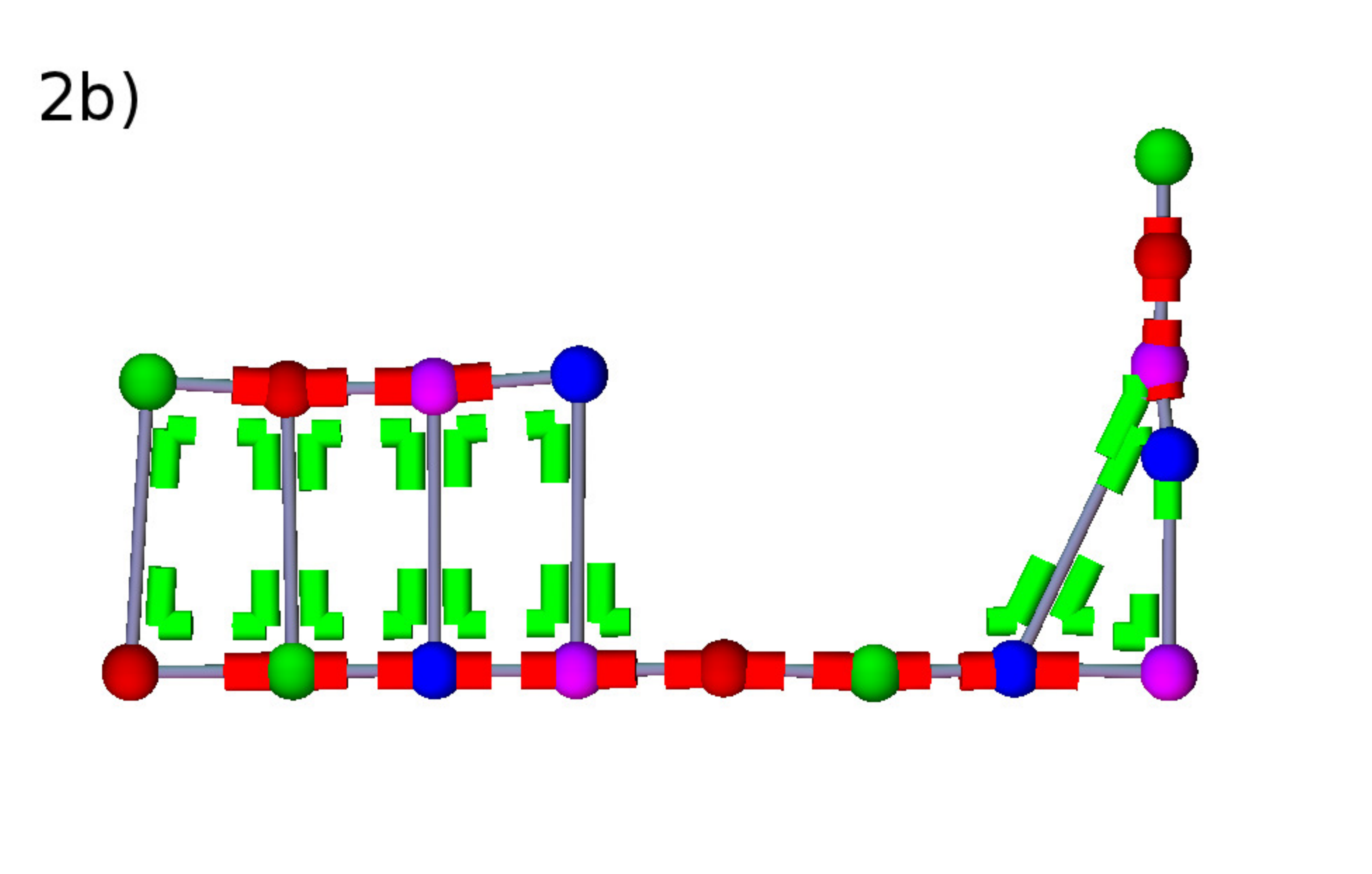}\includegraphics[bb=0bp 0bp 976bp 622bp,width=0.33\columnwidth]{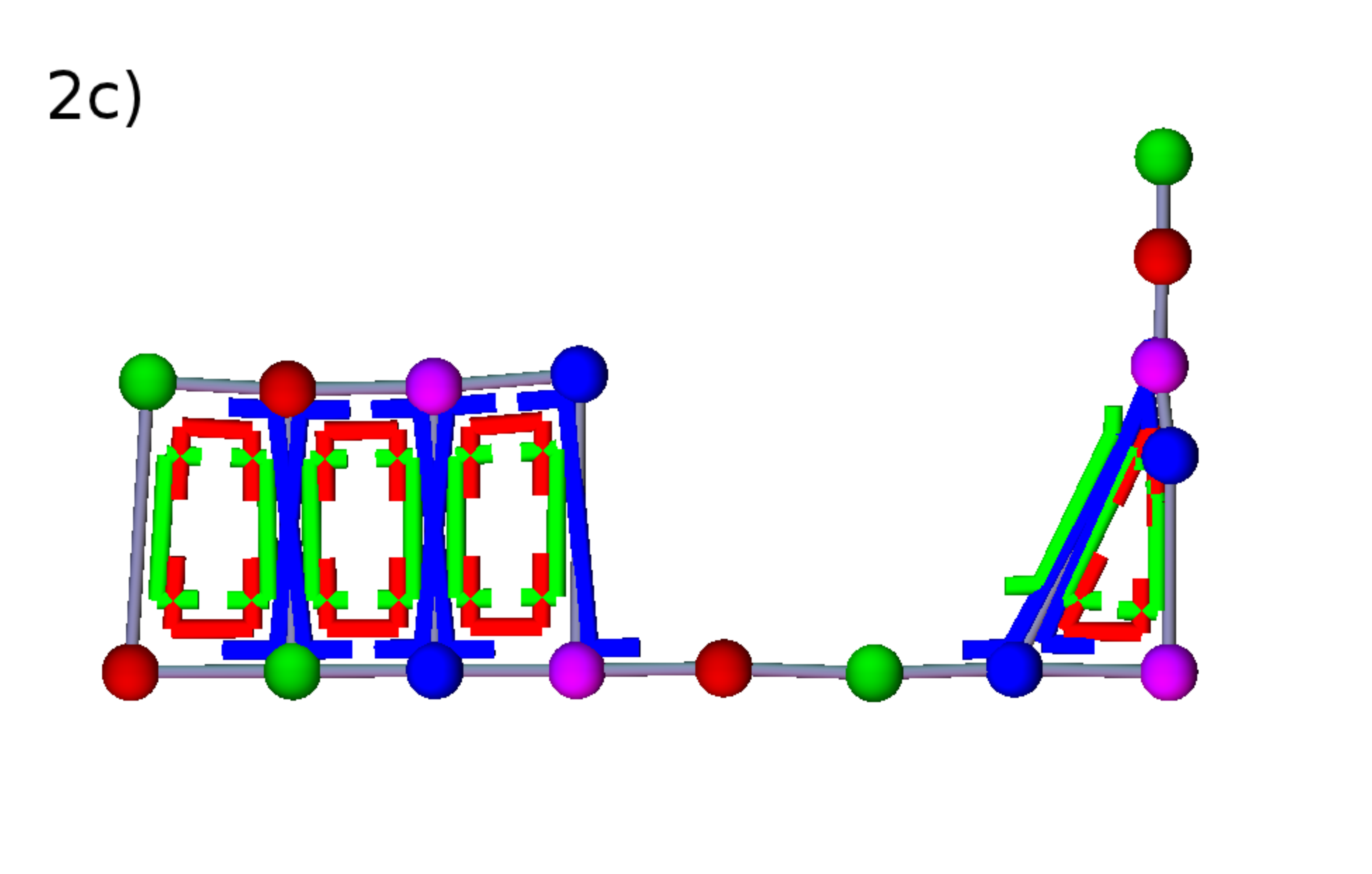}

\includegraphics[bb=0bp 0bp 976bp 622bp,width=0.33\columnwidth]{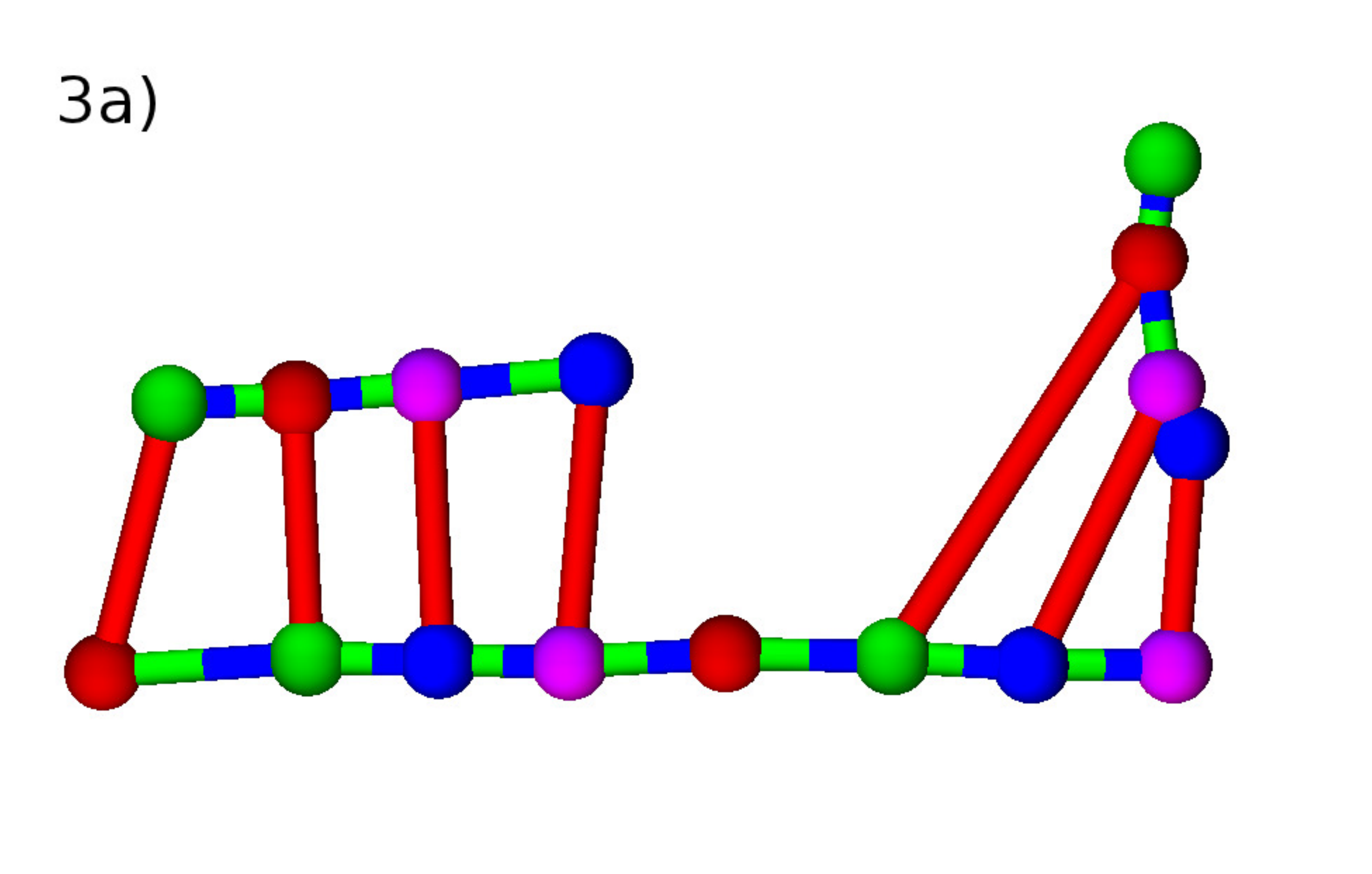}\includegraphics[bb=0bp 0bp 976bp 622bp,width=0.33\columnwidth]{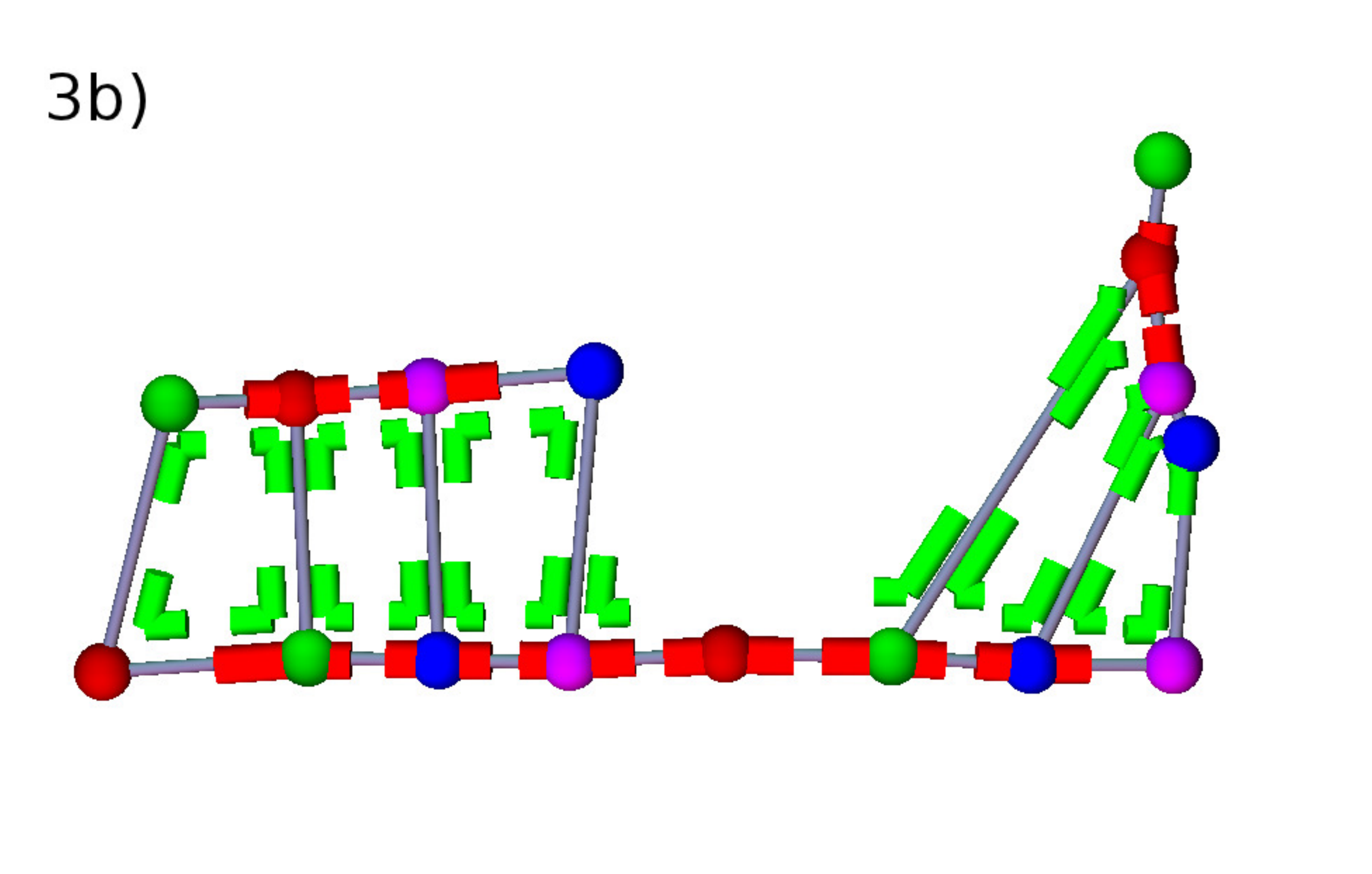}\includegraphics[bb=0bp 0bp 976bp 622bp,width=0.33\columnwidth]{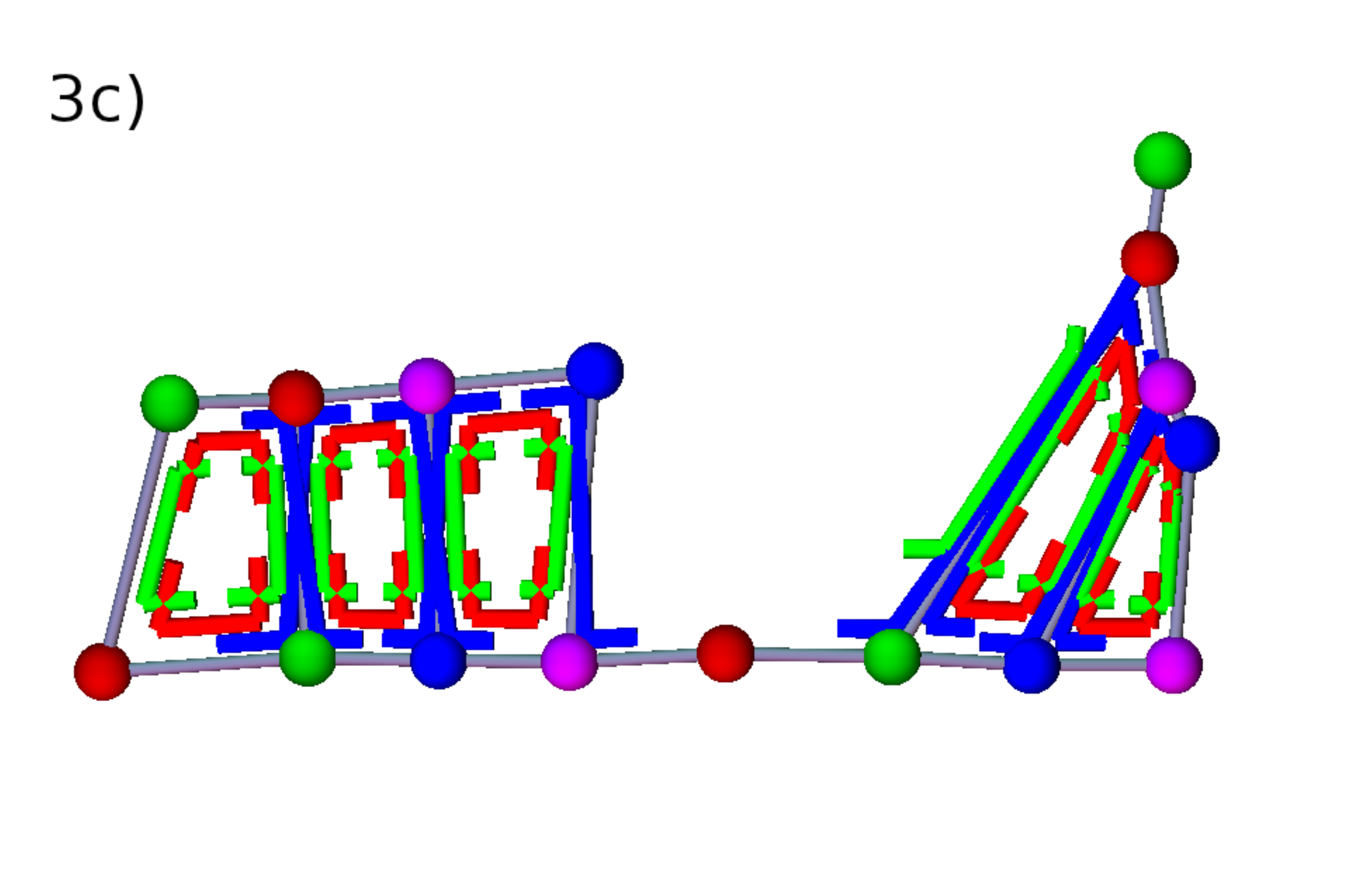}

\includegraphics[bb=0bp 0bp 976bp 622bp,width=0.33\columnwidth]{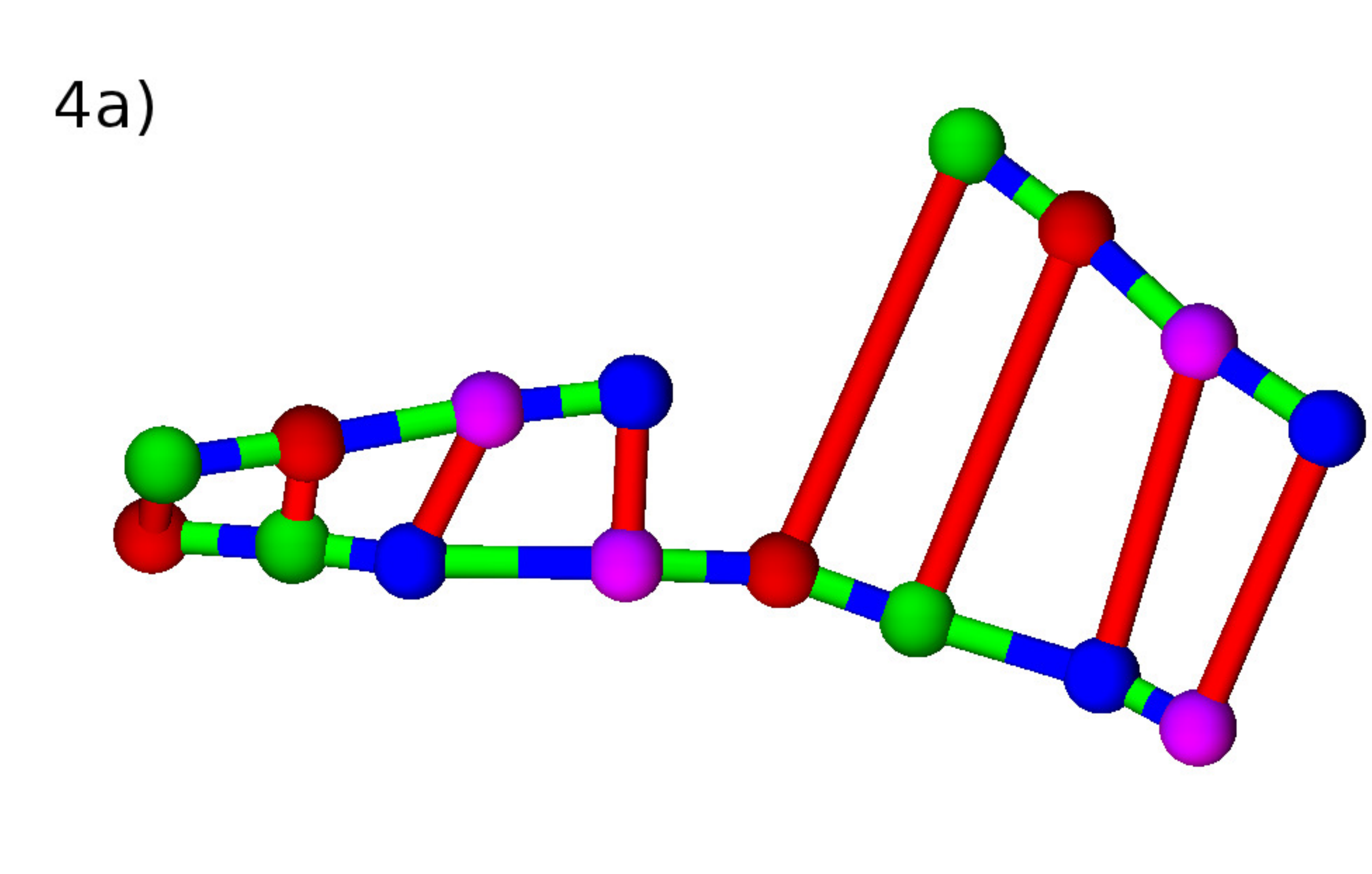}\includegraphics[bb=0bp 0bp 976bp 622bp,width=0.33\columnwidth]{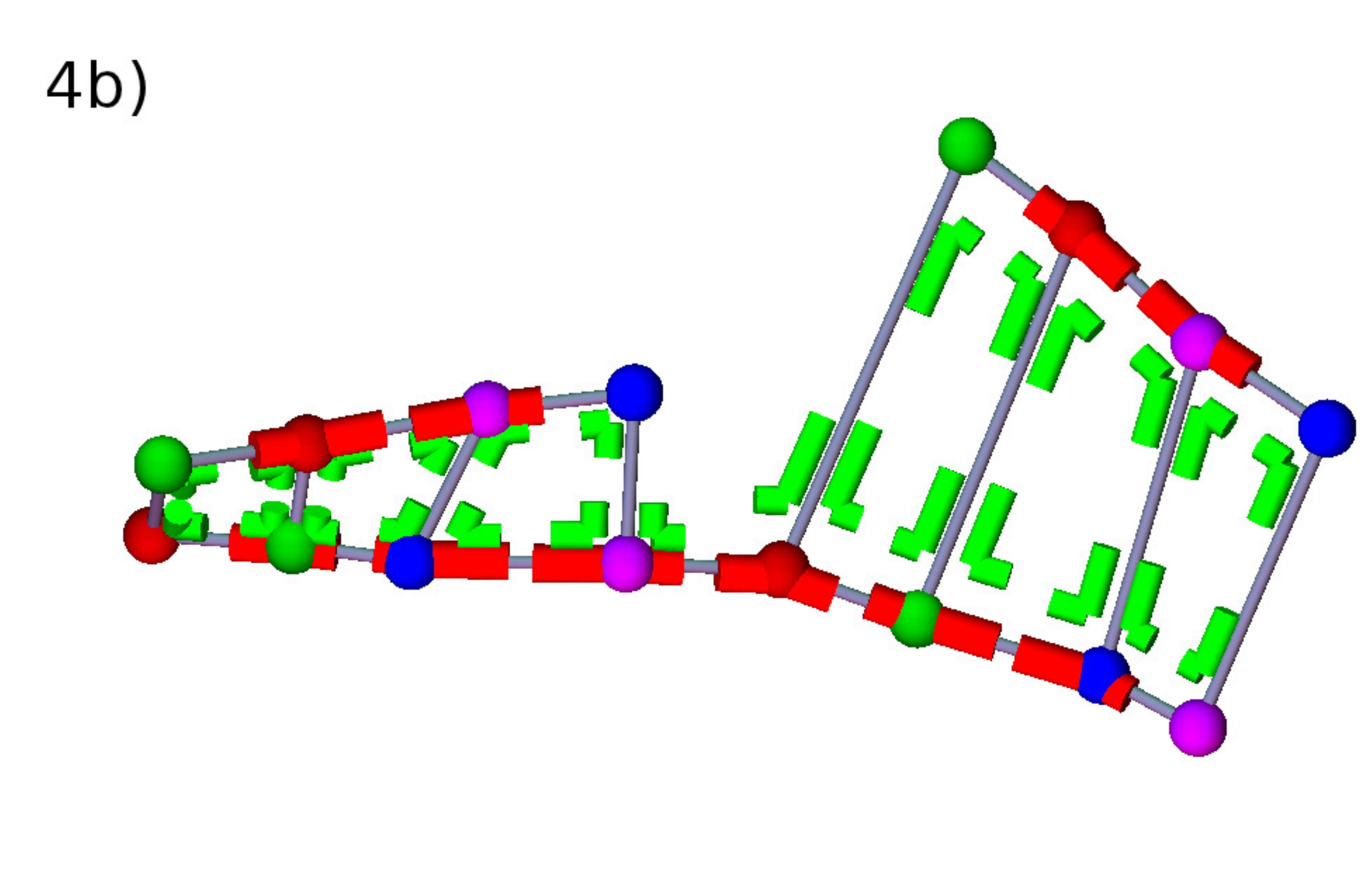}\includegraphics[bb=0bp 0bp 976bp 622bp,width=0.33\columnwidth]{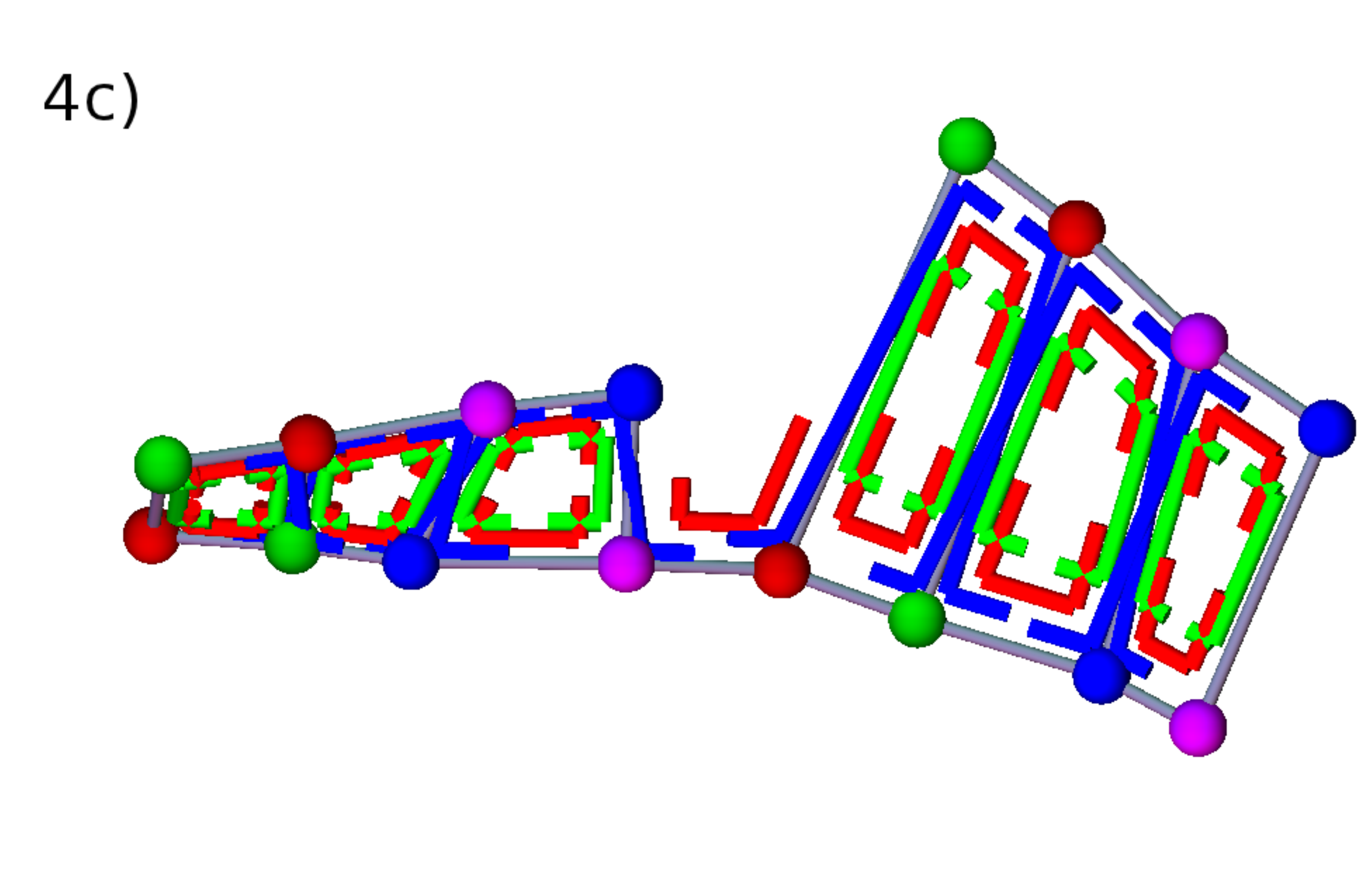}

\caption{\label{fig:example-hybridization}Oligomer - DNA template hybridization
(rows 1-4) showing the dynamics of bond, angular, and dihedral interactions
(columns a-c) for times $t=0$, $0.01\tau$, $0.04\tau$, and $0.23\tau$
into the simulation. Bead and interaction colors match those in fig.
\ref{fig:Fix-DNA-dynamics}. Note that back bone bond directionality
is only shown in the first row for simplicity.}
\end{figure}
To illustrate the dynamic bonding framework with the DNA model, we
simulate a $5'-ATCGATCG-3'$ template in the presence of two $3'-TAGC-5'$
oligomers. The first oligomer is already hybridized with the template,
while the second is placed in the vicinity of the template. Fig. \ref{fig:example-hybridization}
shows snapshots along the trajectory where the remaining oligomer
hybridizes with the template. The top left visualization shows the
initial designed configuration. The blue-green pattern of the hybridized
oligomer backbone shows it has 3' 5' direction, while the green-blue
pattern of the template backbone shows the 5' 3' direction. The top
center visualization shows the angular interactions of the initial
configuration. The back bone stiffness is controlled by the red angular
interactions between back bone bond pairs, which promote a straight
back bone configuration. The green angular interactions promote hybridization
bonds that are perpendicular to the strand axis. The top right visualization
shows the dihedral interactions of the initial configuration. The
hybridized template shows red and green dihedral interactions which
are promote cis arrangement of stacked bead pairs, while the blue
dihedral interaction promotes trans arrangement. Together they stabilize
the ladder-like structure of the double strand. Without the bond directionality,
we would have no way to distinguish between green and blue dihedral
interactions, and hence control over the stiffness of the double strand
relative to that of the single strands.

As we let the simulation run (left column top to bottom) initially
two hybridization bonds are introduced between the two beads at right
most end of the template. Later a third and a fourth hybridization
bond are also introduced. Along with the hybridization bonding dynamics,
angular and dihedral interactions (center and right columns) are also
created. The angular interactions cause the free oligomer to align
with the template, while the dihedral interactions creates a torque
that ensures that the alignment is anti-parallel.
\begin{figure}
\includegraphics[bb=0bp 0bp 862bp 434bp,width=0.33\columnwidth]{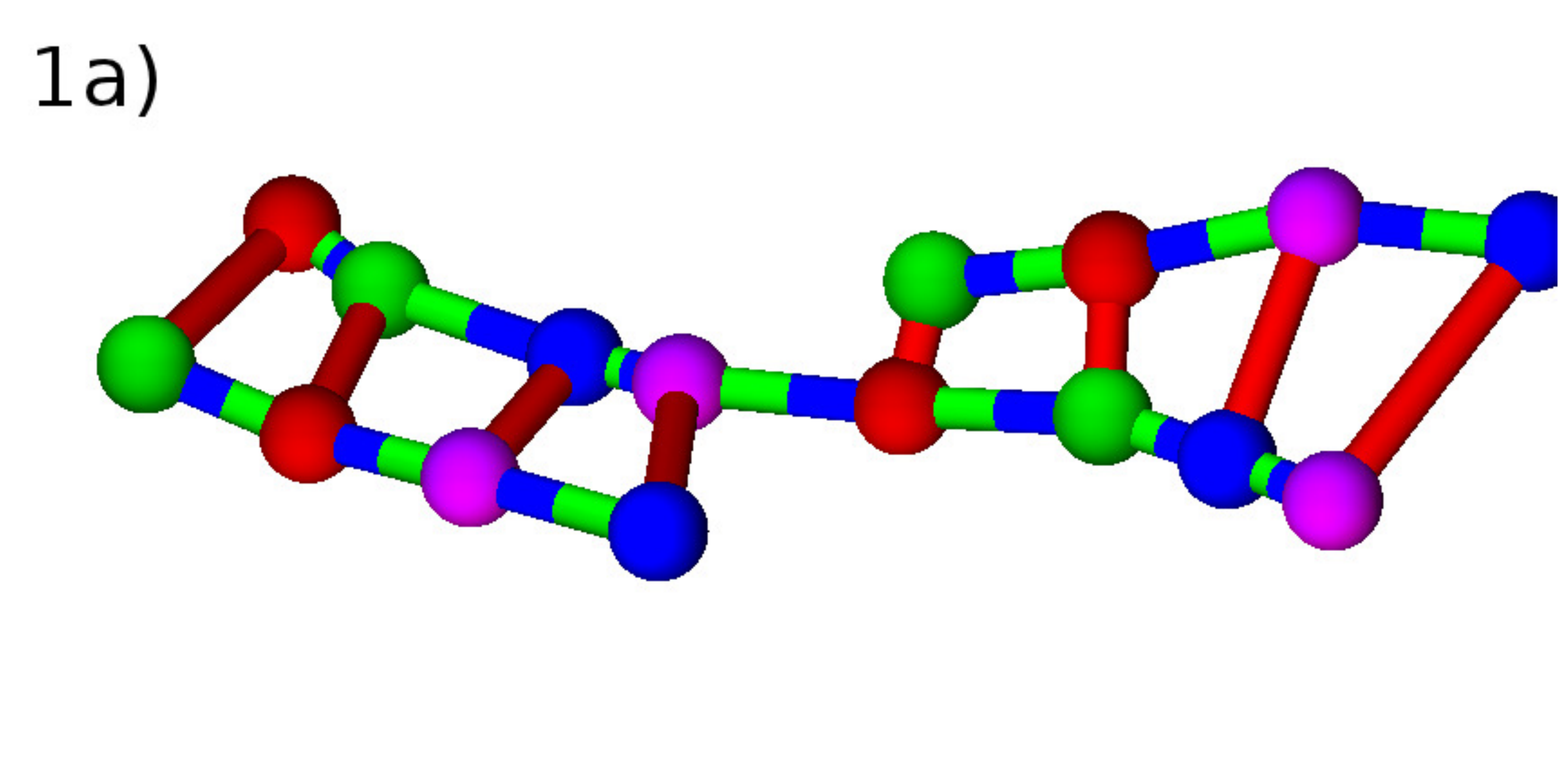}\includegraphics[bb=0bp 0bp 862bp 434bp,width=0.33\columnwidth]{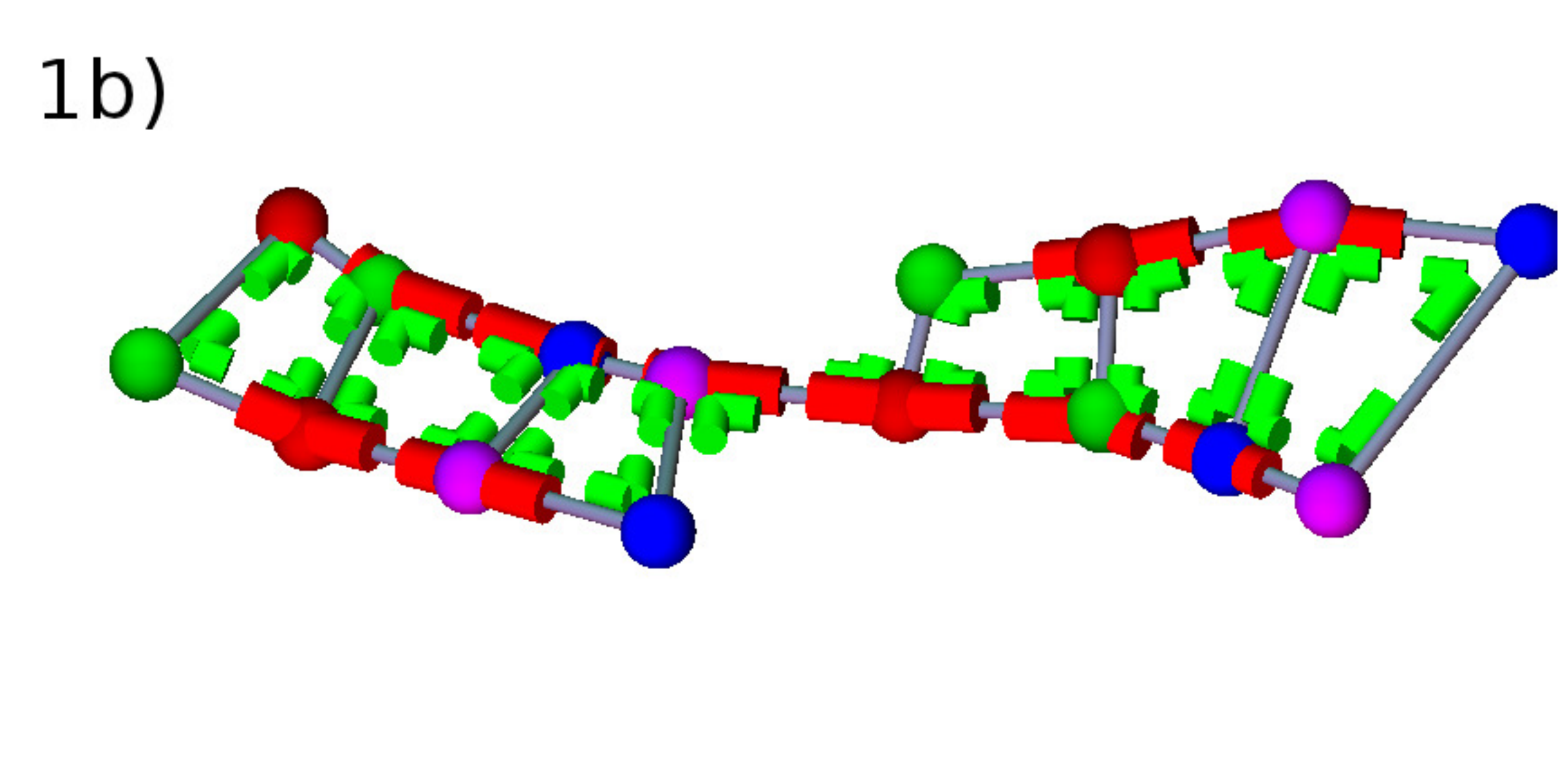}\includegraphics[bb=0bp 0bp 862bp 434bp,width=0.33\columnwidth]{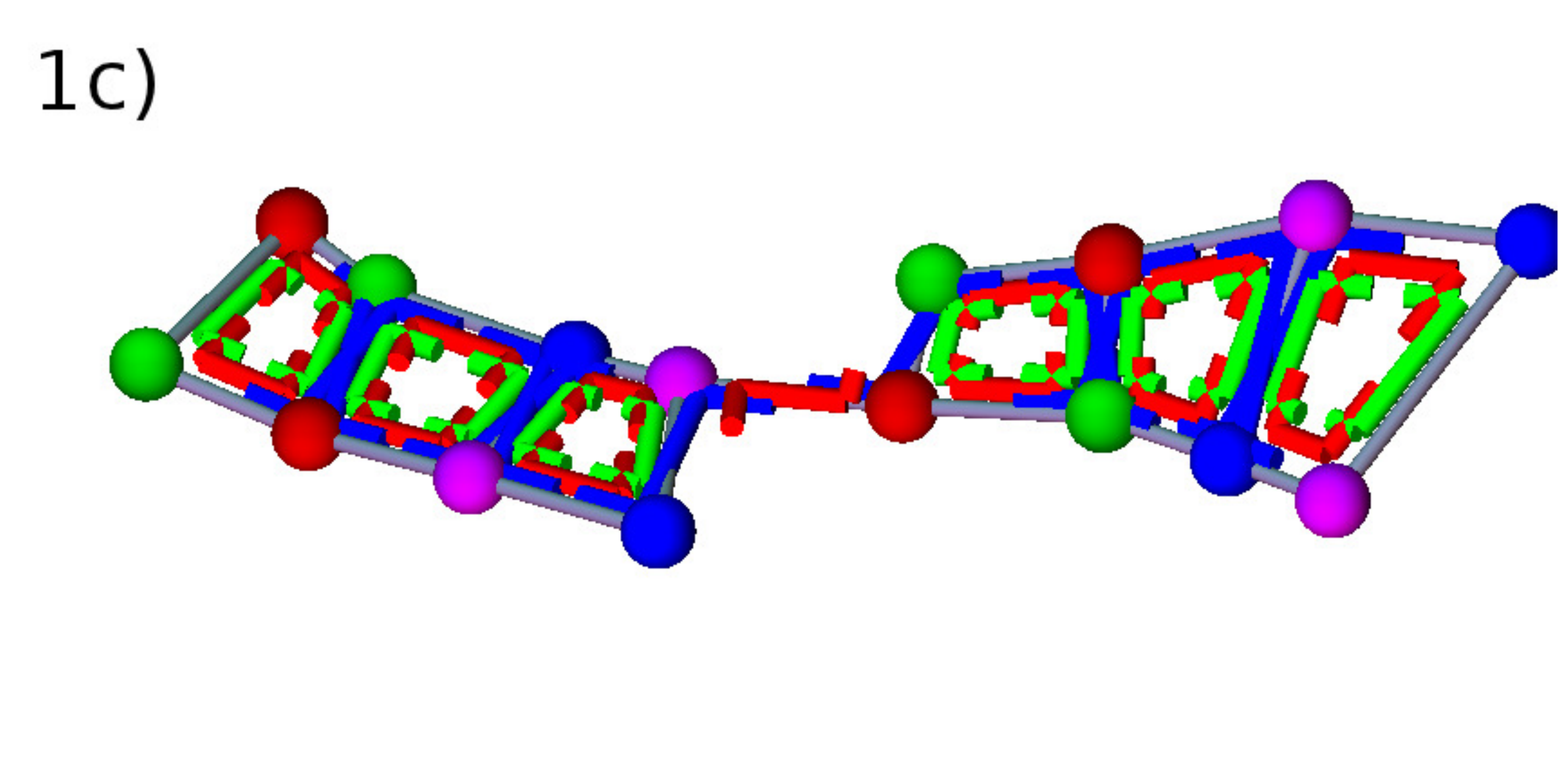}

\includegraphics[width=0.33\columnwidth]{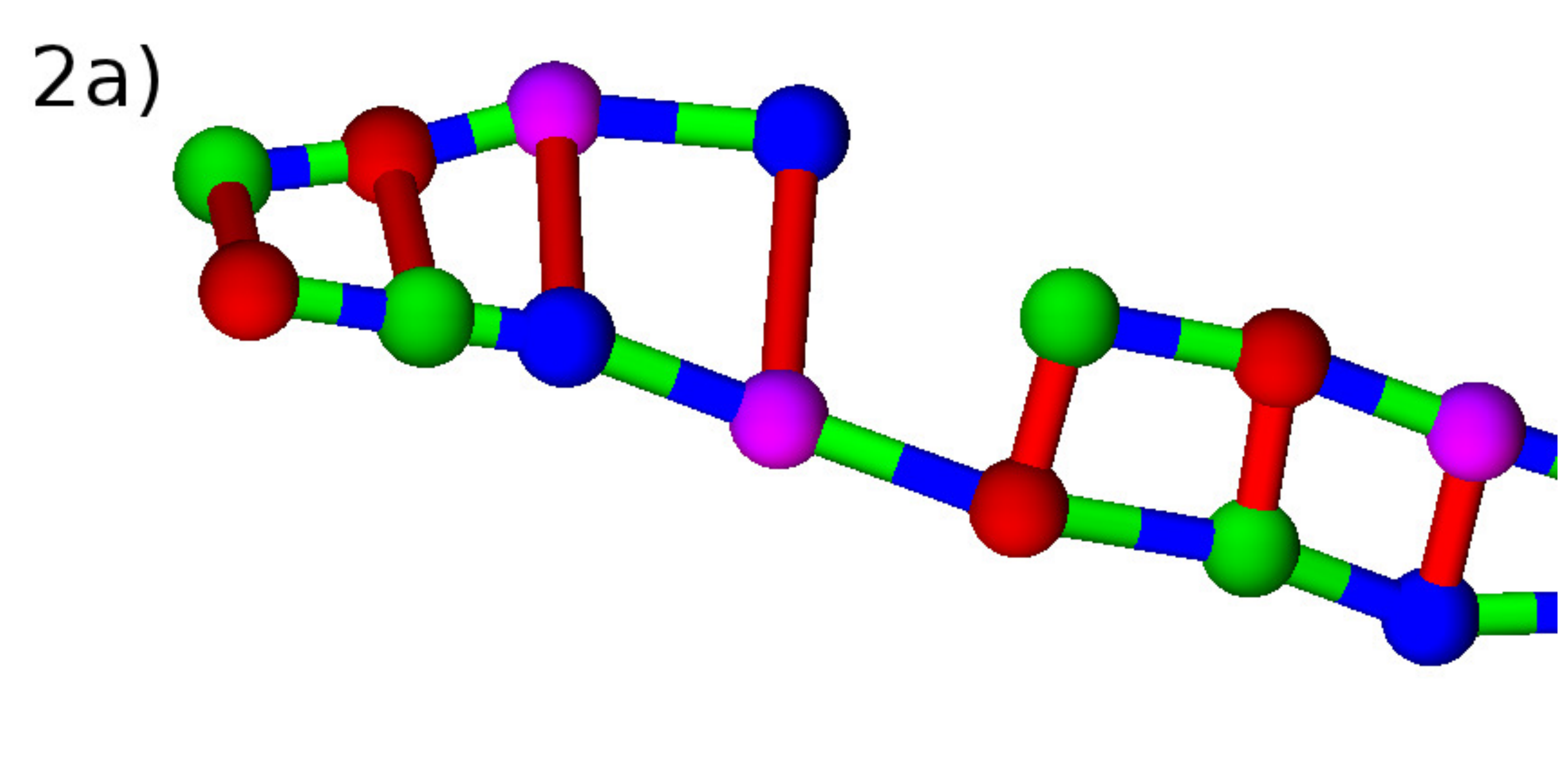}\includegraphics[width=0.33\columnwidth]{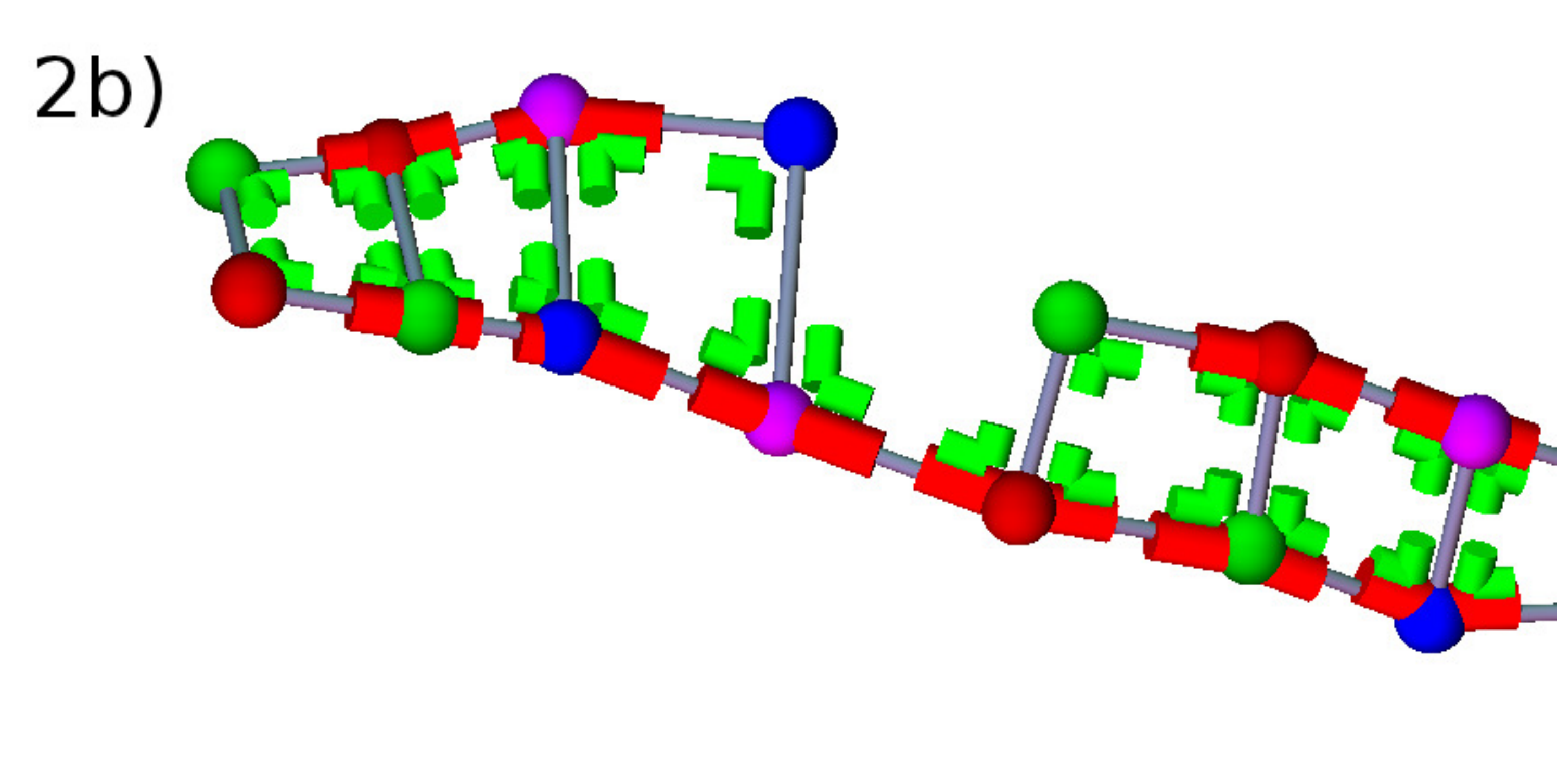}\includegraphics[width=0.33\columnwidth]{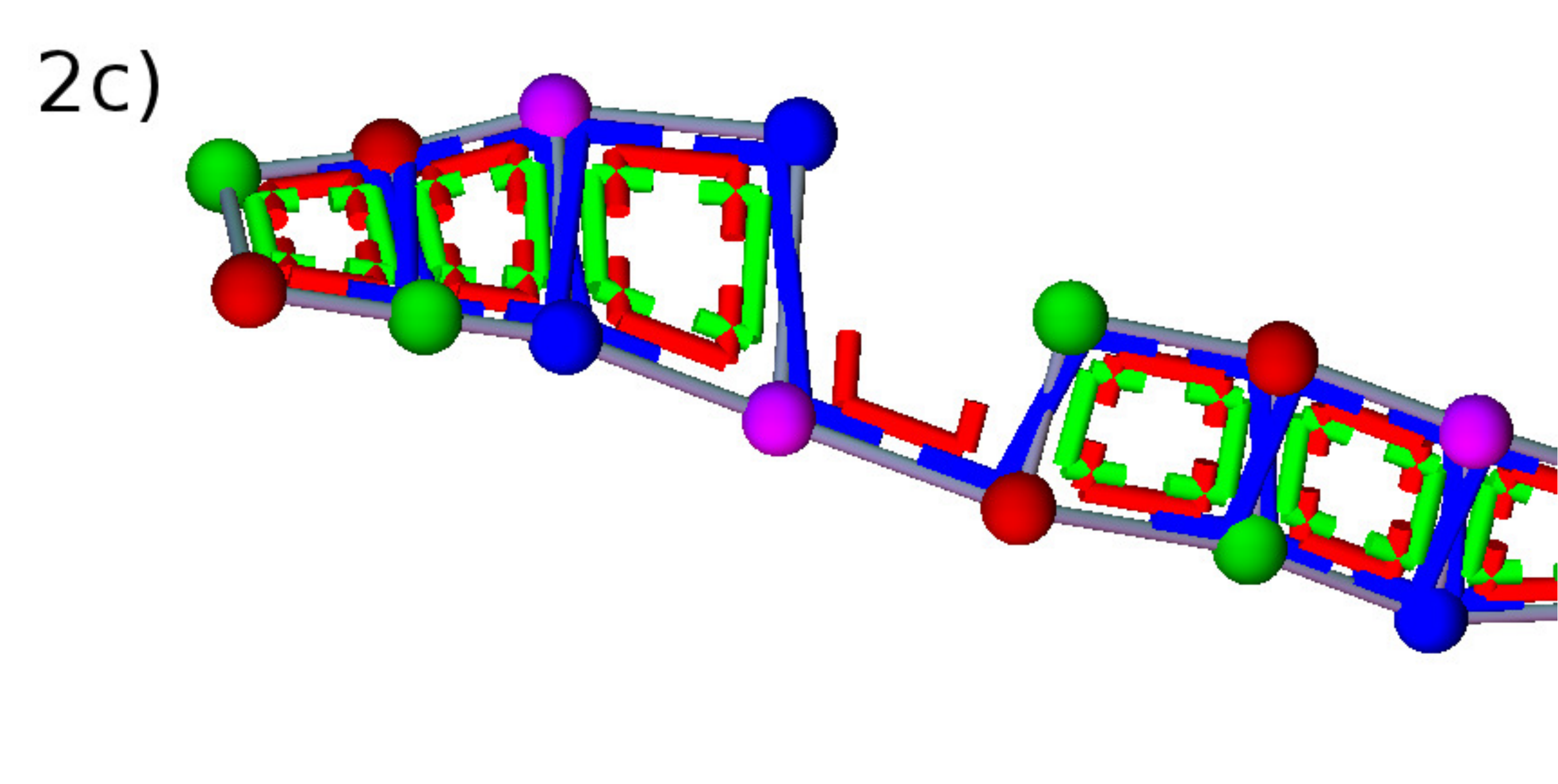}

\includegraphics[width=0.33\columnwidth]{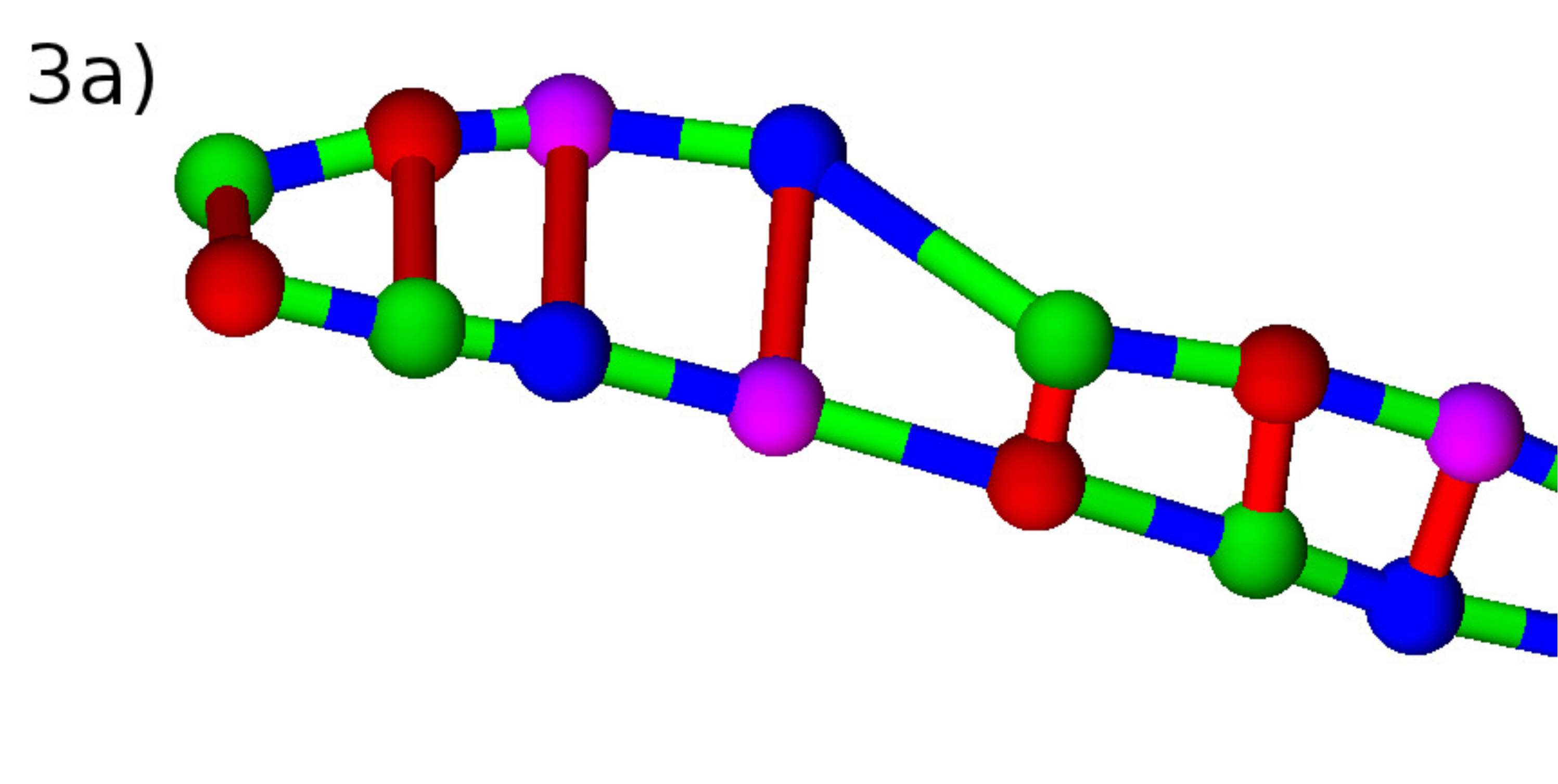}\includegraphics[width=0.33\columnwidth]{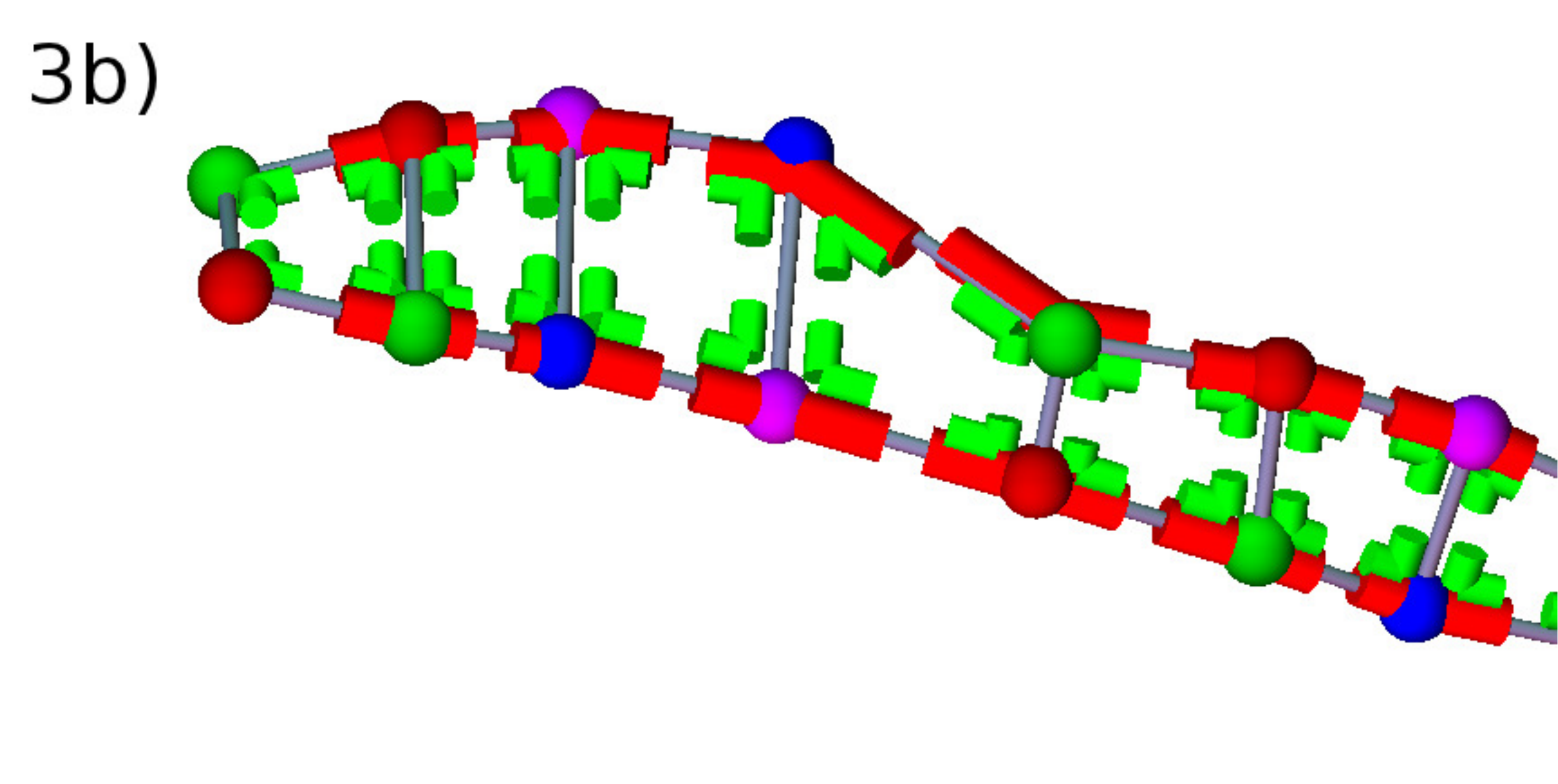}\includegraphics[width=0.33\columnwidth]{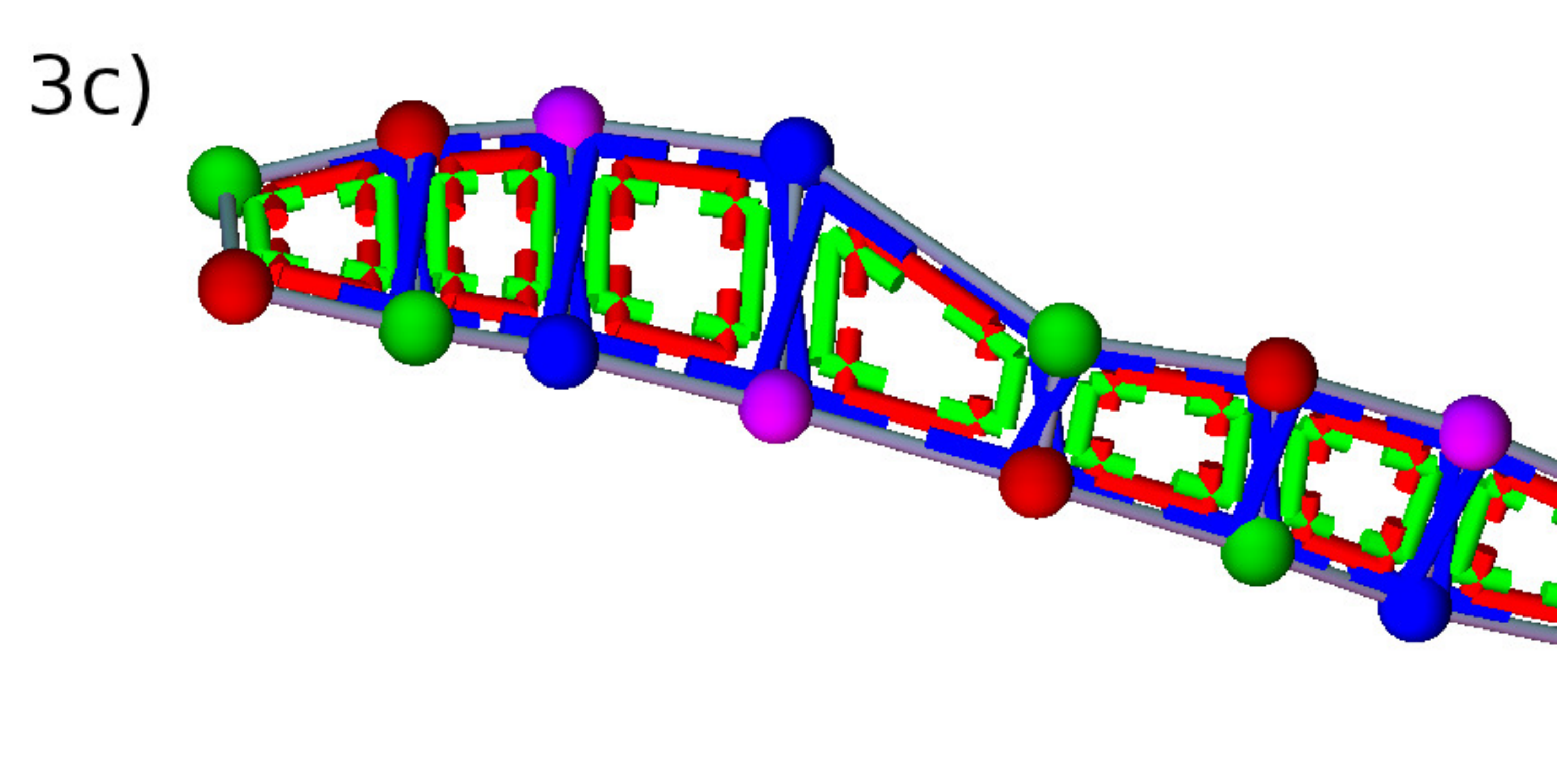}

\caption{\label{fig:example-backbone}Back bone ligation reaction by addition
of directional back bone bond (rows 1-3) showing dynamic of bond,
angular and dihedral interactions (columns a-c) for the simulation
in fig. \ref{fig:example-hybridization} continued to times $11.50\tau$,
$12.46\tau$, and $12.48\tau$, respectively.}
 
\end{figure}

Fig. \ref{fig:example-backbone} shows how the nick in the DNA molecule
is closed by forming a back bone bond. The interactions between the
two oligomers and the template ensures that they are both aligned
anti-parallel to the template backbone axis. The single red dihedral
interaction across the nick promotes a cis configuration, and twists
the two oligomers towards the same side of the template. Finally the
missing back bone bond is created following the 3'-5' directionality
of the strand, along with all the angular and dihedral interactions
to produce a double stranded configuration. Together figs. \ref{fig:example-hybridization}
and \ref{fig:example-backbone} simulates a chemical reaction where
a DNA template and two complementary oligomers first hybridize due
to their complementary sequences, and then ligate to produce the complementary
template sequence.

\begin{figure}
\includegraphics[width=0.33\columnwidth]{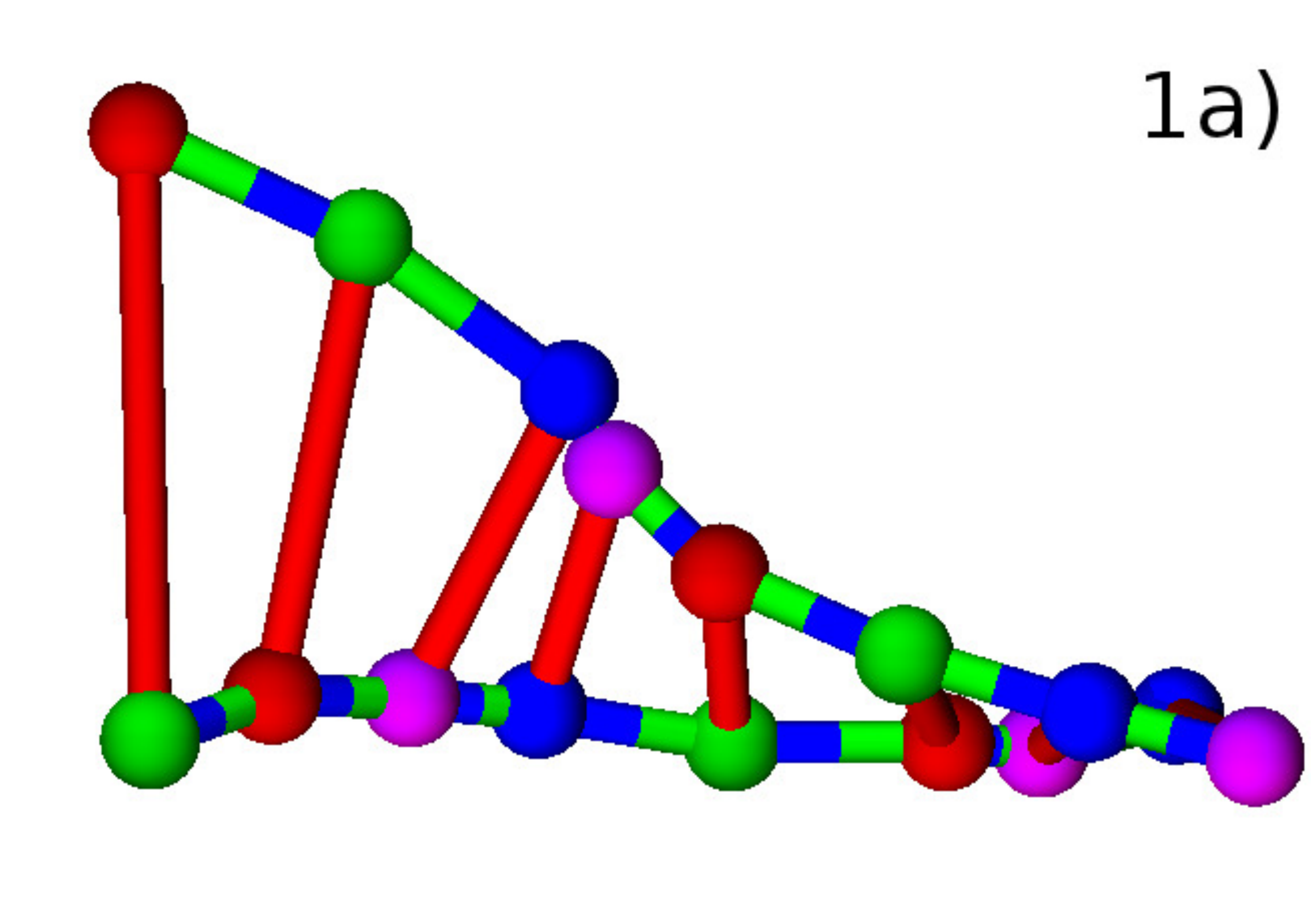}\includegraphics[width=0.33\columnwidth]{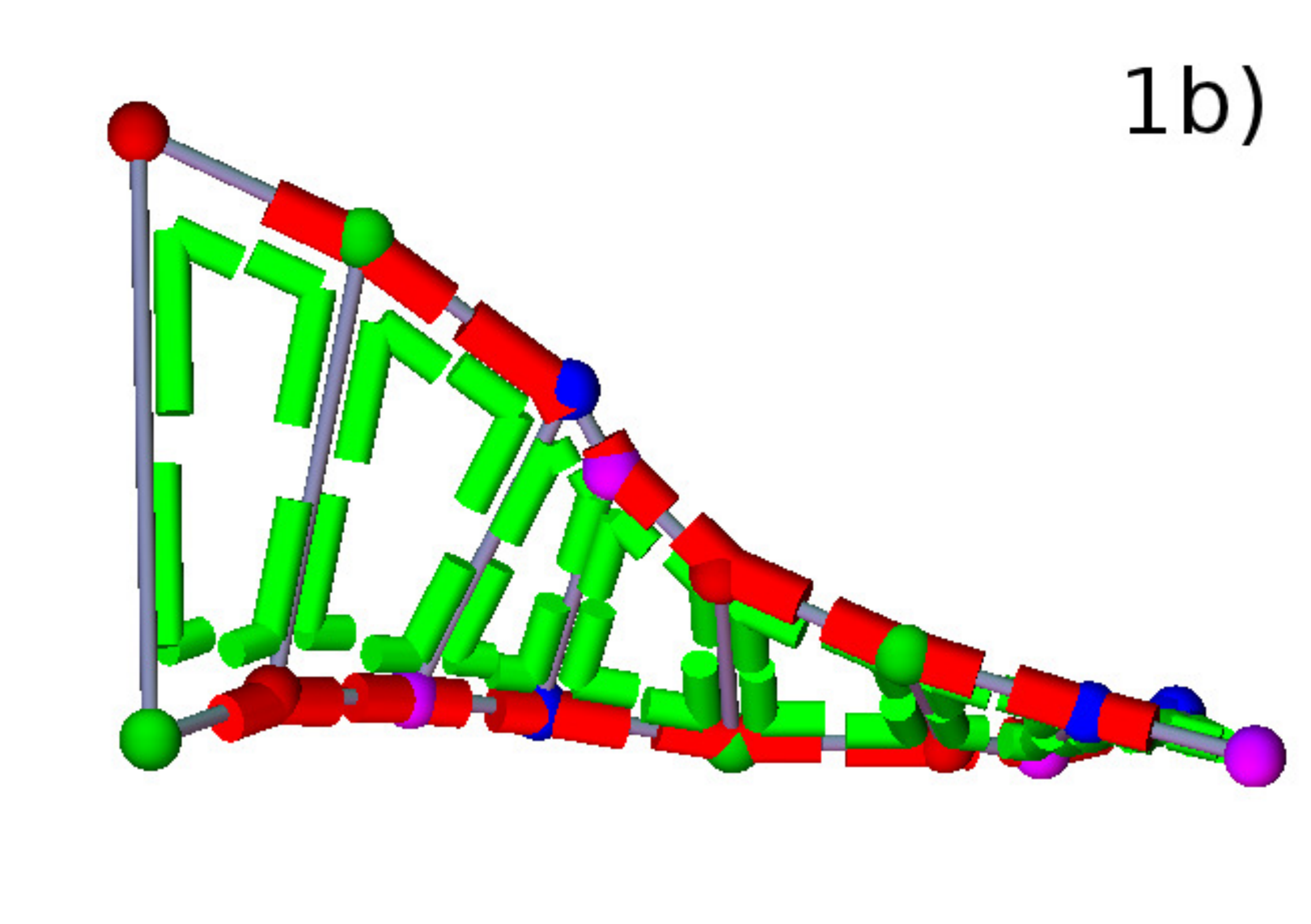}\includegraphics[width=0.33\columnwidth]{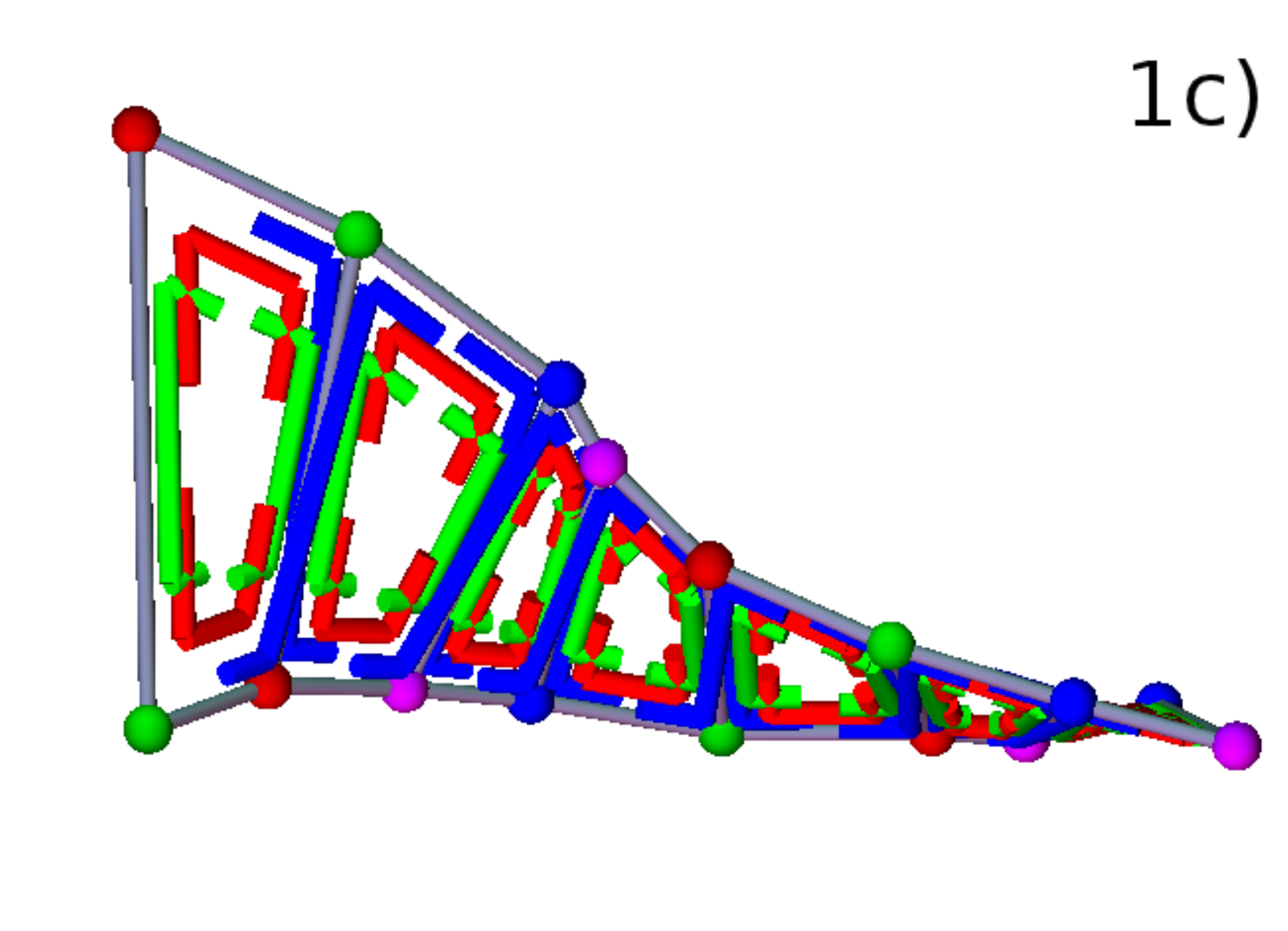}

\includegraphics[width=0.33\columnwidth]{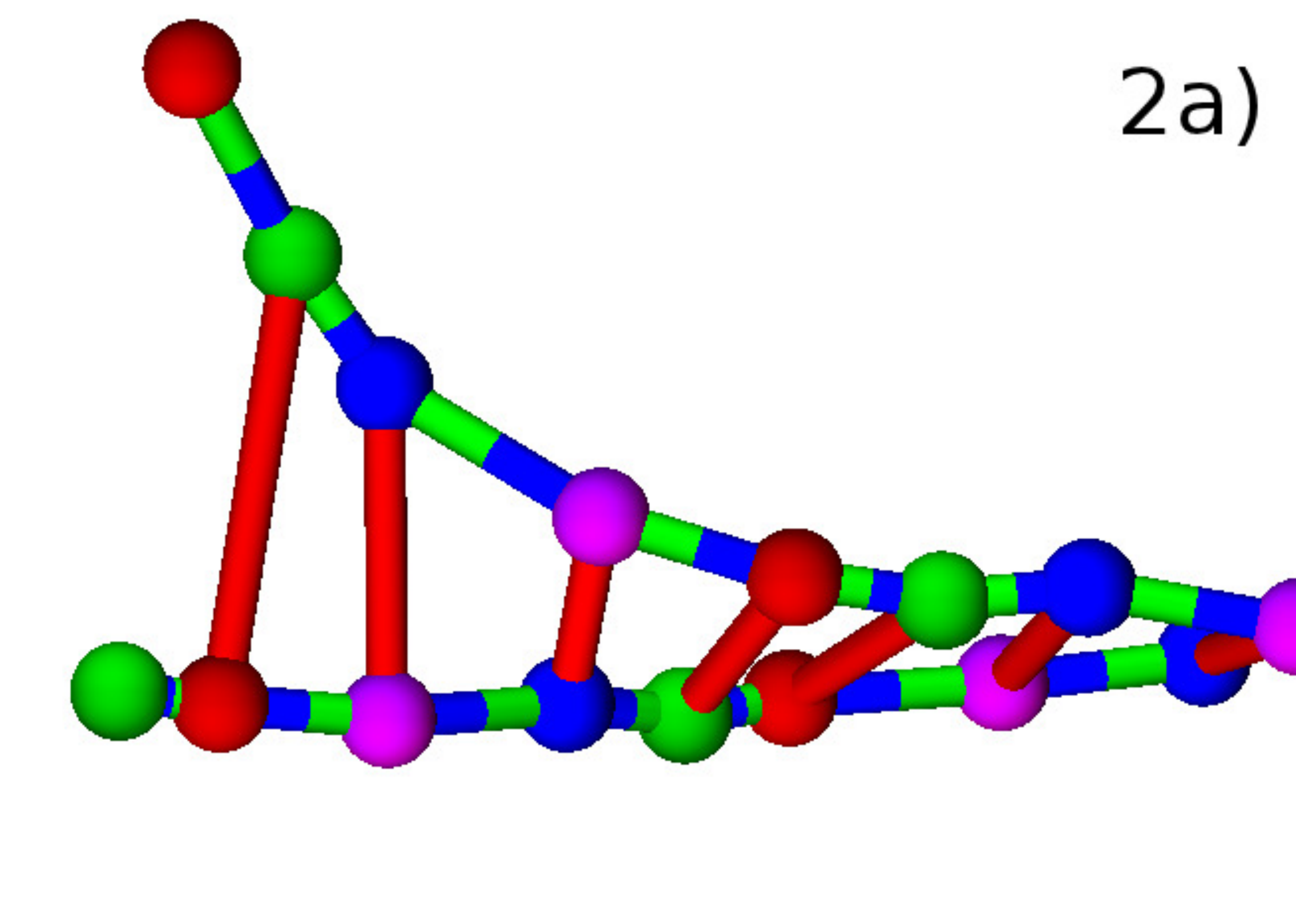}\includegraphics[width=0.33\columnwidth]{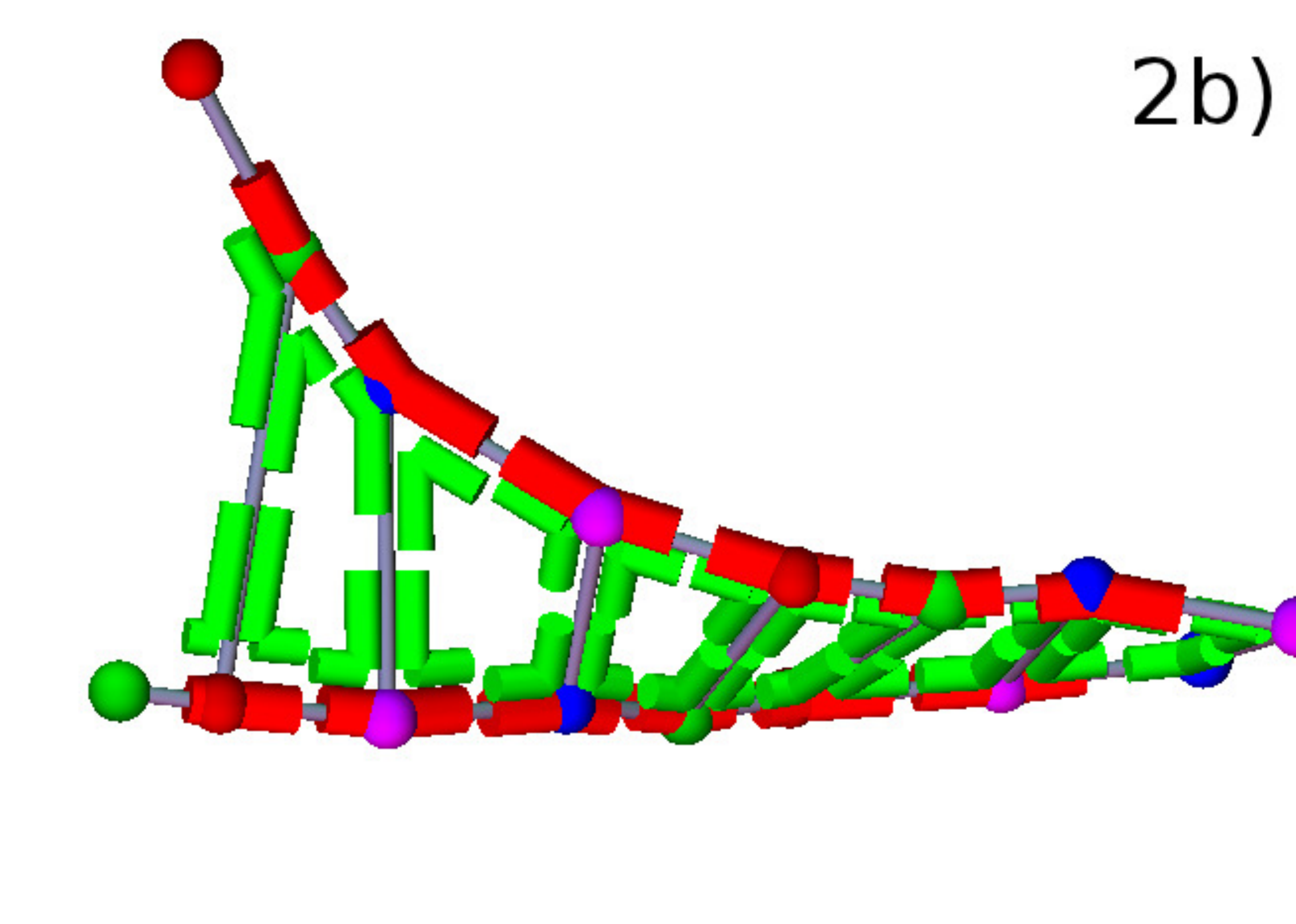}\includegraphics[width=0.33\columnwidth]{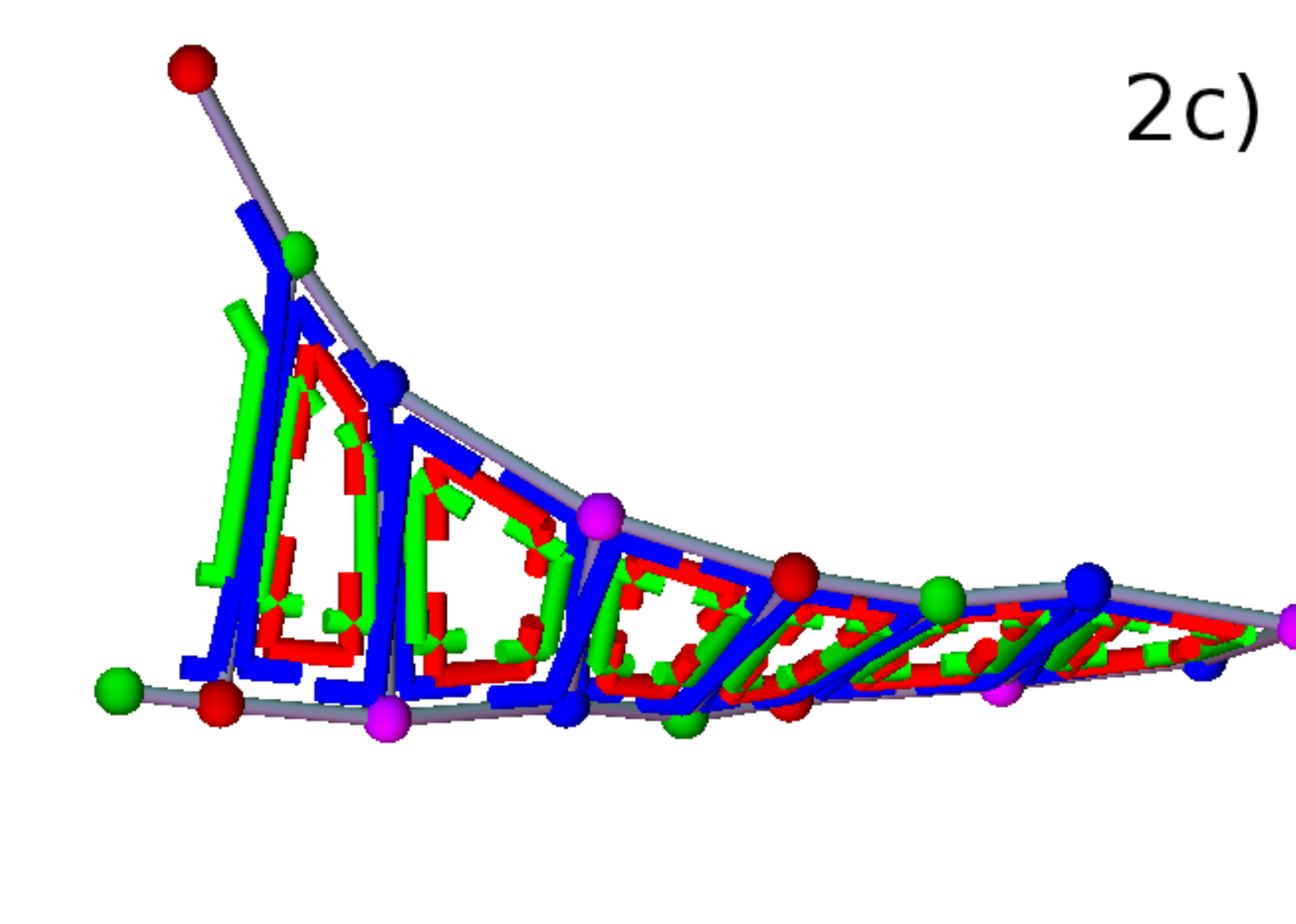}

\includegraphics[width=0.33\columnwidth]{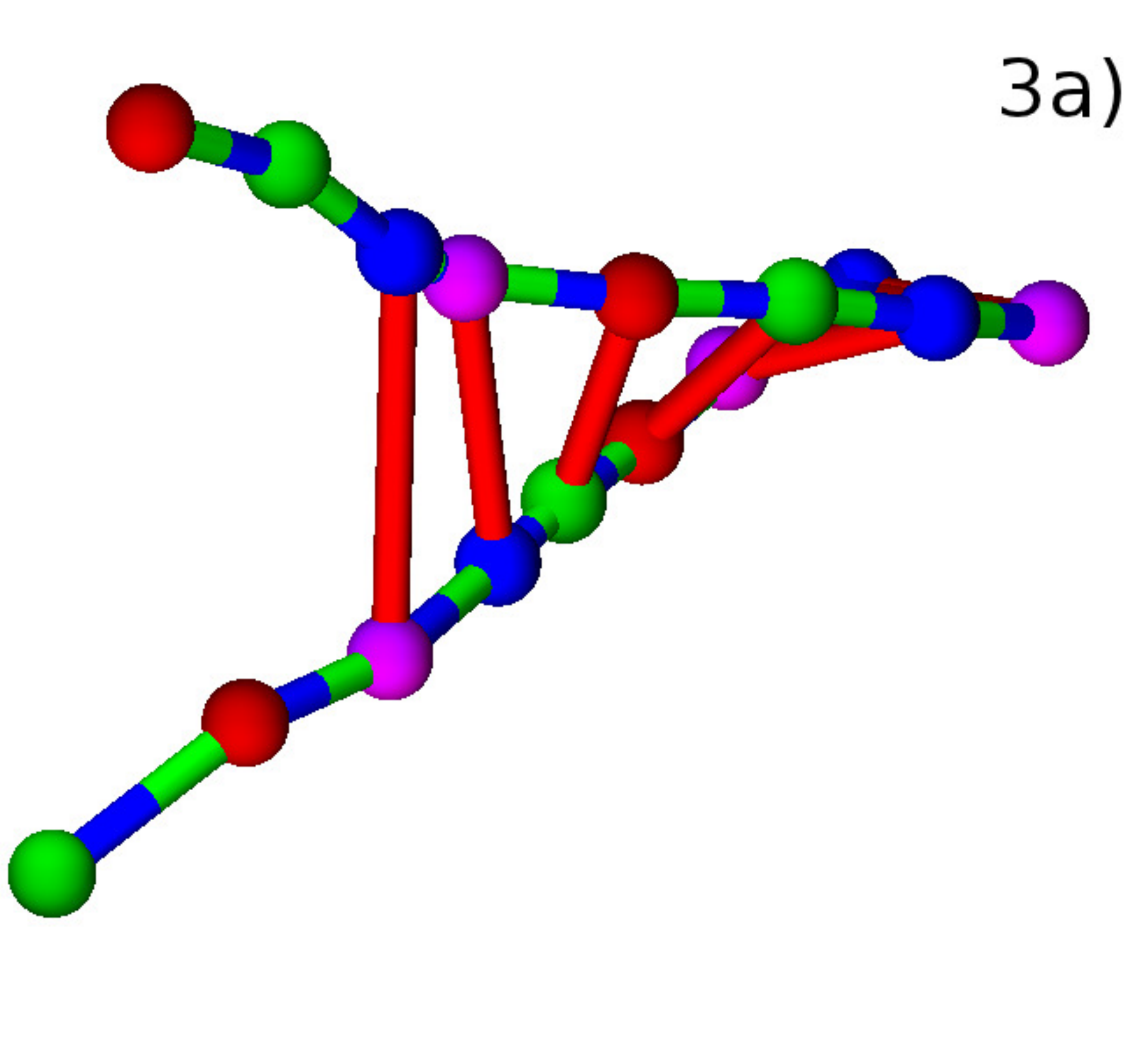}\includegraphics[width=0.33\columnwidth]{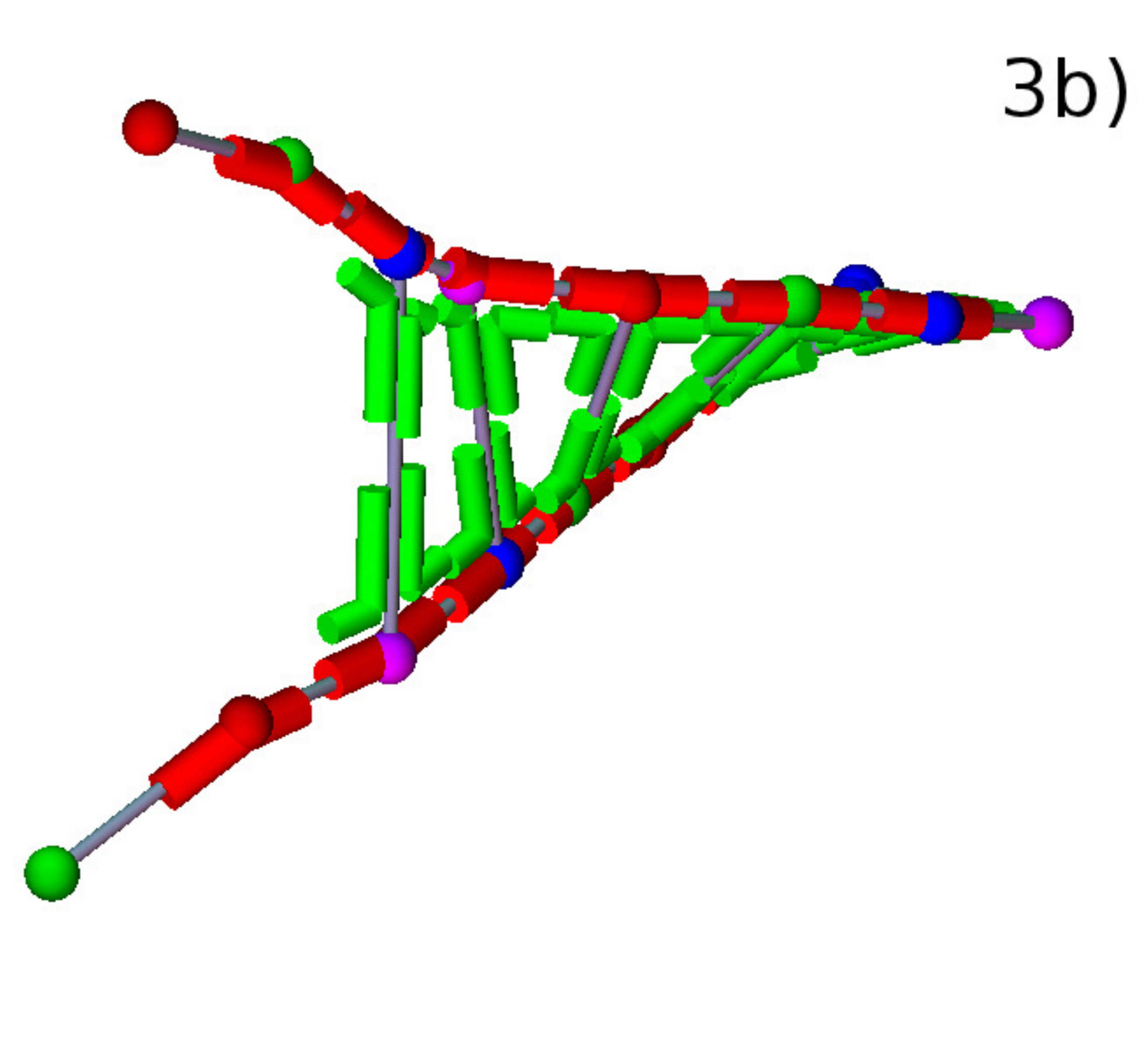}\includegraphics[width=0.33\columnwidth]{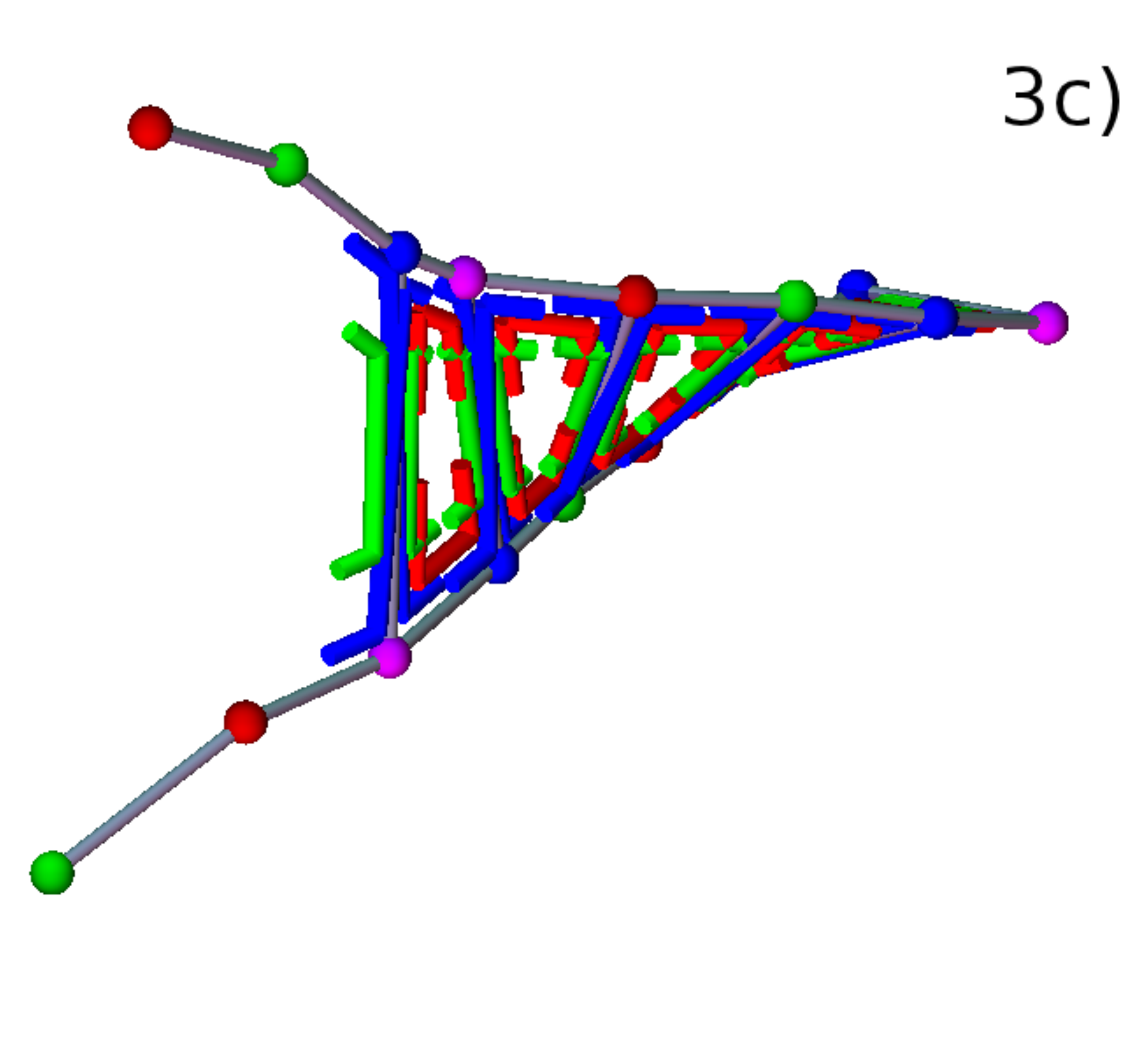}

\includegraphics[width=0.33\columnwidth]{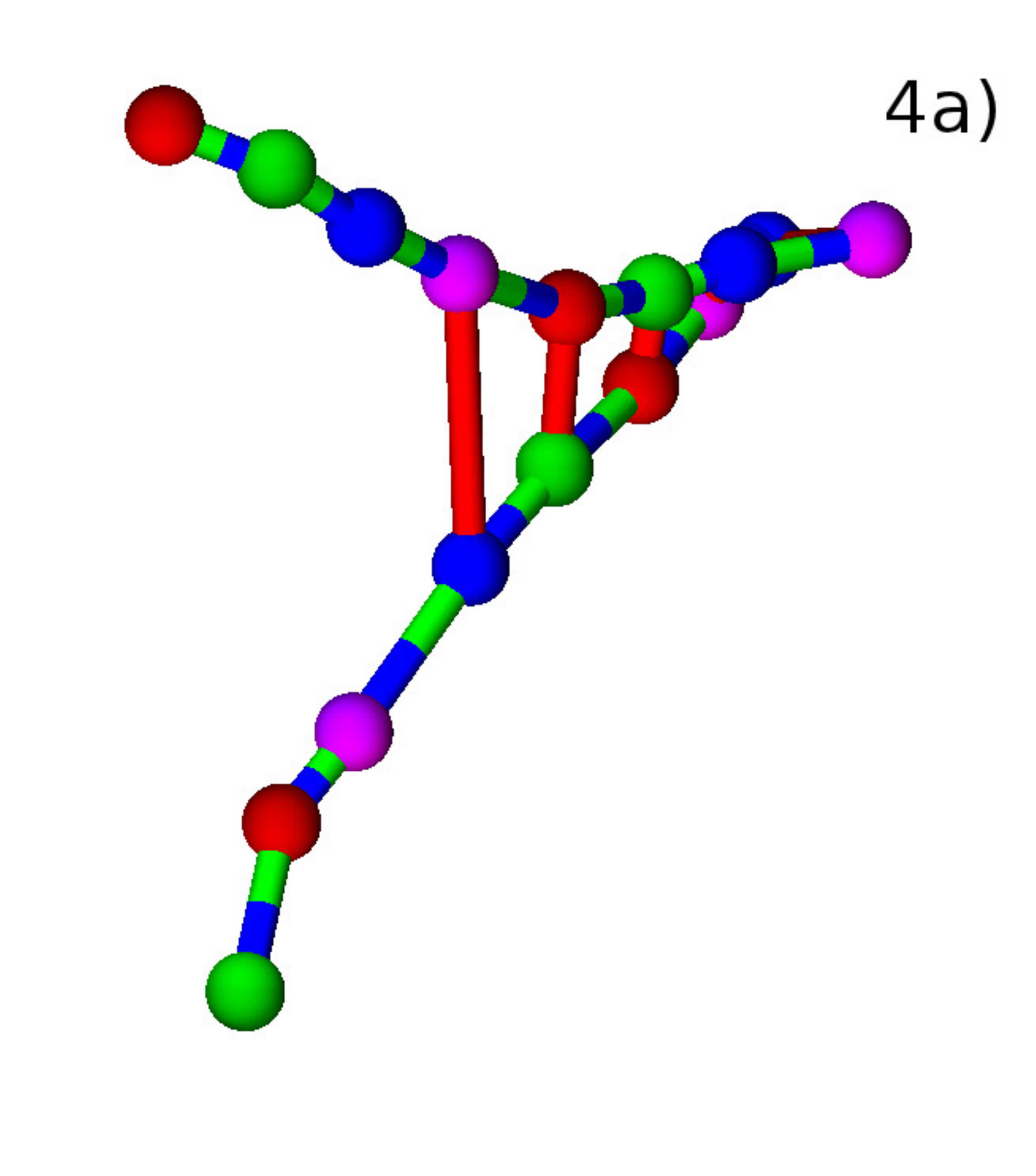}\includegraphics[width=0.33\columnwidth]{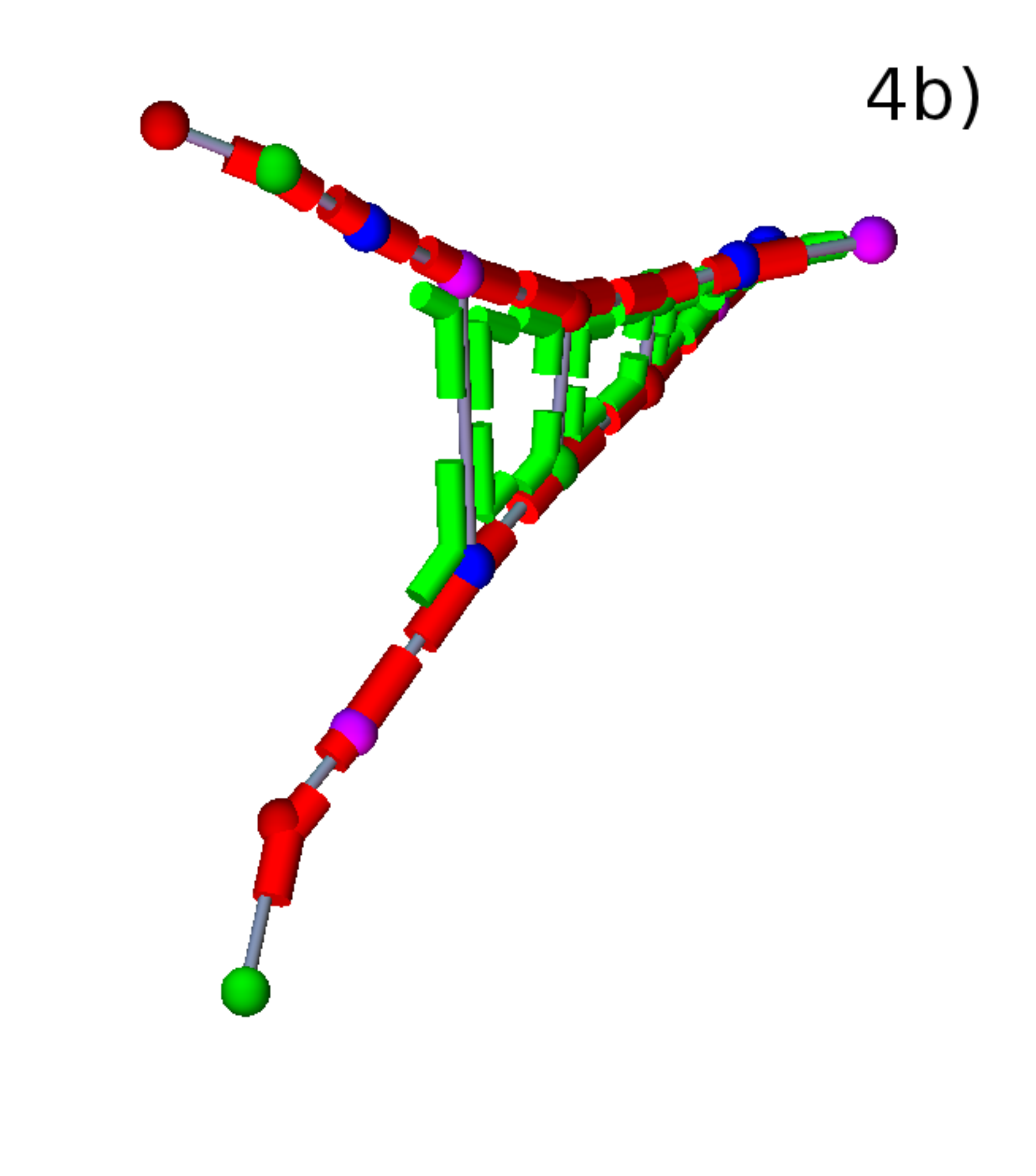}\includegraphics[width=0.33\columnwidth]{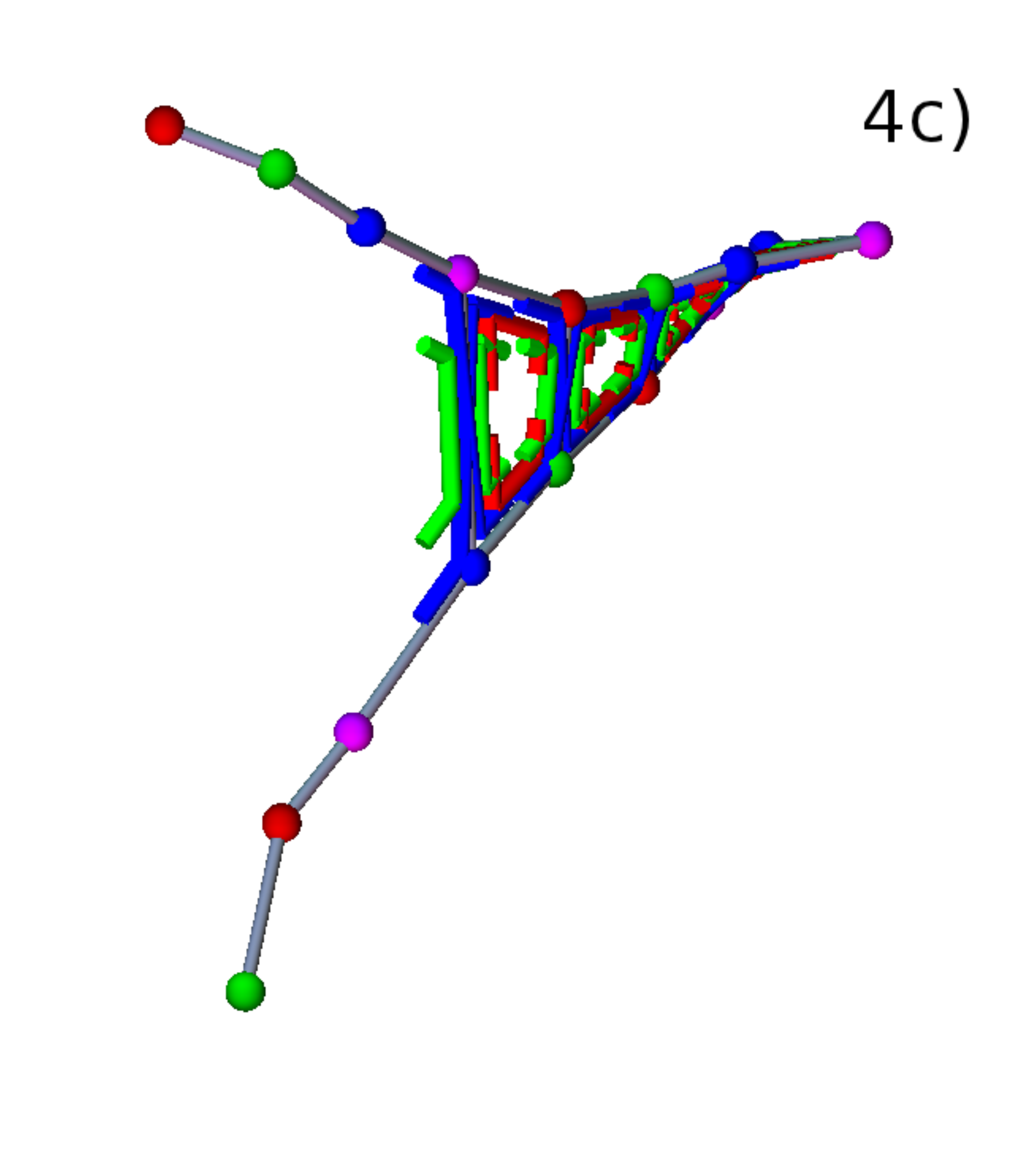}

\caption{\label{fig:example-pulling}DNA unzipping by a weak vertical force
$f=28\epsilon\sigma^{-1}$ applied to the left most bead pair (rows
1-4) for bond, angular and dihedral interactions (columns a-c). The
rows corresponds to times $1.72\tau$, $1.84\tau$, $3.03\tau$, $3.22\tau$,
respectively, starting from a straight double strand conformation
at $t=0\tau$.}
\end{figure}

To melt the double strand, we can e.g. apply an external force to
tear the two strands apart\cite{cocco2002theoretical} or increase
the temperature to let thermal fluctuations do the work. Fig. \ref{fig:example-pulling}
shows the result of applying an external opposing force to left most
nucleotide pair. Progressively the left most hybridization bond snaps.
Along with the breakage of hybridization bonds, we also see the gradual
removal of green angular interactions and all the dihedral interactions.
The external force is opposed by a single left most hybridization
bond along with the angular and dihedral interactions across the gab.
During the unzipping process, often the hybridization bonds are transiently
reformed just after breakage if thermal fluctuations pull them within
the hybridization reaction distance. 

\begin{figure}
\includegraphics[width=0.3\columnwidth]{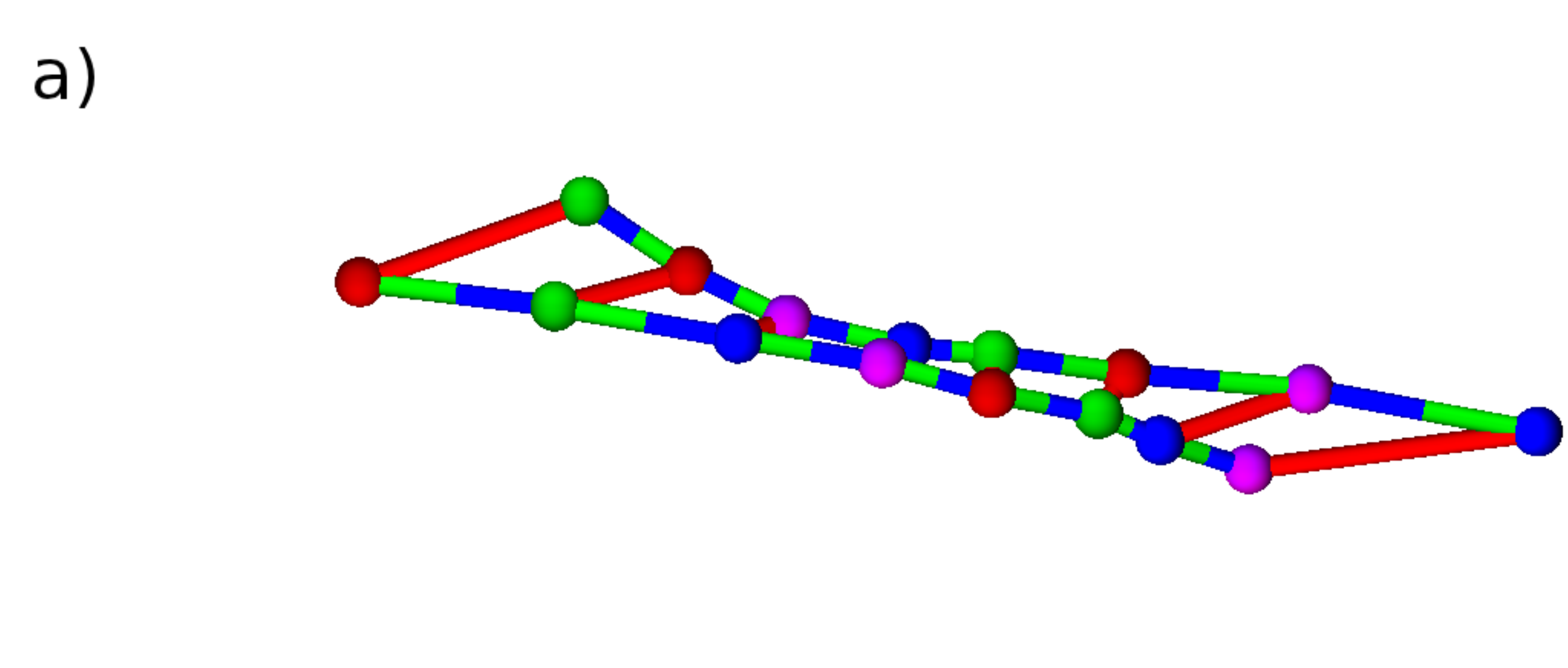}\includegraphics[width=0.3\columnwidth]{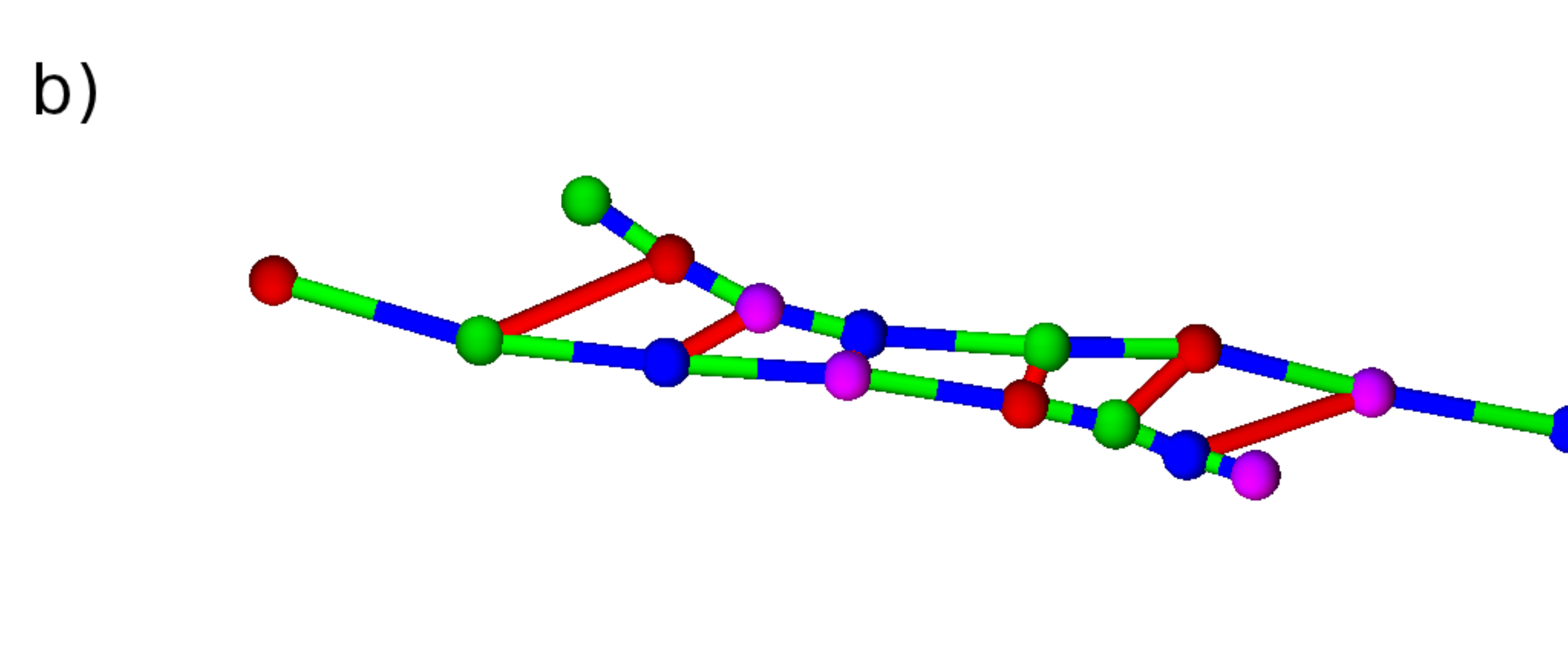}\includegraphics[width=0.3\columnwidth]{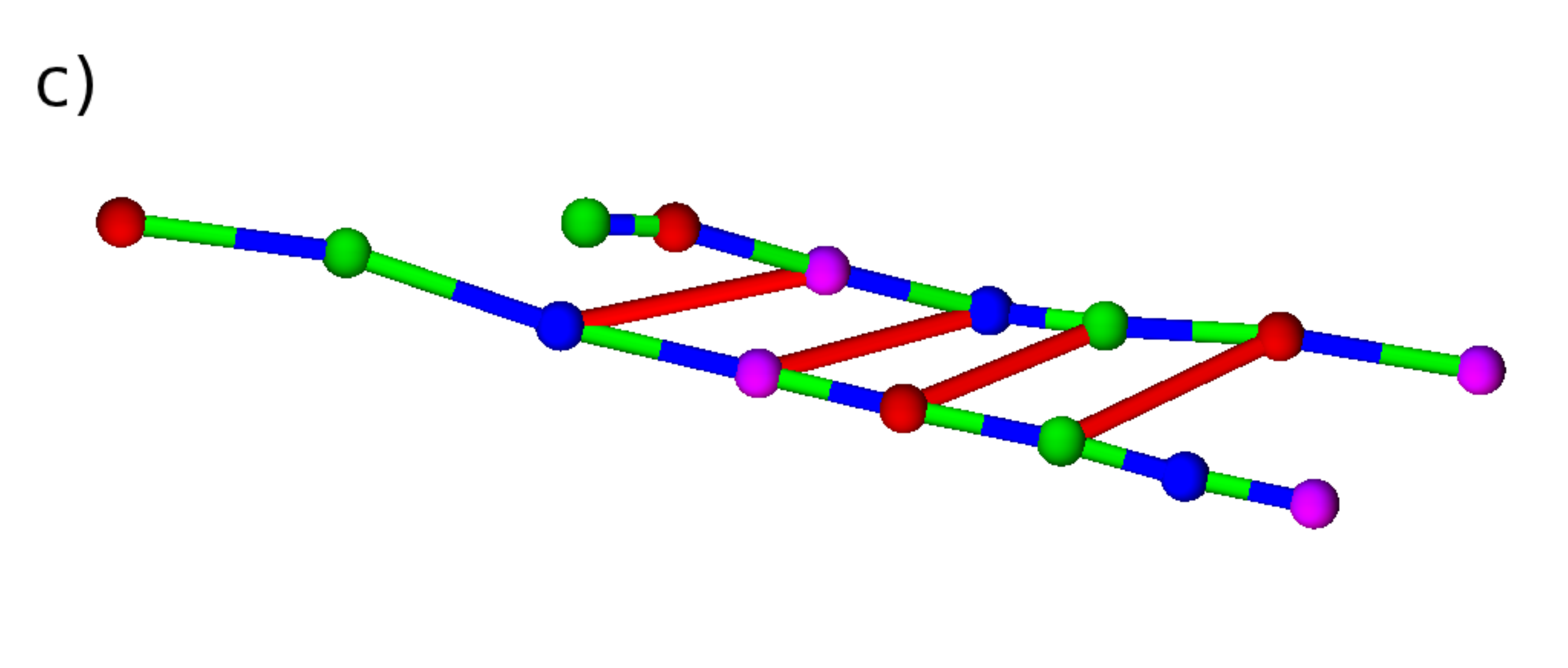}

\includegraphics[width=0.3\columnwidth]{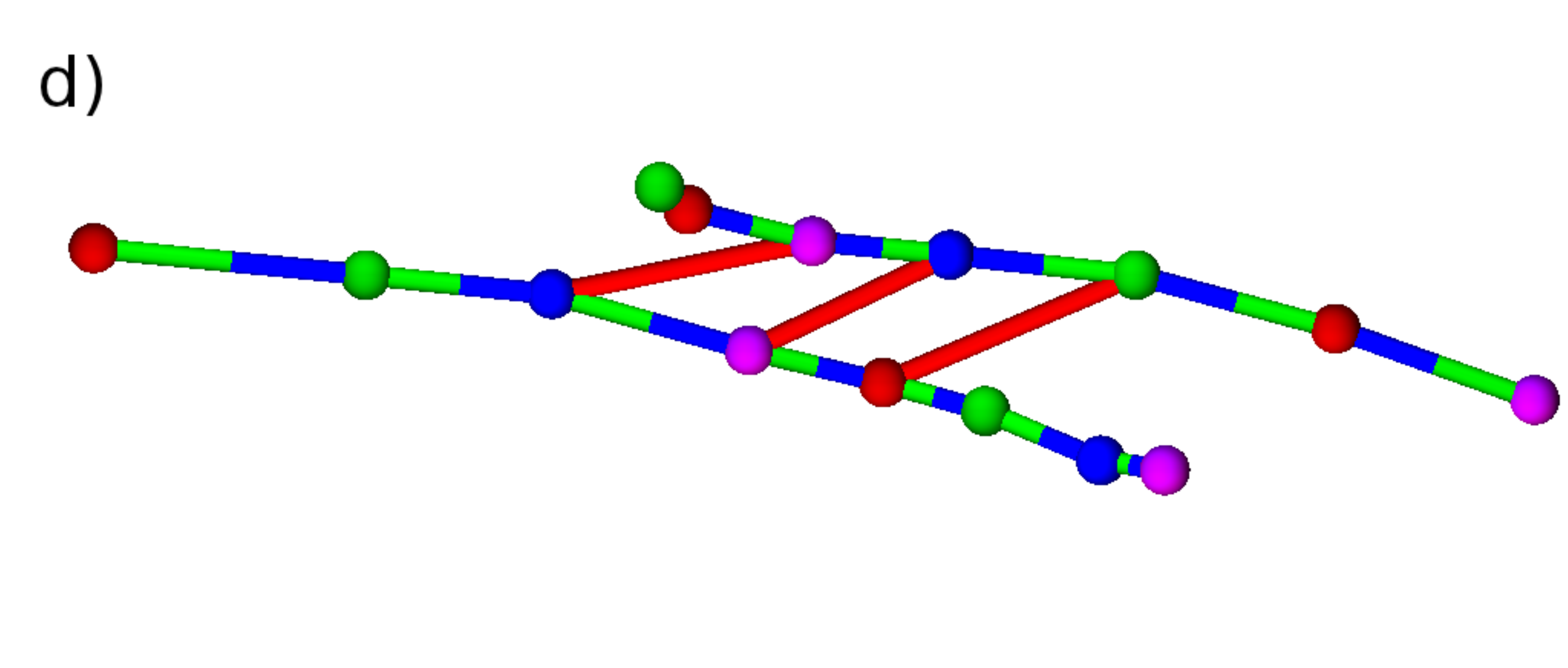}\includegraphics[width=0.3\columnwidth]{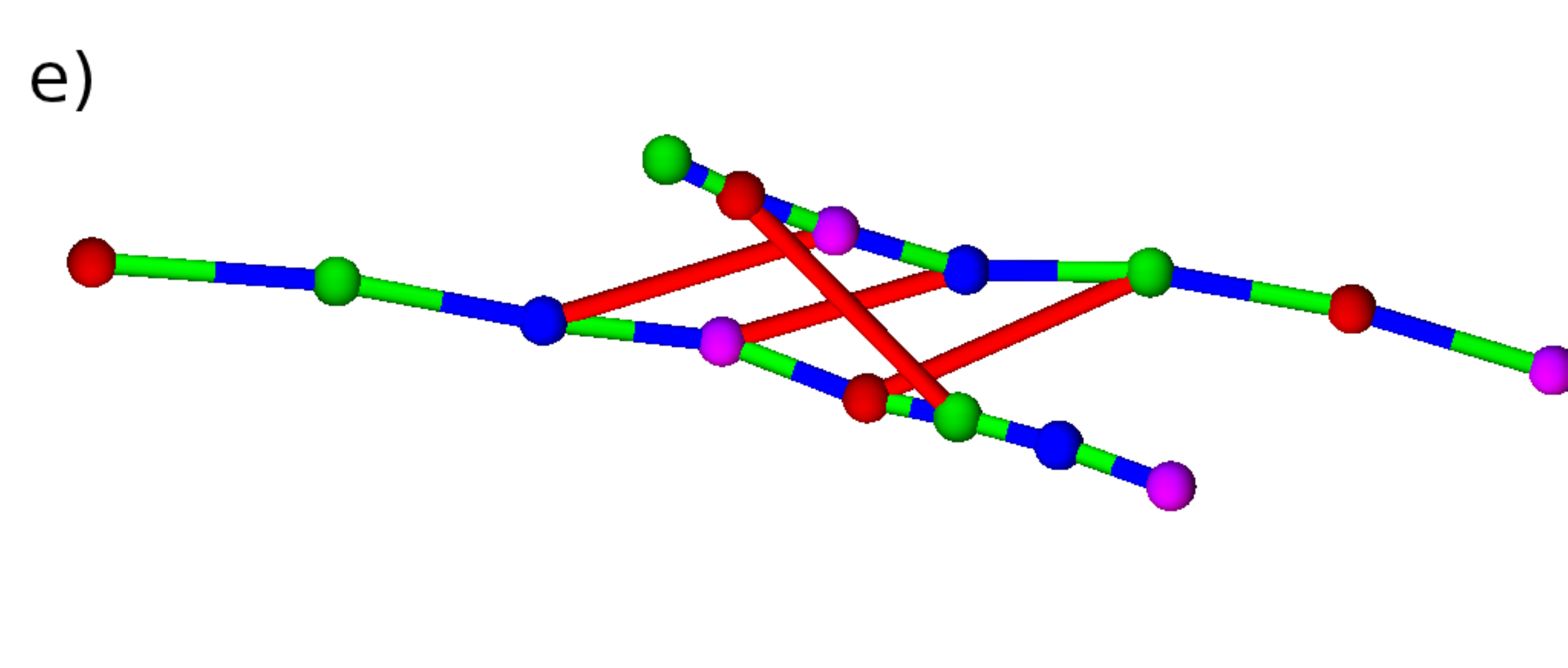}\includegraphics[width=0.3\columnwidth]{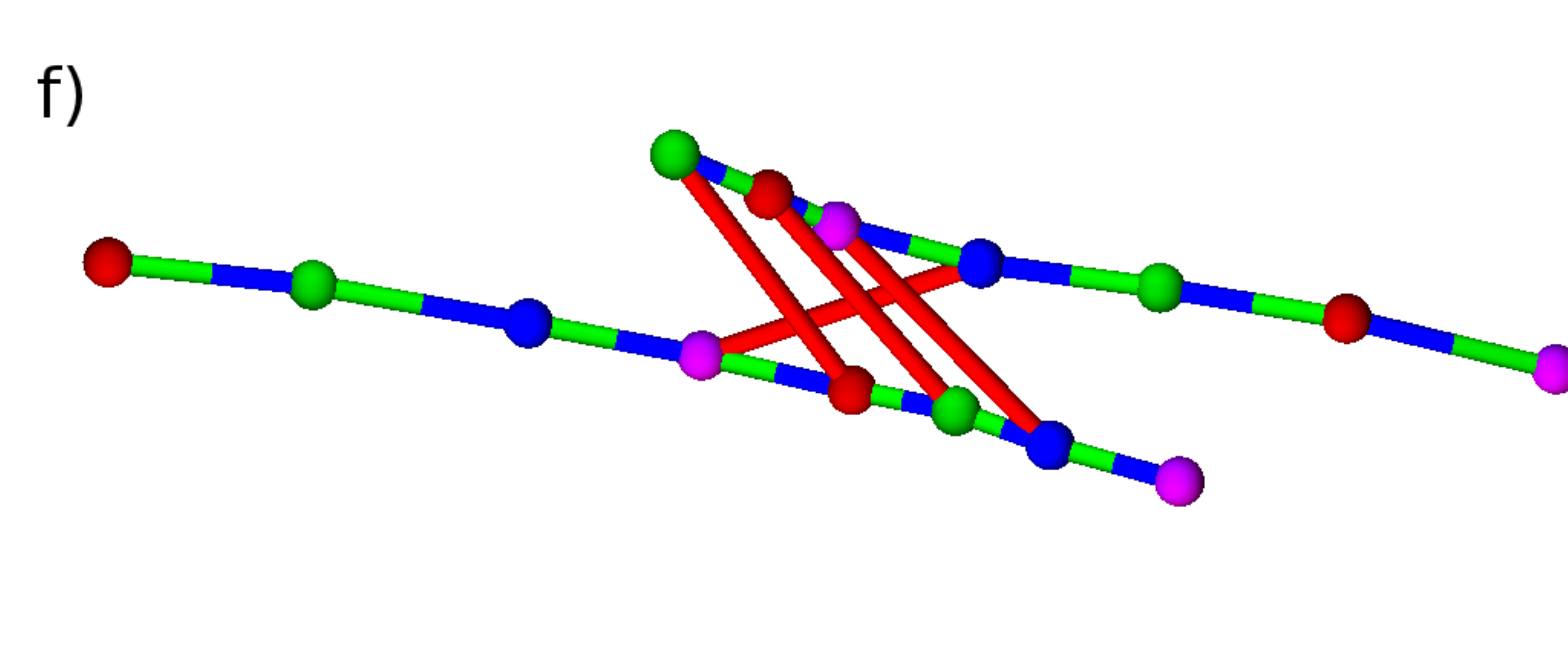}

\includegraphics[width=0.3\columnwidth]{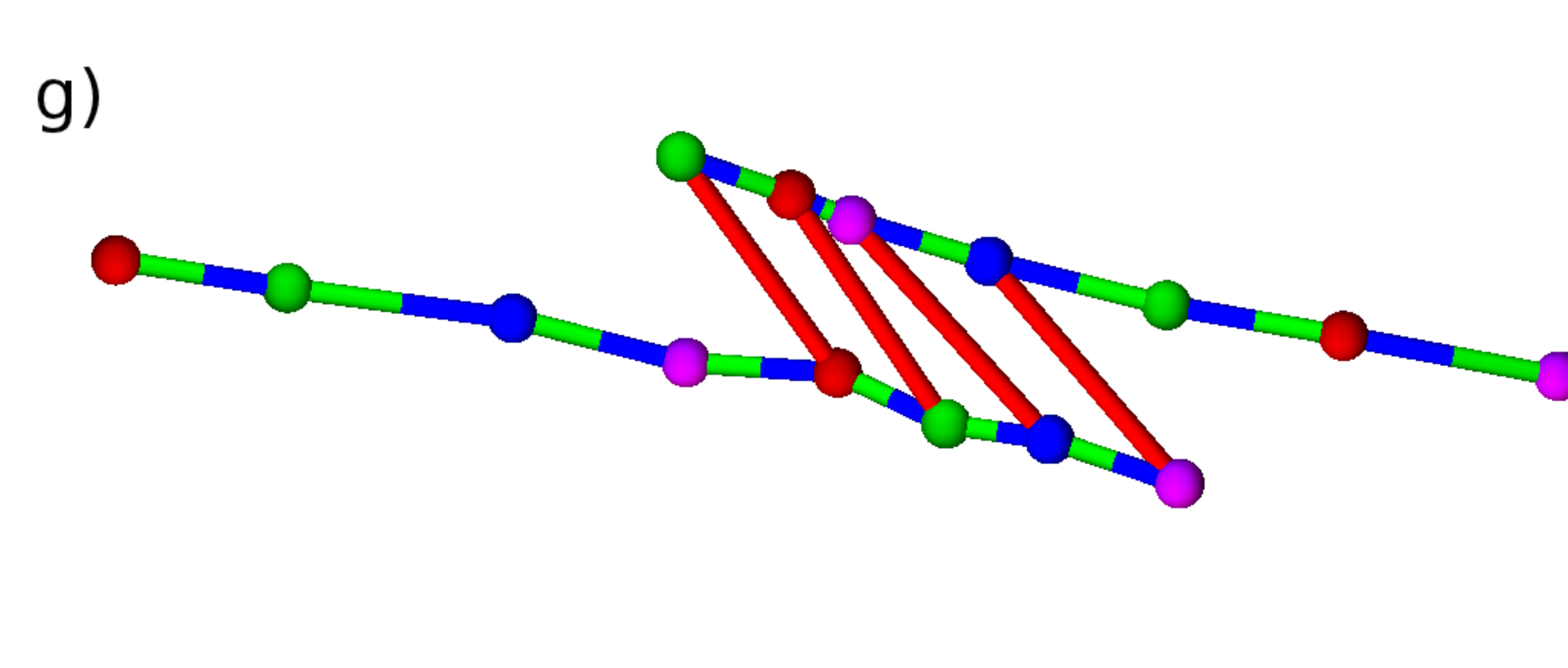}\includegraphics[width=0.3\columnwidth]{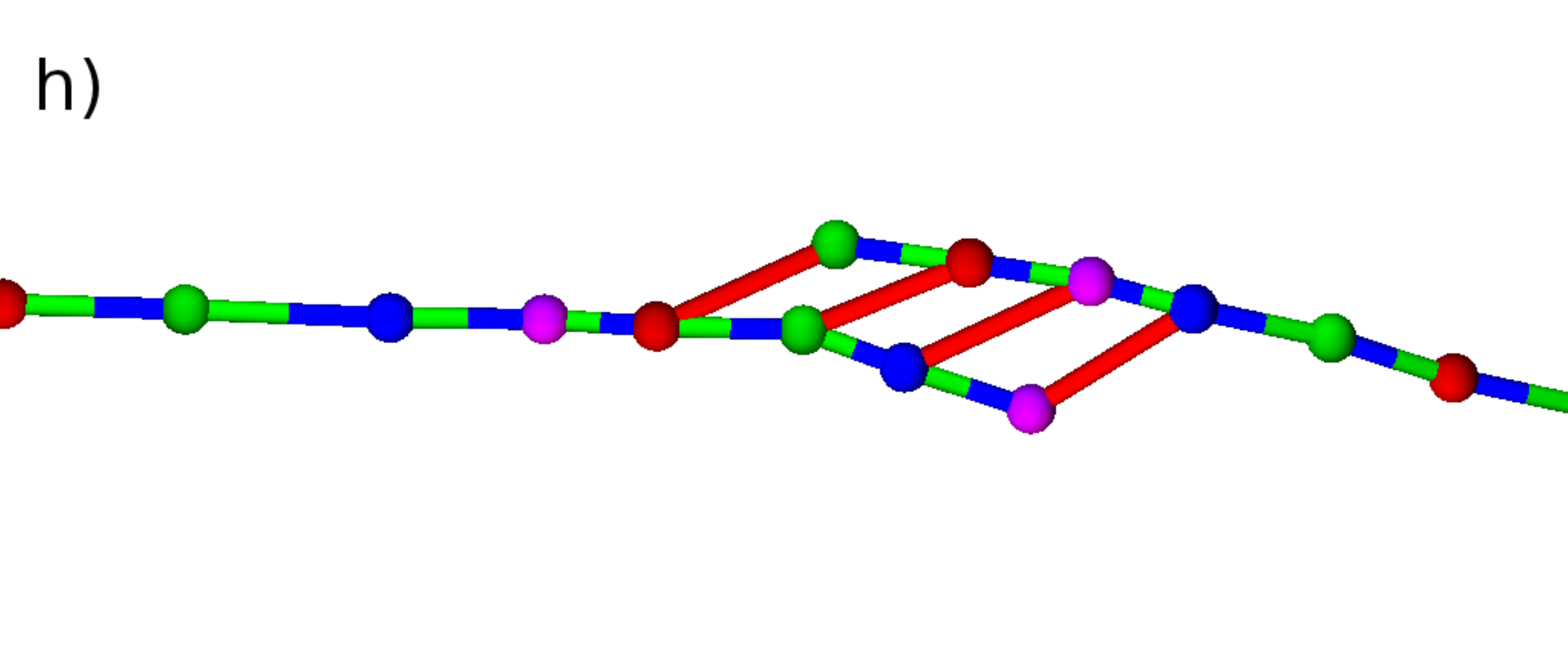}\includegraphics[width=0.3\columnwidth]{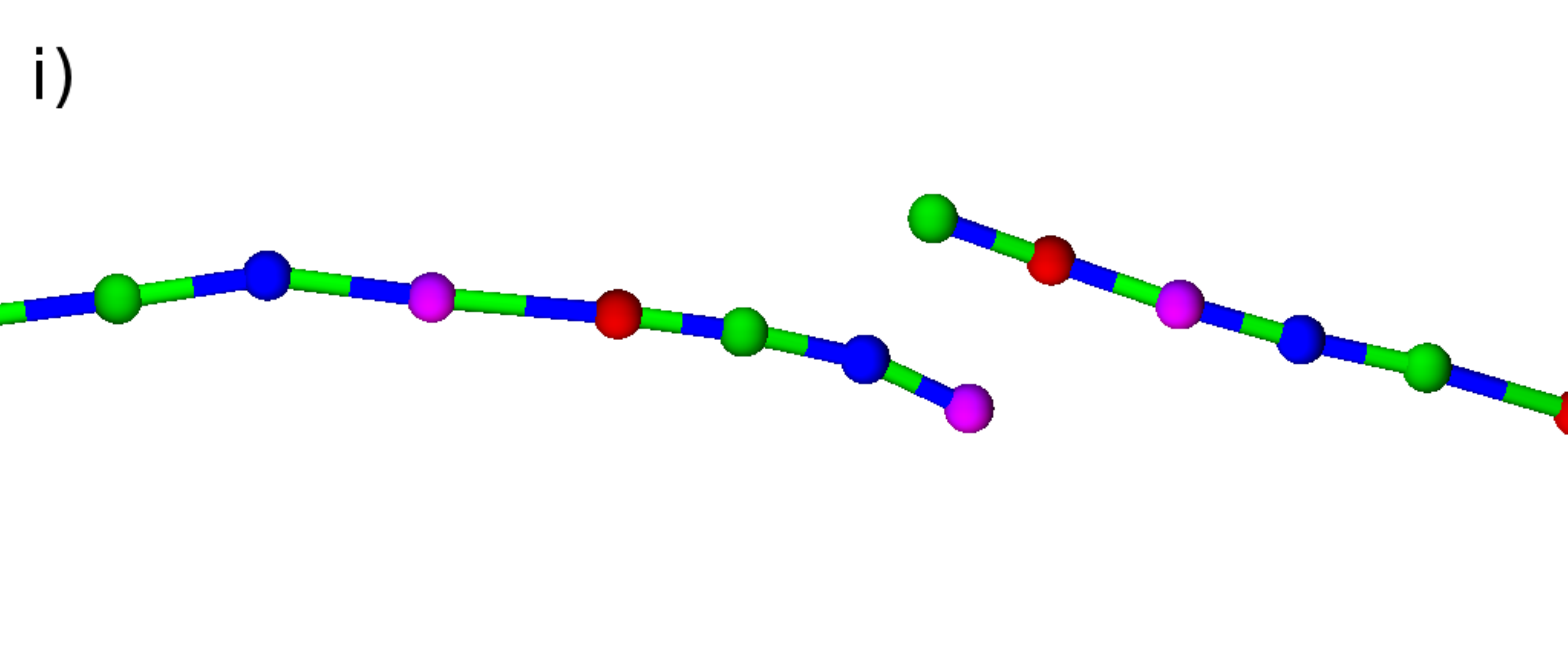}

\caption{\label{fig:example-force-horizontal}Time series of DNA unzipping
by a strong horizontal force $f=100\epsilon\sigma^{-1}$ applied to
the left and right most beads of the two strands (a-h). The snapshots
corresponds to times $0.21\tau$, $0.30\tau$, $0.59\tau$ (top row),
$0.81\tau$, $0.89\tau$, $0.92\tau$ (middle row), and $0.95\tau$,
$1.20\tau$, $1.37\tau$ (bottom row) starting from a straight double
stranded conformation at $t=0\tau$.}
\end{figure}

In fig. \ref{fig:example-force-horizontal} we perform another pulling
experiment, where a much stronger horizontal force is applied to the
left most bottom strand and right most top strand beads of the double
strand. Initially the whole molecule is sheared, as all the green
angular interactions cooperate in opposing the deformation. Gradually
bonds snaps from either end towards the center. Interestingly, since
the two molecules have a 4-nucleotide long repeating sequence, when
the hybridization bonds are broken, they very rapidly reform with
the complementary beads one repeat sequence further down the molecule.
The shear process repeats for the second hybridization sequence until
it too is broken, and two single strands are formed.

\begin{figure}
\includegraphics[width=0.33\columnwidth]{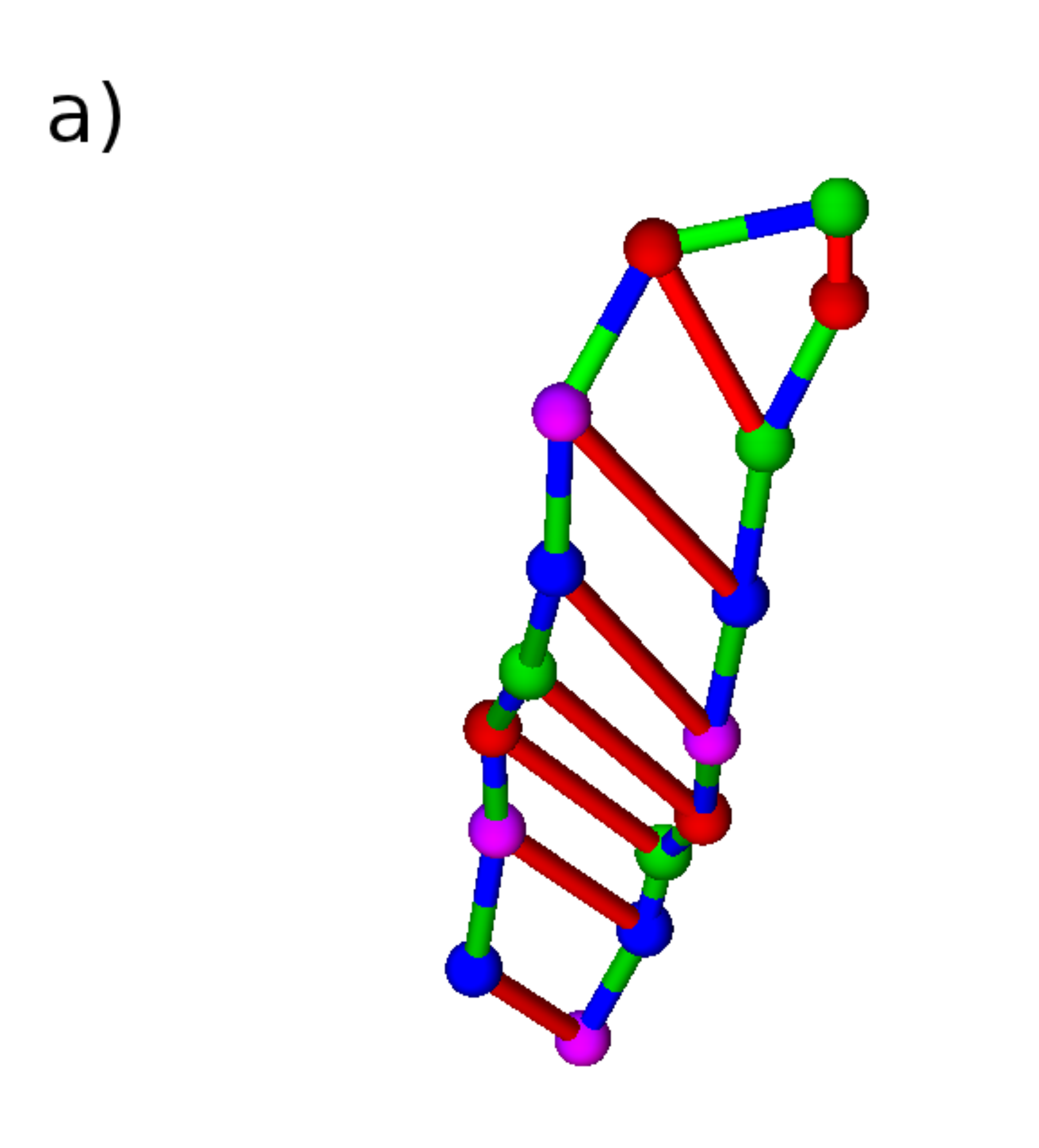}\includegraphics[width=0.33\columnwidth]{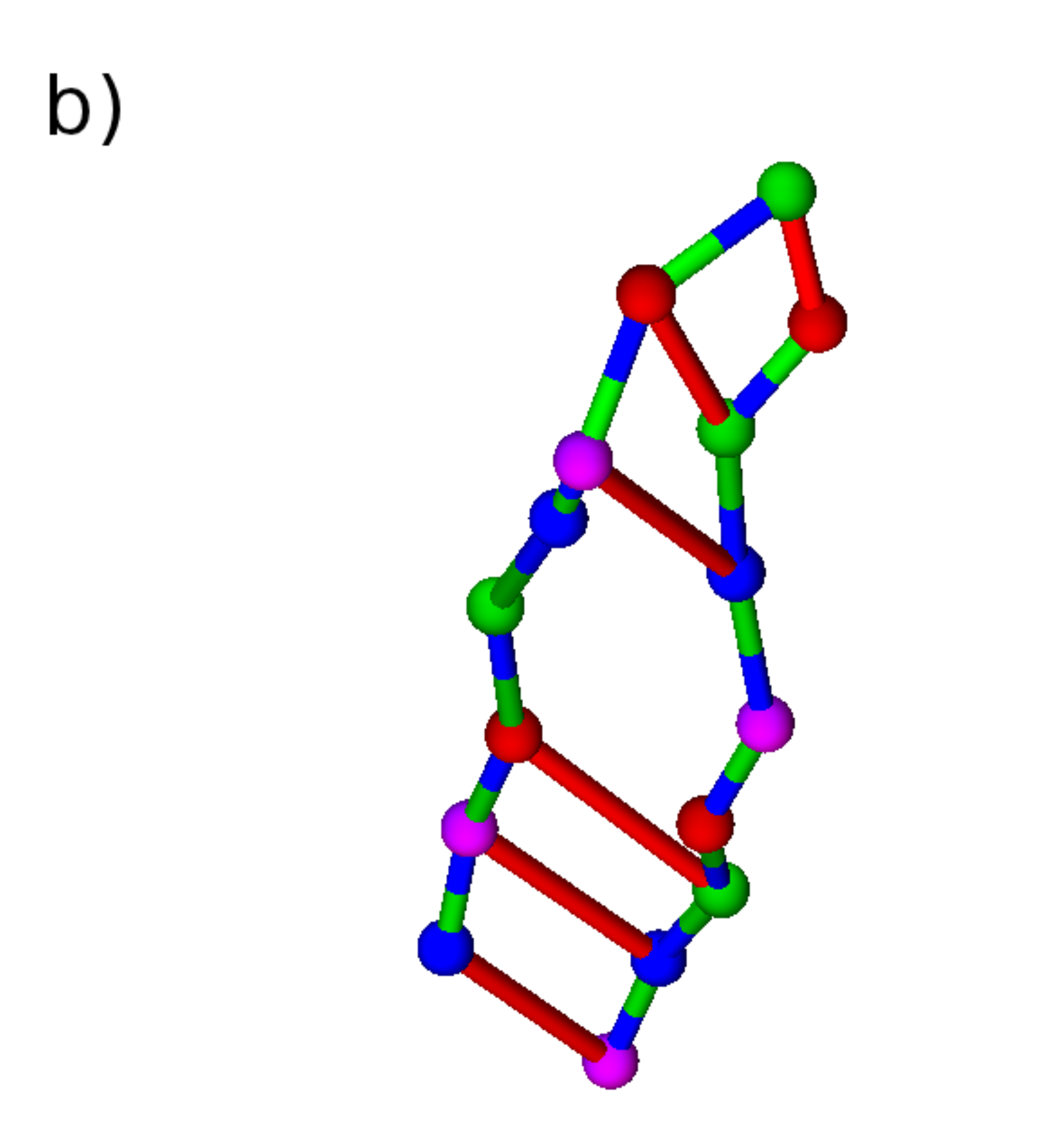}\includegraphics[width=0.33\columnwidth]{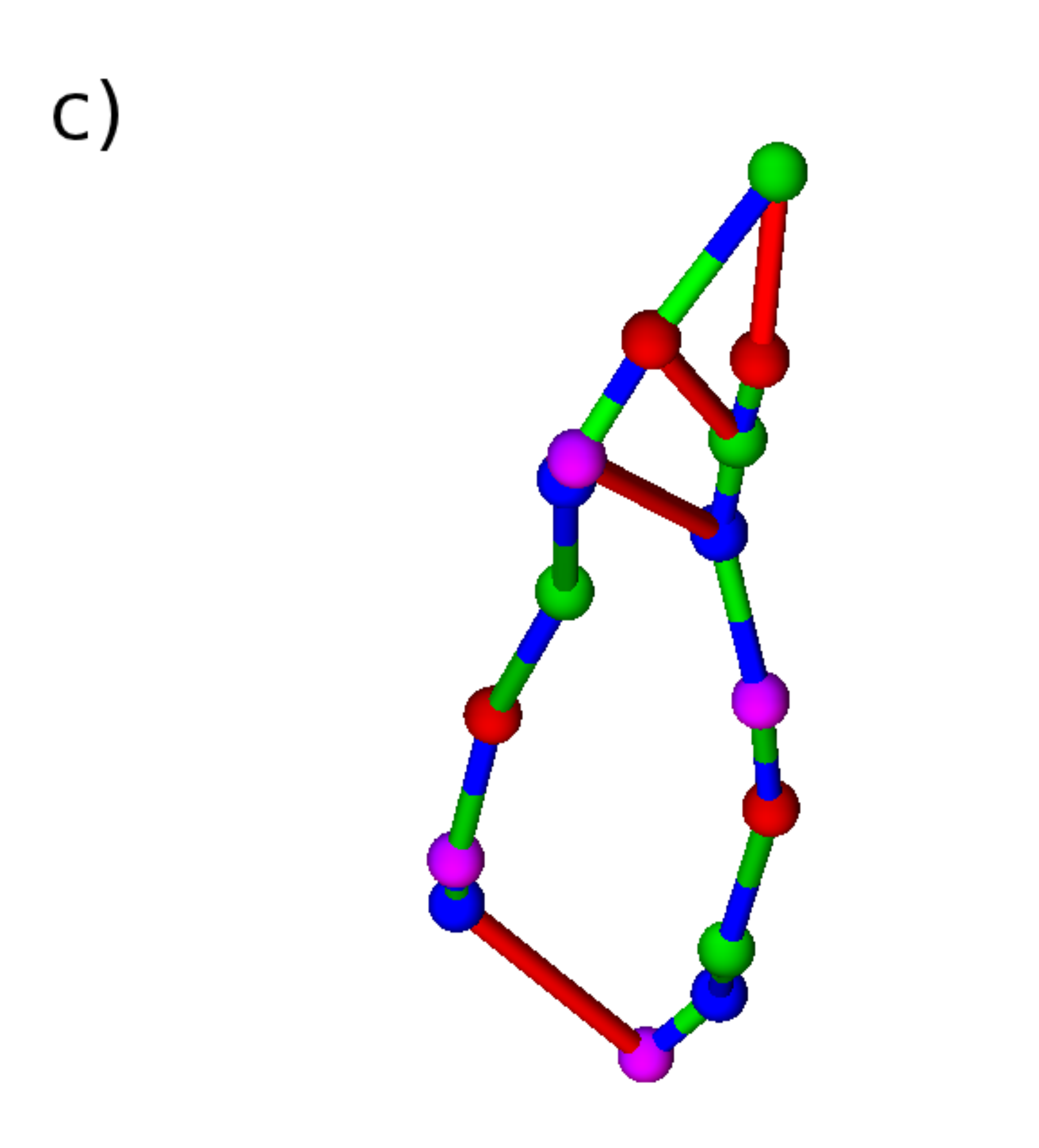}

\includegraphics[width=0.3\columnwidth]{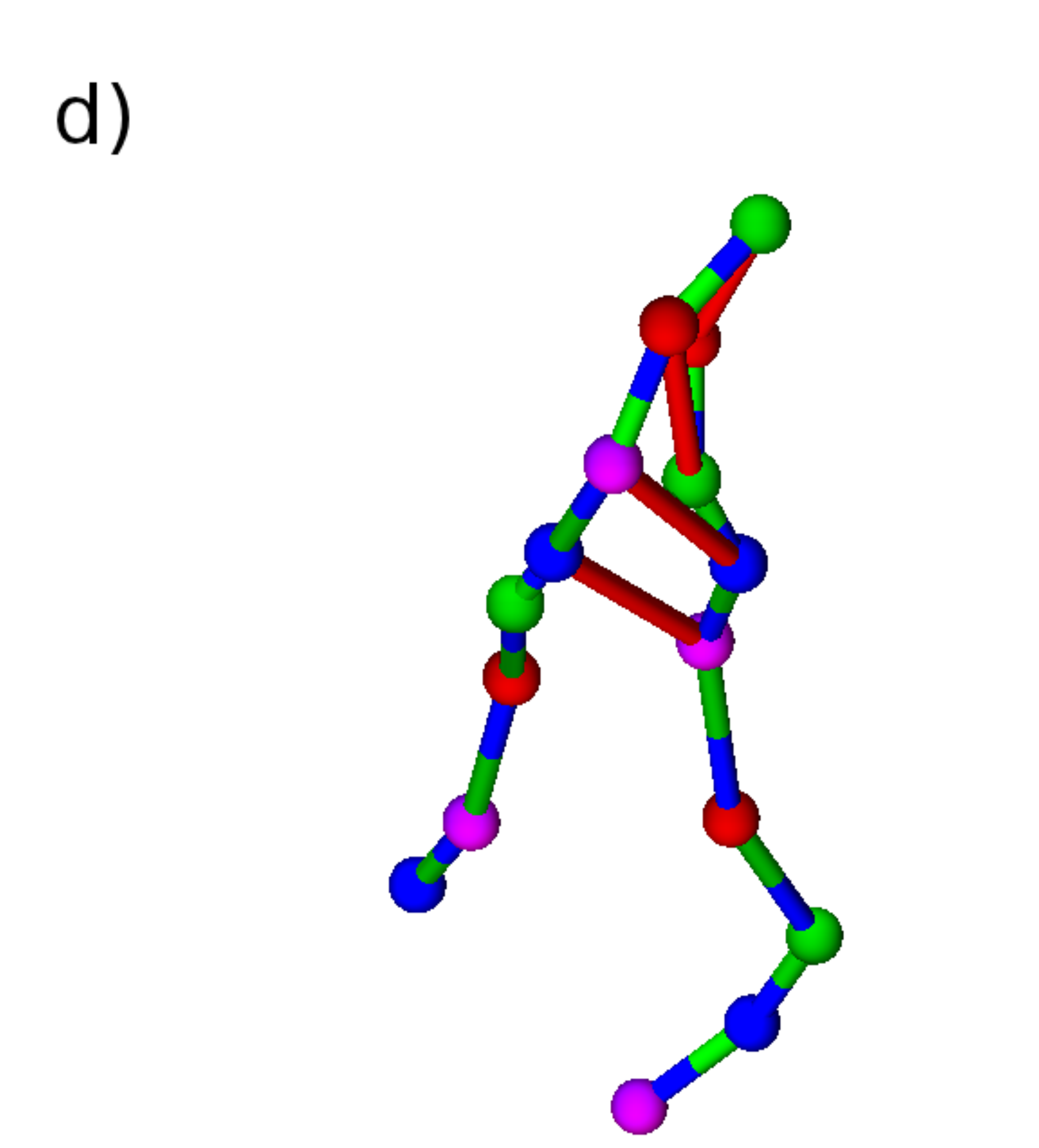}\includegraphics[width=0.33\columnwidth]{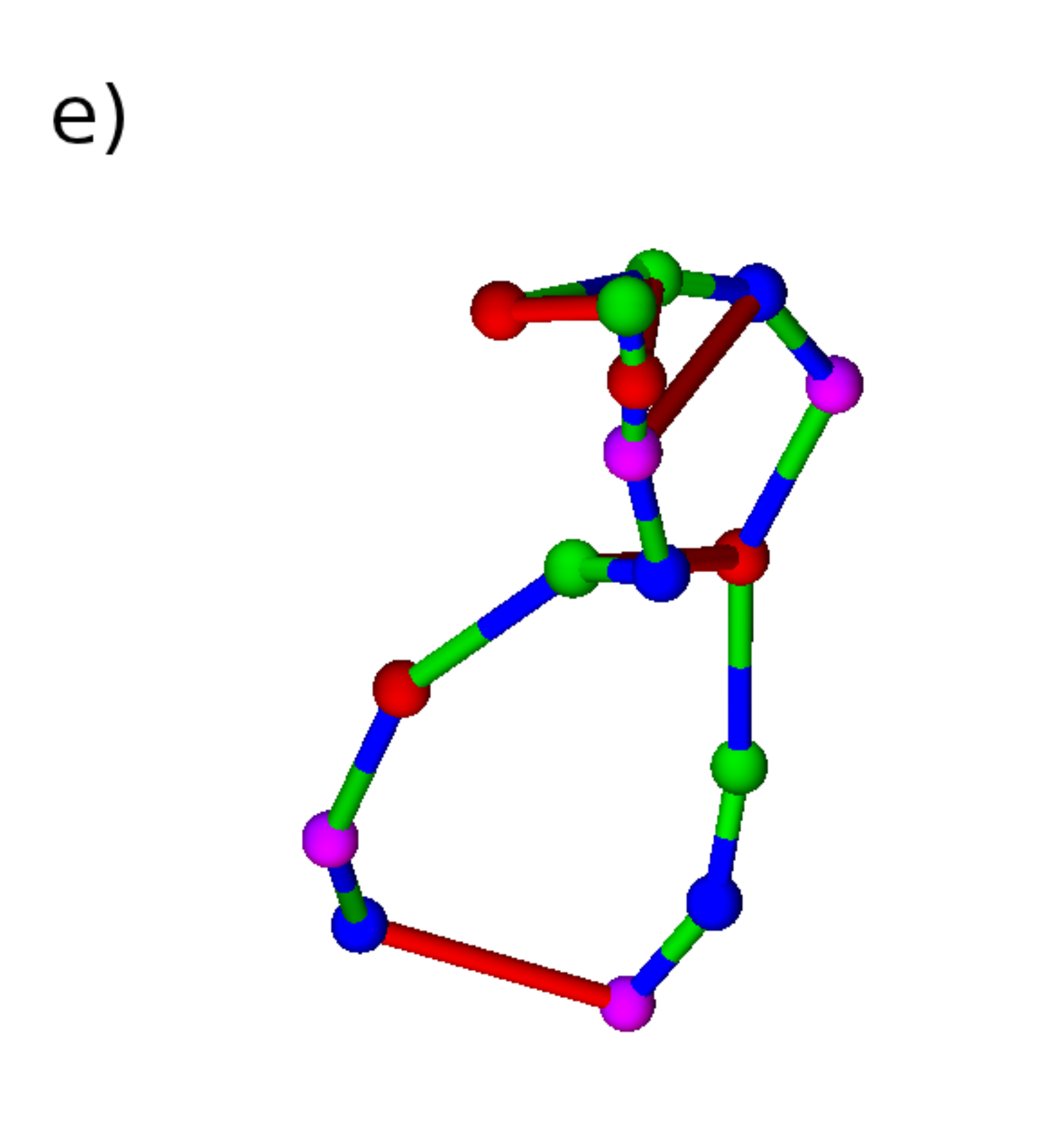}\includegraphics[width=0.33\columnwidth]{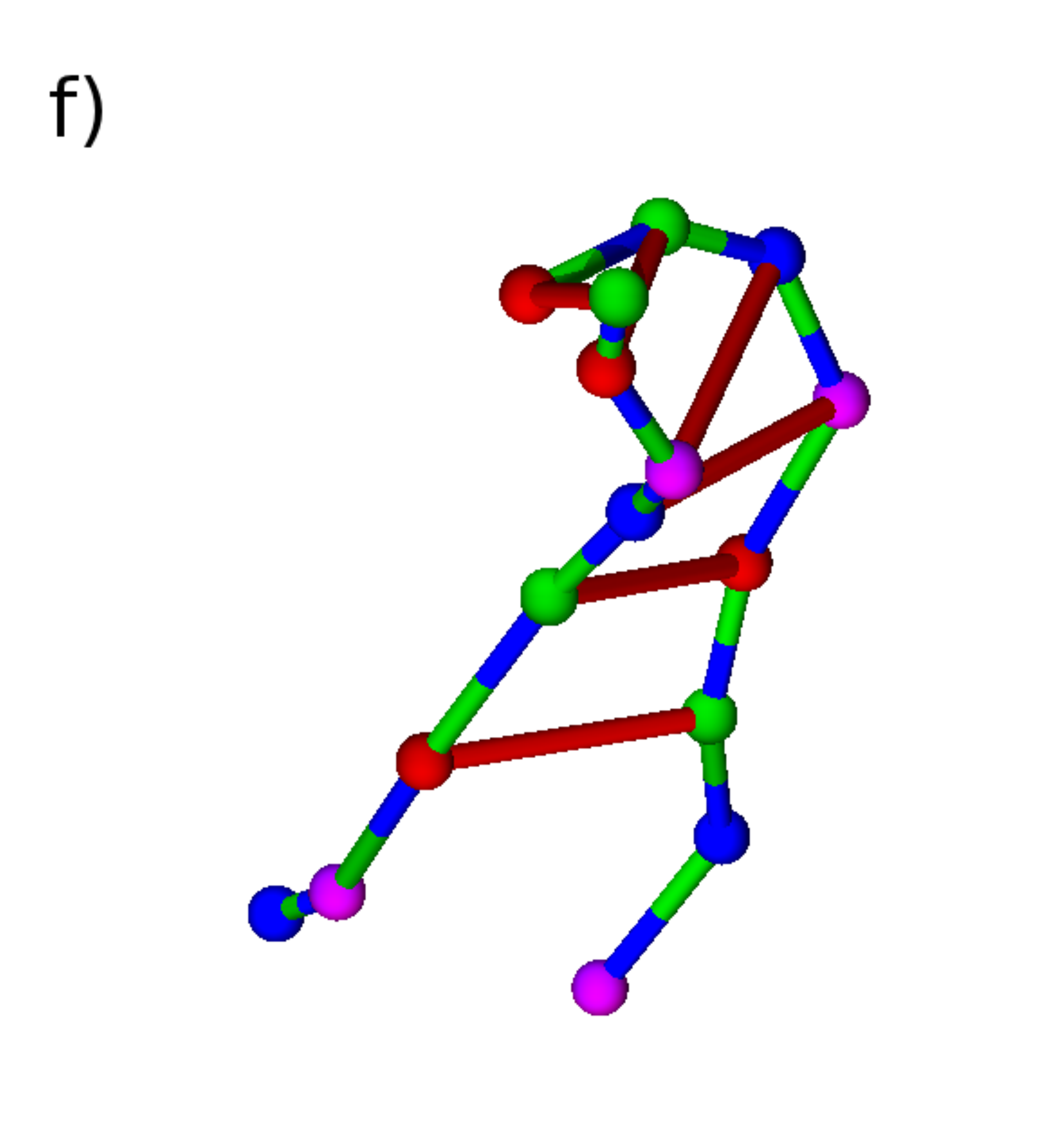}

\includegraphics[width=0.33\columnwidth]{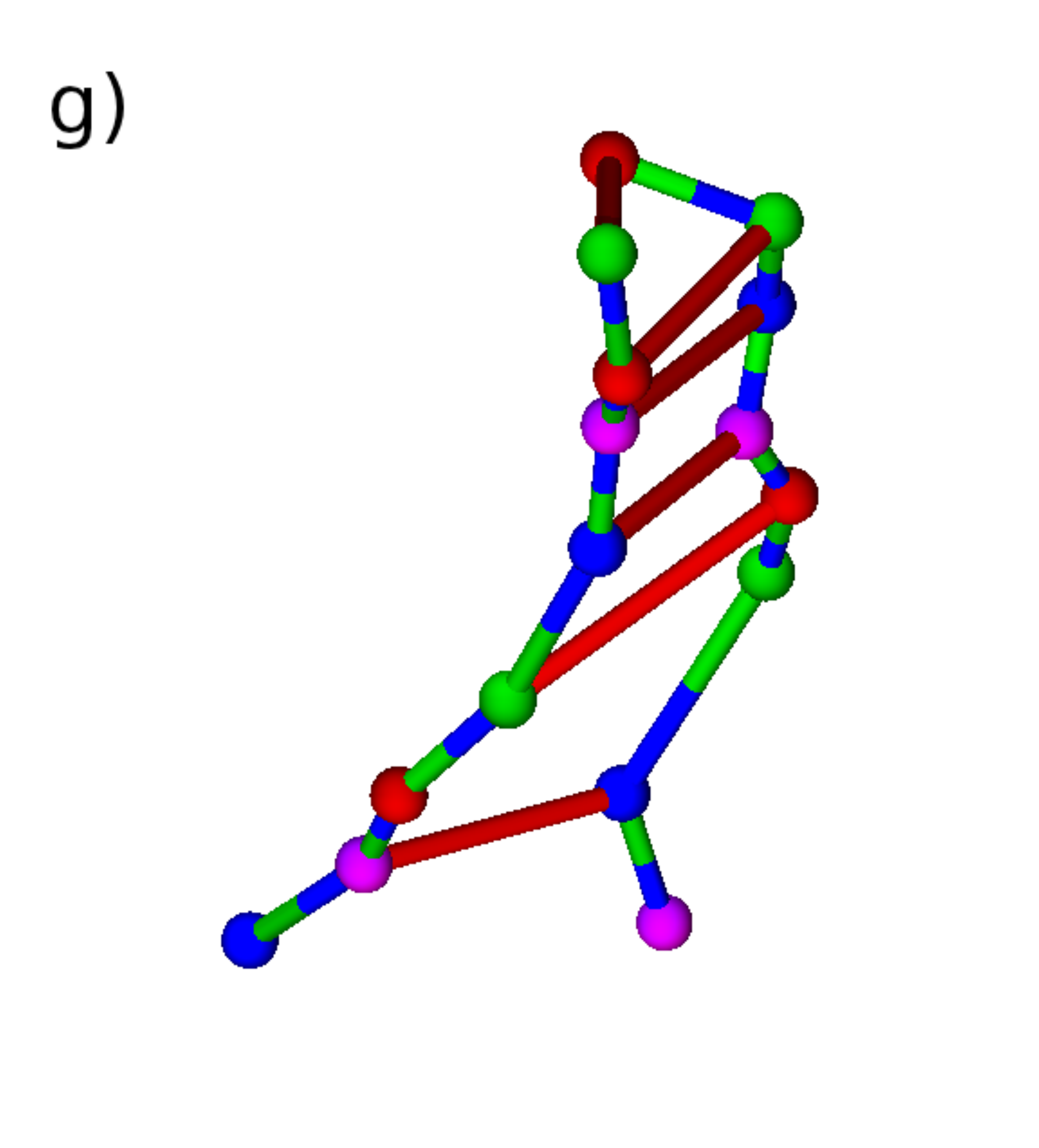}\includegraphics[width=0.33\columnwidth]{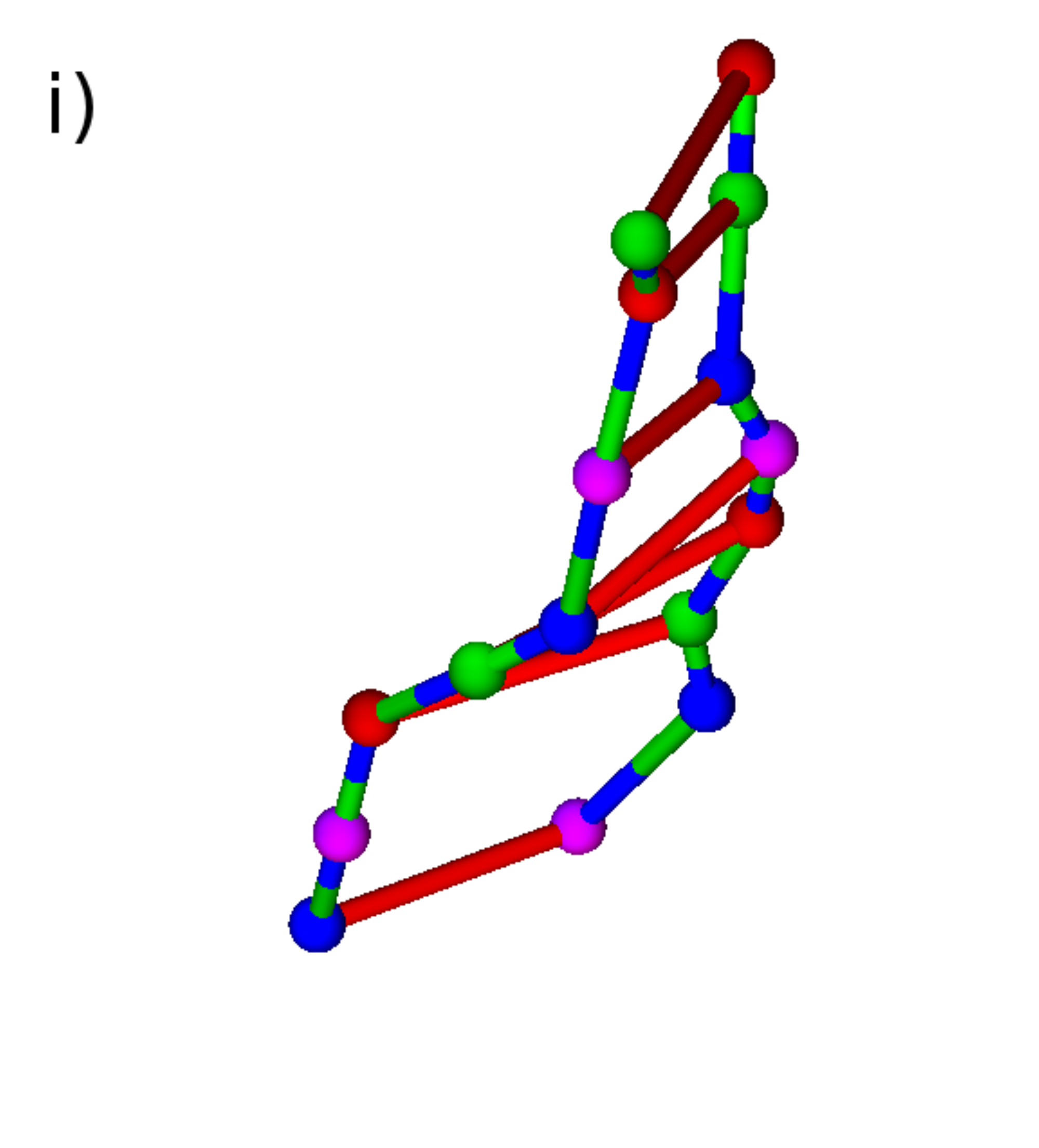}\includegraphics[width=0.33\columnwidth]{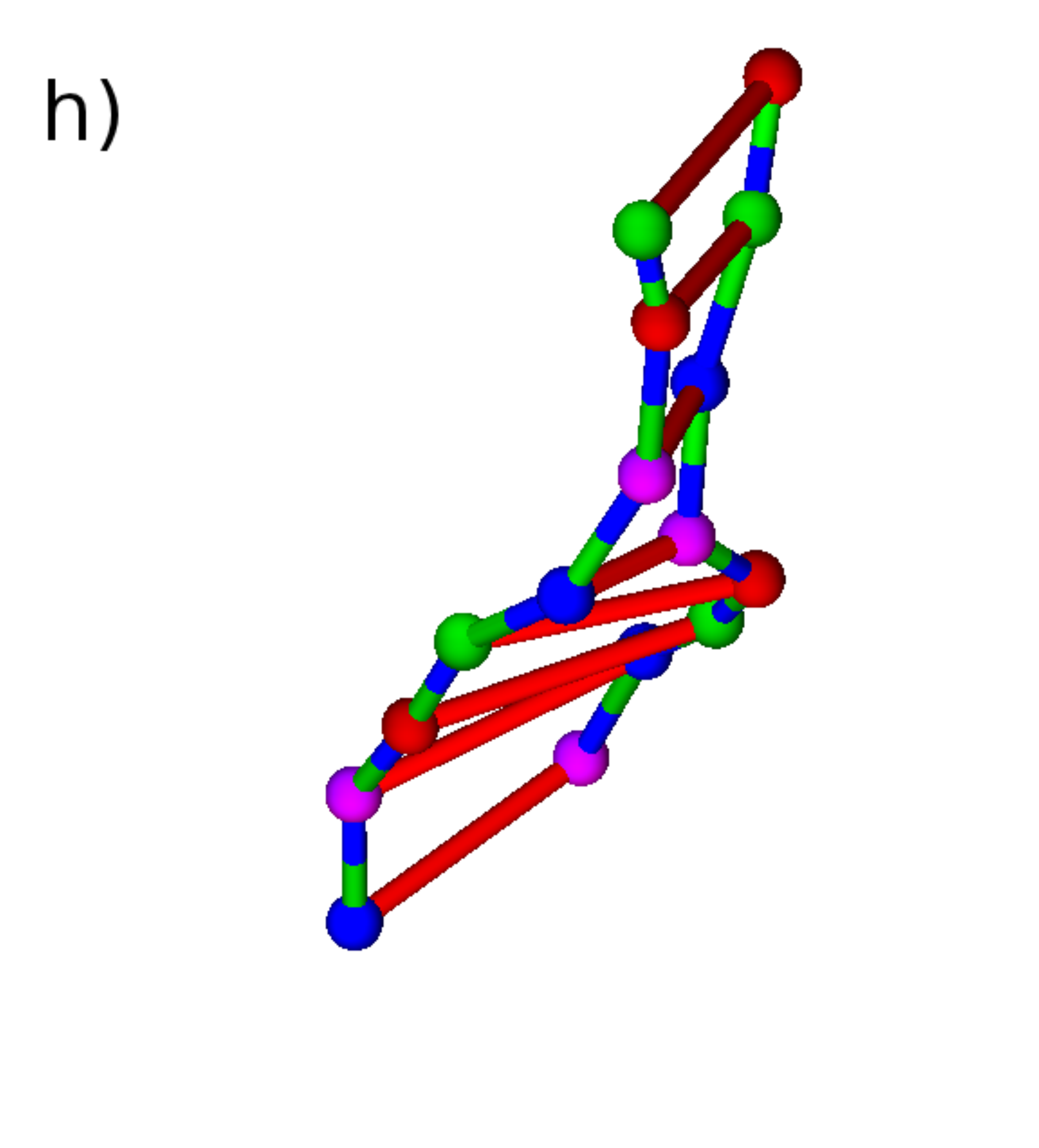}

\caption{\label{fig:example-bubble}Time series showing bubble opening and
closing dynamics for DNA at an elevated temperature $T=5\epsilon$
(a-h). The snapshots are from times $t=55.40\tau$, $55.44\tau$,
$55.48\tau$ (top row), $55.55\tau$, $55.85\tau$, $55.89\tau$ (middle
row), and $55.96\tau$, $56.05\tau$, $56.09\tau$ (bottom row) starting
from a straight double stranded conformation at $t=0\tau$.}
\end{figure}

DNA can be molten by raising the temperature. The melting temperature
depends on the sequence, the length of the strands as well as the
strand concentration.\cite{poland1966phase,owczarzy1997predicting}
Prior to melting, bubbles of open nucleotide sequences appear since
they contribute configuration entropy and hence lower the free energy
similar to vacancies in crystals. At increased temperatures, the number
and the size of these bubbles grow and cause the two strands to melt.\cite{PhysRevLett.91.148101,metzler2005dynamic,rapti2006lengthscales,ambjornsson2007master}
In fig. \ref{fig:example-bubble} we show a time series of a bubble,
that is created by breaking a single hybridization bond, the bubble
grows until it breaks the last hybridization bond. However, the two
frayed strands form a hybridization bond at the end, and progressively
the bubble closes again. Simulating the chain for sufficiently long
time at an elevated temperature will cause the double stands to melt
with a transition very much like the one shown in fig. \ref{fig:example-bubble}.

\section{Conclusions\label{sec:Conclusions}}

We have implemented a versatile framework for studying the effects
of dynamic bonding of ordinary and directional bonds in coarse-grained
models within the context of the Large-scale Atomic/Molecular Massively
Parallel Simulator (LAMMPS)\cite{Lammps}. The dynamic
bonding framework ensures that angular and dihedral interactions are
kept consistent during bond breakage and creation. The code has been
parallelized and optimized to the case where the bond formation or
breakage probability for each bead is relatively low. Since the dynamic
bonding code is very modular it will be easy to extend with other
types of bonding rules. The dynamic bonding framework was written
with the aim of developing a new type of coarse-grained models of
DNA dynamics. We have illustrated a dynamic bonding DNA model using
DNA hybridization and ligation, as well as two geometries of force
induced unzipping and bubble dynamics. Clearly the present DNA model
is very simple, nonetheless it qualitatively captures some of the
fundamental phenomena of DNA molecules. The dynamic bonding framework
will allow us to build DNA models, that we expect will provide quantitative
predictions as good as the Poland-Scheraga model\cite{poland1966phase,jost2009unified},
while we can use these DNA models as components in Molecular Dynamics
and Dissipative Particle Dynamics simulations of hybrid materials
containing both soft-condensed matter and DNA molecules.

\section{Acknowledgements}

C.S. gratefully acknowledges discussions with H. Fellermann, R. Everaers,
P.-A. Monnard, M. Hanczyc, and S. Rasmussen.

\part*{Appendix\label{sec:Appendix-A}}

\section*{Implementation details}

When Newtons 3rd law is not applied to bonded interactions, LAMMPS
has an bond interaction table for each bead listing the other beads
it is bonded to and the type of the bond. Similar angular and dihedral
interaction tables exists for each bead. LAMMPS also has a neighbor
structure where bonded neighbors, next nearest neighbors, and third
nearest neighbors are stored. This information is derived from the
bonding structure, and used to enable or disable non-bonded interactions
between beads connected by up to three bonds.

Initially when LAMMPS reads the control file to set up a simulation,
the dynamic bonding fix is called to parse the entire set of rules
such as those in fig. \ref{fig:Fix-DNA-dynamics}. The rules and their
parameters are sanity checked and stored internally in the fix. When
the simulation is initialized, the dynamic bonding framework starts
by having each simulation domain count how many bonds of each type
each reactive bead has. 

Then at a specified frequency the code does:
\begin{enumerate}
\item Communication. Forward communication of ghost particle positions to
neighboring nodes and the table of bond counts. This is required for
testing distances and for applying maximum rules.
\item Creation nomination. Each reactive bead can nominate a single preferred
bonding partner. The search for partners is performed over all beads
in the reactive group and each creation rule is tested in succession.
The test of rules is done in the order they are specified, and if
more than one rule match the same bead pair, the last matching rule
will apply. The search is over all non-bonded beads and optionally
over beads 2 or 3 bonds away from the current bead. For each bead
pair and creation rule, their types are tested and if they within
the maximum reaction distance. Beads that already have the maximal
number of bonds of the type, that would be produced by the current
rule are discarded. Of all the potential bonding partners, the closest
partner in the same simulation domain (if any) is nominated for bonding.
\item Bond breakage nomination. Each reactive bead can nominate a single
preferred partner to break an existing a bond to. The search for partners
is performed over all beads in the reactive group and each bond break
rule is tested in succession. The test of rules is done in the order
they are specified, and if more than one rule match the same bead
pair, the last matching rule applies. For each bead pair and bond
breakage rule, it is tested if the bond between them has the specified
type, and if they are further apart than the minimum bond breakage
distance. Of all the potential bond breakage partners, the partner
most distant in the same simulation domain (if any) is nominated for
bond breakage. Bond conversion is internally represented as a bond
pair that nominates each other for a bond breakage and creation of
the new bond. Hence bond conversion over rules both bond breakage
and creation in case they occur simultaneously.
\item Communication. The nominated partners are distributed to and aggregated
across neighboring simulation domains and the closest partner is chosen
for creation and the most distant partner is chosen for bond breakage.
Information about which rule lead the nomination of each partner are
also distributed along with a random number for stochastic bond breakage
and a random number for stochastic bond creation.
\item Bond breakage. If any killbond rules are defined, all beads check,
if they are part of a bond longer than the cut-off distance, and if
that is the case then the bond is marked for removal. If two bonds
nominate each other as bond breakage partners, then bond breakage
is attempted. Each bead contributes a uniform random number for bond
breakage, these are averaged and compared to the specified bond breakage
probability. In case the random number is smaller than the probability,
the bond is marked for removal. This ensures that beads on different
simulation domains makes the same random choice. When bonds are marked
for removal the bond type in the corresponding entry in the bond interaction
tables is set to -1. If a maximum rule applies to that particular
bond and bead type, the table of bead functionalities is also updated.
The outdated neighbor structure is retained.
\item Removing angular and dihedral interactions. To ensure parallelism,
each reactive bead is alone responsible for all its angular and dihedral
interactions. If a bond has been broken in its local neighborhood,
the bead has to remove any angular and dihedral interactions involving
that bond. This is done by generating all non-cyclic paths of length
three and four either starting at or crossing the present bead using
the outdated neighbor structure (which still contains the broken bonds).
The beads checks each path for bond breakage events (using the bond
interaction tables, which shows if a bond has been marked for breakage).
If a path involves a broken bond, then the bead removes the corresponding
entry in its angular and dihedral interaction tables, if they exist.
\item The LAMMPS neighbor structure is updated, and the broken bond entries
are removed from the bond interaction tables. If no bonds are to be
created, we can jump directly to 10.
\item Bond creation. If two bonds nominate each other as bond creation partners,
then an attempt is made at creating the bond. Each bead contributes
a uniform random number for bond creation, these are averaged and
compared to the specified bond creation probability. Again this ensures
the same random choice for beads residing in different simulation
domains. The new bond is added to the bond interaction table for the
bead. The neighbor structure is also updated. If a maximum rule applies
to the bond and bead type, the table of bead functionalities is also
updated.
\item Creating angular and dihedral interactions. Again each reactive bead
is responsible for determining if a bond was created in their local
neighborhood. This is done the same way as angular and dihedral interactions
are removed. Since the neighbor structure now contains the new bonds,
we can generate non-cyclic paths of length three and four starting
at or crossing the present bead using the updated neighbor structure
(which now contains the new bonds). Each path is checked for bond
creation events using the bond interaction tables. If the bead determines
that it is part of a new triplet or quartet of beads, then it compares
the bead types and directional bond types with all the angular and
dihedral creation rules. If a match is found, then the bead adds the
corresponding interaction to its interaction table.
\item Statistics. Distribution of statistics of the total number of bonds,
angles, dihedrals introduced and removed in the current time step.
\end{enumerate}
Since bond creation requires a distance check, the LAMMPS pair communication
distance should be at least the longest reaction distance, otherwise
bonds will only be created between bead pairs within the communication
distance from each other. Since the implementation also depends on
all beads knowing about all their bonded, angular, and dihedral interactions,
it will not work without Newtons 3rd law being disabled for bonded
interactions. This is also required for the implementation of directional
bonds. The dynamic bonding framework transparently handles symmetric
bonds, hence they are just special cases of directional bonds.

The dynamic bonding code is optimized to the situation, where the
density of reacting beads is so low that at most one bond breakage
and bond creation event is likely to occur per bead per time step.
 For instance, the match making algorithm does not attempt to make
matches between rejected partners, that could still be eligible for
bond breakage or bond creation rules. Nor does the match making algorithm
attempt to pick the most likely of multiple possible reaction path
ways. For instance, if multiple bond creation rules applies to a single
bead, then only the last nominated bond creation partner is stored.
Hence a creation rule with a low reaction probability can overwrite
the bonding partner nominated by a prior creation rule with much higher
reaction probability. In this case, the high probability reaction
will never happen. Similar issues apply when multiple bond break rules
involve the same bead. Since the bond conversion rules are implemented
as bond deletion followed by bond creation, these can interfere with
both bond creation and bond breakage rules. killbond rules are completely
safe, since they are not implemented using the match making algorithm.
For the DNA model, none of these caveats apply.

\pagebreak


{\bf PROGRAM SUMMARY}

\begin{small}
\noindent
{\em Manuscript Title: LAMMPS Framework for Dynamic Bonding and an Application Modeling
DNA.}                                       \\
{\em Authors: Carsten Svaneborg}                                                \\
{\em Program Title: LAMMPS Framework for Directional Dynamic Bonding}                                          \\
{\em Journal Reference:}                                      \\
{\em Catalogue identifier:}                                   \\
{\em Licensing provisions: GPL}                                   \\
{\em Programming language: C++}                                   \\
{\em Computer: Single and multiple core servers}                                               \\
{\em Operating system: Linux/Unix/Windows}                                       \\
{\em RAM: 1Gb }                                           \\
{\em Number of processors used: Single or Parallel}                              \\
{\em Keywords: Dynamic bonding, directional bonds, molecular dynamics.} \\
{\em Classification: 16.11 Polymers, 16.13 Condensed-phase Simulations}                                         \\
{\em Nature of problem: Simulating coarse-grain models capable of chemistry e.g. DNA hybridization dynamics.}\\
{\em Solution method: Extending LAMMPS to handle dynamic bonding and directional bonds.}\\
{\em Unusual features:Allows bonds to created and broken while angular and dihedral interactions are kept consistent.}\\
{\em Running time:hours to days}\\
    \\

\end{small}

\pagebreak

\end{document}